\documentclass[aps,pra,singlecolumn,superscriptaddress,longbibliography, 12pt]{revtex4-1}%
\usepackage{graphicx}
\usepackage{float}
\usepackage{dcolumn}
\usepackage{bm,color}
\usepackage{amsmath,amssymb,dsfont,amstext,amsfonts}
\usepackage{extarrows}
\usepackage[colorlinks=true,linkcolor=blue,urlcolor=blue,citecolor=blue]{hyperref}
\usepackage{xcolor}
\usepackage{wasysym}
\usepackage{mathtools}
\usepackage{bbold}

\usepackage[normalem]{ulem} 
\usepackage{color}

\def\mod{\:\mathrm{mod}\:}
\def\bs#1{\boldsymbol{#1}}

\begin{document}
\title{Observation of non-Abelian topological semimetals and their phase transitions}

\author
{Bin Jiang $^{1\ast}$, Adrien Bouhon$^{2, \ast, \dagger}$, Zhi-Kang Lin$^{1,\ast}$, \\
Xiaoxi Zhou$^{1}$, Bo Hou$^{1}$, Feng Li$^{1,5}$, Robert-Jan Slager$^{3, \star \dagger }$, Jian-Hua Jiang$^{1, 6 \star \dagger}$\\\vspace{2em}
\normalsize{$^{1}$School of Physical Science and Technology and }\\
\normalsize{Collaborative Innovation Center of Suzhou Nano Science and Technology,}\\
\normalsize{Soochow University, 1 Shizi Street, Suzhou 215006, China}\\
\normalsize{$^{2}$Nordic Institute for Theoretical Physics (NORDITA),}\\
\normalsize{Stockholm, Sweden}\\
\normalsize{$^{3}$TCM Group, Cavendish Laboratory, University of Cambridge,}\\
\normalsize{J. J. Thomson Avenue, Cambridge CB3 0HE, United Kingdom}\\
\normalsize{$^{5}$ School of Physics, Beijing Institute of Technology, Beijing 100081, China}\\
\normalsize{$^{6}$ Key Lab of Advanced Optical Manufacturing Technologies of Jiangsu Province}\\
\normalsize{$\&$ Key Lab of Modern Optical Technologies of Ministry of Education, Soochow University, Suzhou 215006, China.}\\
\normalsize{$^\dagger$To whom correspondence should be addressed:}\\
\normalsize{adrien.bouhon@su.se (AB), rjs269@cam.ac.uk (RJS) and jianhuajiang@suda.edu.cn (JHJ).}\\
\normalsize{$\ast,\star$ These authors contributed equally.}
}

\date{\today}

\begin{abstract}
{\bf
Topological phases of matter lie at the heart of physics, connecting elegant mathematical principles to real materials that are believed to shape future electronic and quantum computing technologies. To date, studies in this discipline have almost exclusively been restricted to single-gap band topology because of the Fermi-Dirac filling effect. Here, we theoretically analyze and experimentally confirm a novel class of multi-gap topological phases, which we will refer to as  non-Abelian topological semimetals, on kagome geometries. These unprecedented forms of matter depend on the notion of Euler class and frame charges which arise due to non-Abelian charge conversion processes when band nodes of different gaps are braided along each other in momentum space. We identify such exotic phenomena in acoustic metamaterials and uncover a rich topological phase diagram induced by the creation, braiding and recombination of band nodes. Using pump-probe measurements, we verify the non-Abelian charge conversion processes where topological charges of nodes are transferred from one gap to another. Moreover, in such processes, we discover symmetry-enforced intermediate phases featuring triply-degenerate band nodes with unique dispersions that are directly linked to the multi-gap topological invariants. Furthermore, we confirm that edge states can faithfully characterize the multi-gap topological phase diagram. Our study unveils a new regime of topological phases where multi-gap topology and non-Abelian charges of band nodes play a crucial role in understanding semimetals with inter-connected multiple bands.}
\end{abstract}
\maketitle


With the advent of topological phases, the past few years have seen rapid progresses in classifying topological states of matter. Upon analyzing the manner in which bands transform at high-symmetry momenta, one can deduce compatibility relations \cite{Clas3} conveying how many different classes of gapped band structures exist on a lattice geometry with a given set of symmetries (see Fig. 1A for schematic illustration). Moreover, by comparing to real space, efficient schemes were proposed that can discern which of these classes can be deemed topological by absence of an atomic limit \cite{Clas4, Clas5}.

Here, we consider an exotic type of topological phases beyond the above paradigms that, instead, depend on topological charge conversion processes when band nodes are braided with respect to each other in momentum space or recombined over the Brillouin zone. The braiding of band nodes is in some sense the reciprocal space analog of the non-Abelian braiding of particles in real space and is manifested by a new topological invariant, the Euler class, $\xi$~\cite{Wu1273,BJY_nielsen, bouhon2019nonabelian}.
Considering a kagome model and its realization in acoustic metamaterials, we uncover a rich interplay between these concepts and crystalline symmetries~\cite{Clas1,Clas2,Wi2, Chenprb2012, Codefects2,Cornfeld_2021} and further experimentally confirm the above physics in acoustic metamaterials by tuning their geometry. The rich phase diagram features unavoidable intermediate phases featuring band degeneracies at high symmetry point with unconventional dispersion features that have a topological origin that is not captured by representation theory alone \cite{yang2014classification, Bradlynaaf5037}. In addition, we reveal the edge physics associated with the non-Abelian band braiding within the above context. 


\begin{figure*}
	\centering\includegraphics[width=1\linewidth]{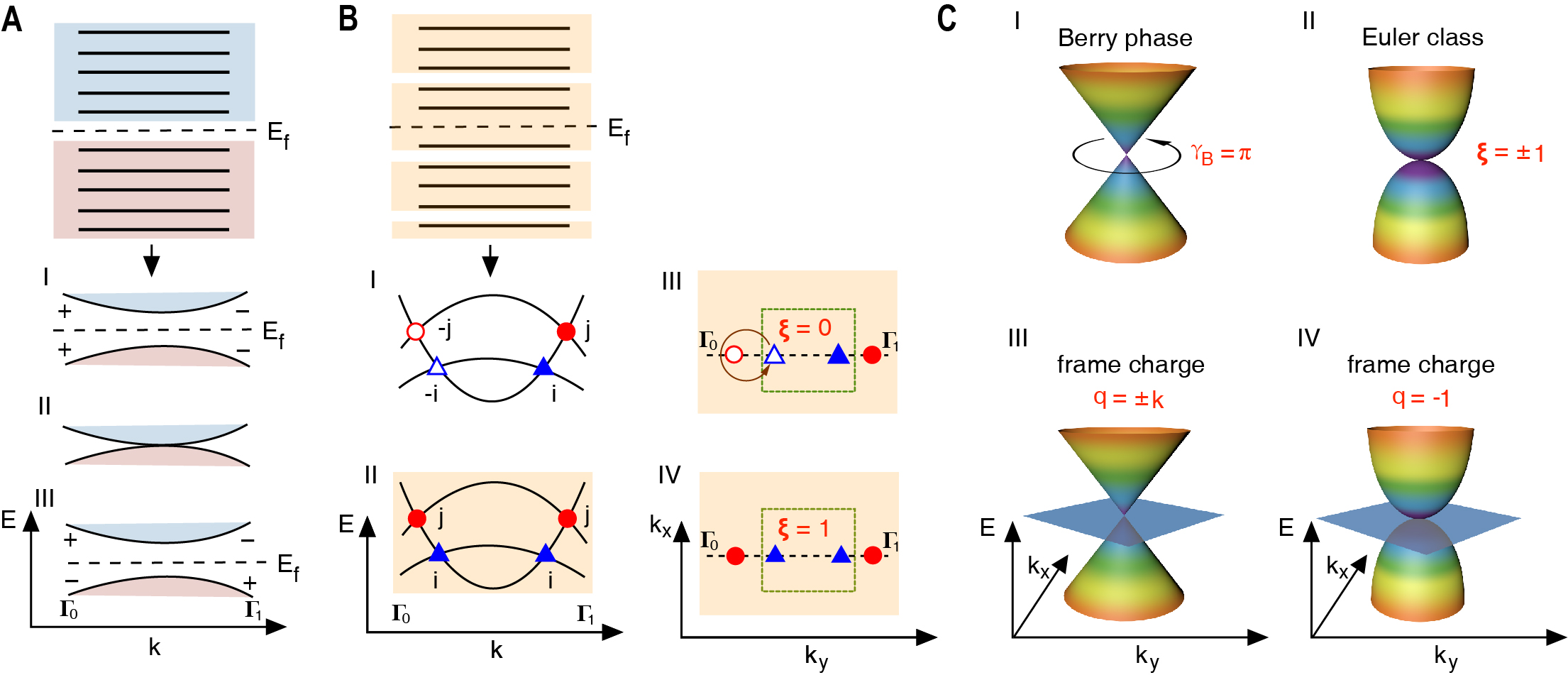}[h]
	\caption{{\bf Conventional topological insulators (TIs) versus multi-gap topology of Euler class.} {\bf A}, In the usual TI setting, the spectrum is divided in a conduction and valence band sector (red and blue, respectively). Starting from a trivial insulator (I), a TI (III) can be obtained from a band inversion (II), which changes the character of the band when going from one high-symmetry momentum $\Gamma_0$ to another $\Gamma_1$. This change is schematically denoted as $\pm$, although this can be a general property of the irreducible band representation at the high-symmetry momenta. {\bf B}, In the Euler case, one considers a refined partitioning of bands, for example a subblock that has two band gaps. Starting from a trivial gapped system, band nodes can be induced in both gaps (I). These nodes are oppositely charged (open/closed symbols) within a gap and behave as quaternions $\pm i$ and $\pm j$ with respect to the charges in the other gap (red circles vs. blue triangles). As a result, braiding a band node of the first gap around one in the second gap (III) along the brown trajectory changes its charge by a factor of $-1$, resulting in a spectrum exhibiting nodes with similarly-valued frame charges (II). The Euler class $\xi$ integrated over Brillouin zone patches (green boxes) then quantifies the stability of the nodes (III and IV). {\bf C}, (I) A Dirac cone with a Berry phase $\gamma_B=\pi$. (II) A quadratic band touching point stabilized by the patch Euler class $\xi=\pm 1$. (III) A linearly-dispersed triply-degenerate point with a topological frame charge $q=\pm k$. (IV) A parabolically-dispersed triply-degenerate point with a topological frame charge $q=-1$. The latter can intuitively understood as the combination of a double $i,j$ charge with a third band.}
	\label{Fig1}
\end{figure*}

\noindent {\bf Euler class and multi-gap topology}\\
While the study of topological materials and their interplay has resulted in a thriving field on both the  theoretical and experimental front, Euler class provides a new paradigm of topological phases. Akin to Chern numbers, systems that feature combined twofold rotation and time reversal symmetry, $C_2\cal{T}$, or combined parity and time reversal symmetry, $\cal{PT}$, can be described by a real Hamiltonian that can exhibit this topological invariant~\cite{Wu1273,BJY_nielsen, bouhon2019nonabelian}. From a more physical perspective, these new forms of topology can be best understood as being the consequence of a perspective that takes into account multiple band gap conditions \cite{bouhonGeometric2020}.
For instance, one can revert to a minimal Euler model that arises in a three-band system featuring two gapless bands and a gapped band, as detailed in Fig.~\ref{Fig1}. The Euler invariant manifests itself by the presence of a number of $2\xi$ irremovable band nodes between the two gapless bands. Conventionally, semi-metals host nodes with opposite topological charges that can be pairwise annihilated as the consequence of the celebrated Nielsen-Ninomiya theorem. In the Euler case, however, the nodes have an additional frame charge that is the {\it same} for the irremovable nodes.  The according inability to annihilate the band nodes reveals the topological obstruction that the topological Euler number $\xi$ imposes.


The above counter-intuitive possibility is intimately tied to both the real-valued form of the Hamiltonian and the concept of orientability.
Namely, the reality condition ensures that the set of Eigenstates can be interpreted as an oriented orthonormal frame, or {\it dreibein} when restricting to three-band systems as in this work. Decomposing the 3D rotation matrix along paths encircling stable nodes we can then directly compute the acquired angle of the monopole, or frame charge, reflecting the geometry of SO(3) rotations \cite{Sjoeqvist_2004,bouhon2019nonabelian}. Then, as a component of the spectral decomposition of the Hamiltonian, this frame is only defined up to a sign for each Eigenstate. Hence, the  frame charges of the nodes are related to the flag variety $\mathsf{SO}(3)/D_2$
\cite{Wu1273,bouhon2019nonabelian}. 
As a consequence, the band nodes correspond to $\pi$-disclinations in a biaxial nematic liquid crystals~\cite{Kamienrmp, Prx2016, volovik2018investigation} and the charges $q$ associated to nodal points within three bands take values in the conjugacy classes $\pm q$ of the quaternion group, i.e.~with $q\in \{1,i,j,k,-1\}$. Specifically, in our case $\pm i$ and $\pm j$ characterize single nodes in each of the respective gaps, $\pm k$ represents a pair of nodes in each gap, and $-1$($=i^2=j^2$) captures the stability of two nodes of the same gap \cite{Wu1273,Tiwari}.  

Using the fact that frame charges behave as quaternions, we then directly uncover a geometric relation between the braiding process involving the two gaps \cite{Wu1273,BJY_nielsen, bouhon2019nonabelian}, as explained in Fig.~1B, and the Euler invariant. Specifically, one can start from a trivially gapped system and create pairs of band nodes in the two gaps having charges $\pm i$ and $\pm j$, respectively. Each gap hosts a pair of oppositely-valued band node charges, which may be gapped out in a reverse process to arrive back at the trivial system (i.e., Euler class $\xi=0$). However, braiding a node residing in the first band gap, say $-i$, along a node in the second gap, say $-j$, produces in each gap a pair of nodes that cannot be annihilated upon collision (non-zero Euler class), which can effectively be phrased in terms of the conversion of the non-Abelian charges of the braided nodes rendering, e.g., the combination $\{+i,+i,+j,+j\}$ \cite{Wu1273,BJY_nielsen, bouhon2019nonabelian}, see Fig.~\ref{Fig1}. We note that an additional trivial band can thus trivialize the Euler model by the reverse braid process, showing that this enigmatic topology is formally a form a fragile topology, albeit rather distinct from the symmetry eigenvalue indicated variant \cite{Ft1,bouhon2019wilson, song2019fragile,Peri797, Song794}. 

In the above we have implicitly applied the concept of \textit{patch Euler class}, defined for a patch in the Brillouin zone that contains the nodes of pairs of bands isolated from all other bands by an energy gap \cite{bouhon2019nonabelian}. The patch Euler class is thus computed independently of the potential presence of nodes connecting the two bands under consideration with other bands \textit{outside} the patch.


\noindent {\bf Dirac Strings}\\
The concrete setup that we will present shortly, however, takes the above insight to a next level and features a profound role of so-called Dirac strings in one band gap that affect the charges in the other. Dirac strings naturally appear when considering smooth representations of real eigenvectors, marking the necessary discontinuity of the gauge sign upon the creation of any pair of nodes \cite{BJY_nielsen}. They are the 2D analogue within the real context of the Dirac strings of 3D Weyl semimetals within the complex context. It can be shown that by specifying a relative sign as the charge of each node within a gap (which we represent below through open/filled symbols for opposite charges and with distinct symbols for different gaps, e.g., ${\color{red}\boldsymbol{\bullet/\circ}}$, and ${\color{blue}\boldsymbol{\blacktriangle/}\triangle}$, for the first and second gap respectively) as well as the Dirac strings connecting such nodes, we obtain a picture that exhaustively captures the topological content of the band structure.
To this end only a few rules have to be taken into account in order to predict the outcomes of the braiding of nodes. Namely, ($i$) each DS must connect one pair of same-gap-nodes, i.e.~${\color{red}\boldsymbol{\bullet\sim\circ}}$, ${\color{blue}\boldsymbol{\blacktriangle\boldsymbol{\sim}}\triangle}$, ${\color{red}\boldsymbol{\bullet\sim\bullet}}$, etc., ($ii$) the charge of a node switches when crossing the DS of a neighboring (adjacent) gap, i.e.~$[{\color{red}\boldsymbol{\bullet}}{\color{blue}\rotatebox[origin=c]{90}{$\boldsymbol{\sim}$}}\times] \leftrightarrow [\times {\color{blue}\rotatebox[origin=c]{90}{$\boldsymbol{\sim}$}} {\color{red}\boldsymbol{\circ}}]$, $[{\color{blue}\boldsymbol{\blacktriangle}}{\color{red}\rotatebox[origin=c]{90}{$\boldsymbol{\sim}$}}\times] \leftrightarrow [\times {\color{red}\rotatebox[origin=c]{90}{$\boldsymbol{\sim}$}} {\color{blue}\boldsymbol{\triangle}}]$, ($iii$) any couple of Dirac strings in the same gap can be recombined through the permutation of the end nodes, i.e.~$[{\color{red}\boldsymbol{\bullet_1\boldsymbol{\sim}\circ_2}} \,{\scriptstyle\cup}\,
{\color{red}\boldsymbol{\bullet_3\boldsymbol{\sim}\circ_4}}] \leftrightarrow [{\color{red}\boldsymbol{\bullet_1\boldsymbol{\sim} \bullet_3}} \,{\scriptstyle\cup}\, 
{\color{red}\boldsymbol{\circ_2 \boldsymbol{\sim}\circ_4}}] \leftrightarrow [{\color{red}\boldsymbol{\bullet_1\boldsymbol{\sim} \circ_4 }} \,{\scriptstyle\cup}\,
{\color{red}\boldsymbol{\circ_2 \boldsymbol{\sim} \bullet_3}} ]$. We note that these rules are merely different representations of a consistent charge assignment, and refer to Supplementary Note~\ref{SM:DS} for a detailed account on deforming DS configurations. 


\noindent {\bf Interplay with kagome geometry}\\
The above view relating Euler forms and multi-gap topologies can be made rigorous using homotopy theory \cite{bouhonGeometric2020}, culminating in more intricate insulating Euler class topologies. However, it is all we need to put the physics presented here into perspective. Starting from a kagome geometry, we combine a symmetry-eigenvalue analysis together with the above new topological concepts to unravel topological transitions between a rich variety of phases arising via a symmetry-constrained braiding of band nodes. These insights are then accompanied by direct signatures that reveal Euler class physics~\cite{Euler_drive}. 

We now turn to the specific model that will be the center stage to put the above ideas into play.
We depart from $s$-orbitals located at $(\boldsymbol{r}_A,\boldsymbol{r}_B,\boldsymbol{r}_C) = (\boldsymbol{a}_1/2,\boldsymbol{a}_2/2,-\boldsymbol{a}_1/2-\boldsymbol{a}_2/2)$ (see Fig.~2A) and include the NN and the NNN couplings to arrive at the following Hamiltonian,
\begin{equation}
    H(\boldsymbol{k}) = \left(
        \begin{array}{ccc}
            \epsilon_A & h_{a}(\boldsymbol{k}) &
            h_{b}(\boldsymbol{k})\\
            h_{a}(\boldsymbol{k}) & \epsilon_B & h_{c}(\boldsymbol{k})\\
            h_{b}(\boldsymbol{k}) & h_{c}(\boldsymbol{k}) & \epsilon_C
        \end{array}
    \right),
\end{equation}
written in the Bloch basis $\left\{\vert \varphi_i,\boldsymbol{k} \rangle\right\}_{i=A,B,C}$ using the Zak gauge (see Supplementary Note~\ref{sm:model})
, with 
\begin{equation}
\begin{array}{rcl}
    h_{a}(\boldsymbol{k}) &=&  t \cos(k_1/2+k_2/2) + t' \cos(k_1/2-k_2/2),\\
    h_{b}(\boldsymbol{k}) &=&  t \cos(k_2/2) + t' \cos(k_1+k_2/2),\\
    h_{c}(\boldsymbol{k}) &=& t \cos(k_1/2) + t' \cos(k_1/2+k_2).
\end{array}
\end{equation}
The wavevector is given over the Brillouin zone in units of the reciprocal lattice vectors, i.e., $\boldsymbol{k} = \frac{1}{2\pi} ( k_1 \boldsymbol{b}_1  + k_2 \boldsymbol{b}_2)$ with $k_1,k_2\in[-\pi,\pi]$ where $\{\boldsymbol{b}_1,\boldsymbol{b}_2\}$ are the reciprocal primitive vectors dual to $\{\boldsymbol{a}_1,\boldsymbol{a}_2\}$, and $t$ ($t'$) are the NN (NNN) couplings. In the following, we refer to the $\Gamma$ point as $(k_1,k_2) = (0,0)$, the $K=(4\pi/3,-2\pi/3)$ and $K^\prime=(2\pi/3,2\pi/3)$ points, and the $M=(\pi,-\pi)$, $M^\prime=(\pi,0)$ and $M^{\prime\prime}=(0,\pi)$ points.

\begin{figure*}
	\centering\includegraphics[width=0.9\linewidth]{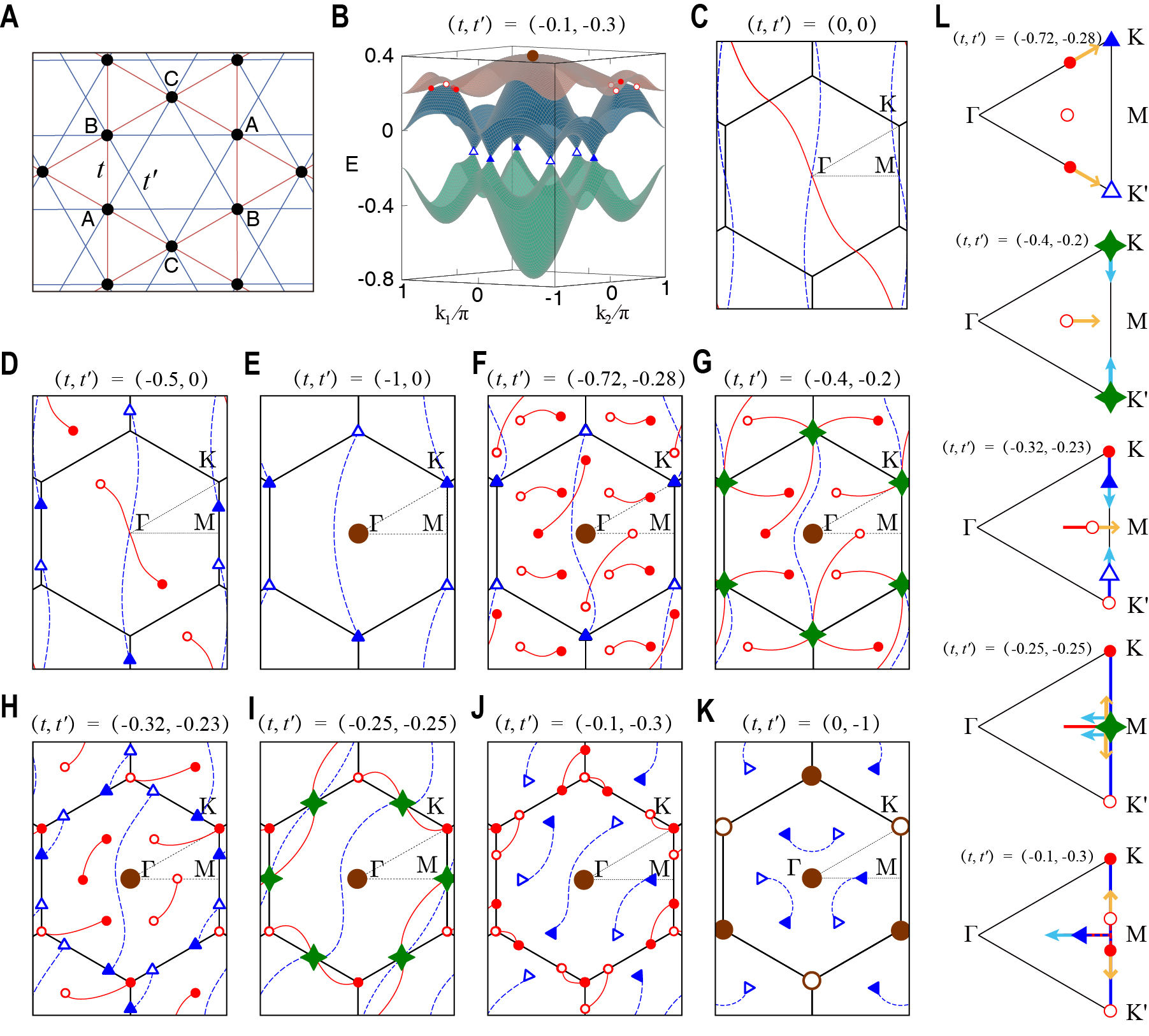}
\caption{{\bf Multi-gap topology in kagome models.} {\bf A}, Illustration of the kagome tight-binding model with the NN ($t$) and NNN ($t^\prime$) couplings in arbitrary energy units. {\bf B}, Band structure of the kagome model with vanishing onsite energy and $(t,t^\prime)=(-0.1,-0.3)$. {\bf C}, Taking $(\epsilon_A,\epsilon_B,\epsilon_C) = (1,0,-1)$ and $(t,t^\prime)=(0,0)$ already gives crossing Dirac strings (DS) in both gaps (blue for first gap, red for second gap). {\bf D} and {\bf E}, As the next step, turning off the onsite potentials and switching on the hopping terms induces band nodes. The nodes in the second gap (filled/empty red circles indicating $\pm$ topological charges) cross the DS in the first gap, forming a stable pair as the double node at at $\Gamma$ (brown circle) in ({\bf E}) which has finite patch Euler class $\xi=1$. Meanwhile, the first gap features nodes at $K$ points (triangles). {\bf E}-{\bf K}, Braiding process of the band nodes from one gap to another. Via the outlined steps depicted in {\bf L}, band nodes and DS strings evolve such that the degeneracy at $K$ in the first gap (blue triangles) is tuned into a double node configuration that has finite patch Euler class in the second gap (brown circles). 
} 
\label{kagome_bulk_charges}
\end{figure*}

\noindent {\bf Topological phase diagram}\\
We start from the simplest gapped phase and then introduce band nodes, to study the evolution and braiding of the nodes along a designed trajectory in parameter space. To this end, consider the gapped atomic insulating phase that breaks $C_6$ symmetry, by setting $(\epsilon_A,\epsilon_B,\epsilon_C) = (1,0,-1)$ and $(t,t')=(0,0)$. This system then has the band energies $E_1=-1$, $E_2=0$, and $E_3=1$, and the corresponding Bloch eigenstates $\vert \psi_1 ,\boldsymbol{k}\rangle = \vert \varphi_C,\boldsymbol{k} \rangle$, $\vert \psi_2 ,\boldsymbol{k}\rangle = \vert \varphi_B,\boldsymbol{k} \rangle$, and $\vert \psi_3 ,\boldsymbol{k}\rangle = \vert \varphi_A,\boldsymbol{k} \rangle$. In the following, we refer to gap I as the (partial) energy gap between bands 1 and 2, and gap II as the (partial) energy gap between bands 2 and 3.

The Berry phase factors for each band $\mathcal{B}_n$ ($n=1,2,3$) over a non-contractible path $l_{\boldsymbol{K}} = [\Gamma+\boldsymbol{K}\leftarrow\Gamma]$ where $\boldsymbol{K}$ is a reciprocal lattice vector, i.e., $\exp(\mathrm{i}\gamma_{\mathcal{B}_n}[l_{\boldsymbol{K}}]) = \langle \psi_{n},\Gamma+\boldsymbol{K} \vert \prod\limits_{\boldsymbol{k}\in l} P_{n}(\boldsymbol{k}) \vert  \psi_{n},\Gamma \rangle$ with the $n$-th band projector $P_{n}(\boldsymbol{k}) = \vert \psi_n, \boldsymbol{k} \rangle \langle \psi_n, \boldsymbol{k} \vert$, are then readily given by 
\begin{equation}
\begin{array}{rcl}
    \mathrm{e}^{\mathrm{i}\gamma_{\mathcal{B}_1}[l_{\boldsymbol{K}}]} &=& \langle \varphi_C,\Gamma+\boldsymbol{K}\vert \varphi_C,\Gamma \rangle = \mathrm{e}^{\mathrm{i} \boldsymbol{K}\cdot \boldsymbol{r}_C},\\
    \mathrm{e}^{\mathrm{i}\gamma_{\mathcal{B}_2}[l_{\boldsymbol{K}}]} &=& \langle \varphi_B,\Gamma+\boldsymbol{K}\vert \varphi_B,\Gamma \rangle = \mathrm{e}^{\mathrm{i} \boldsymbol{K}\cdot \boldsymbol{r}_B},\\
    \mathrm{e}^{\mathrm{i}\gamma_{\mathcal{B}_3}[l_{\boldsymbol{K}}]} &=& \langle \varphi_A,\Gamma+\boldsymbol{K}\vert \varphi_A,\Gamma \rangle = \mathrm{e}^{\mathrm{i} \boldsymbol{K}\cdot \boldsymbol{r}_A},
\end{array}
\end{equation}
where we have used the periodic gauge for the Wilson loop integration \cite{Wi2}. Berry phases on non-contractible path are also referred as Zak phases \cite{Zak2}. Taking $\boldsymbol{K} = \boldsymbol{b}_1,\boldsymbol{b}_2$ and writing the vector $\boldsymbol{\gamma}_{\mathcal{B}_n} = (\gamma_{\mathcal{B}_n}[l_{\boldsymbol{b}_1}] , \gamma_{\mathcal{B}_n}[l_{\boldsymbol{b}_2}])$, we find (modulo $2\pi$)
\begin{equation}
\begin{array}{rclrclrcl}
    \boldsymbol{\gamma}_{\mathcal{B}_1} &=& (\pi,\pi) ,&
    \boldsymbol{\gamma}_{\mathcal{B}_2} &=& (0,\pi) ,&
    \boldsymbol{\gamma}_{\mathcal{B}_3} &=& (\pi,0)  .
\end{array}
\end{equation}
While the total band bundle of the three bands is orientable, i.e., the above Berry phases sum to $0\mod2\pi$ in each direction, the `flag bundle' (where we consider each band separately) is non-orientable as indicated by the non-zero Berry phases~\cite{bouhonGeometric2020}. Each $\pi$-Berry phase indicates the presence of an odd-number of Dirac strings crossing perpendicularly the path of integration \cite{Ahn2018b}. The corresponding configuration of Dirac strings of this non-orientable atomic phase is shown in Fig.~2C where the Dirac string across gap I (DSI) is shown as a blue dashed line, and the Dirac string across gap II (DSII) as a red full line. 
It is remarkable that, as long as $C_2T$ is preserved, the system cannot exhibit a fully trivial and orientable band structure, as will become clear below.

We now deform the model to obtain the simplest kagome model while restoring the $C_6$ symmetry. This is done by setting $t = -1$ and  $\epsilon_A,\epsilon_B,\epsilon_C\rightarrow 0$. During the deformation there is a band inversion in each gap, thus creating two simple nodes of opposite charges in each gap, see Fig.~2D where $(\epsilon_A,\epsilon_B,\epsilon_C) = (0.5,0,-0.5)$ and $(t,t')=(-0.5,0)$. The final configuration shown in Fig.~2E has two simple nodal points at the K and $K^\prime$ points in gap I (small blue triangles) that are connected by DSI (a Dirac string in gap I), and a double nodal point at $\Gamma$ in gap II (large brown circle). DSII is fully resorbed, and DSI has been chosen as to avoid the nodes of gap II. While these degeneracies, at $K^\prime$s and $\Gamma$, correspond to the two-dimensional irreducible representations $K_3$ and $\Gamma_5$ of the elementary band representation of $s$-atomic orbitals at Wyckoff position $3c$ of layer group L80 \cite{ITCA,ITCE}, our analysis reveals the origin of stability of the double node at $\Gamma$ which is independent of the crystalline symmetries. Indeed, in order to go from Fig.~2D to Fig.~2E, DSI must cross one node of gap II and flip its charge (cf.~rule ($ii$) above, and the left column of Fig.~\ref{fig:DS_recombination} in Supplementary Note \ref{SM:DS}). As a consequence, the two nodes in gap II in the vicinity of $\Gamma$ have {\it equal charge} leading to a patch Euler class of $\xi=\pm1$ (equivalently, a frame charge of $-1$), thus indicating that they cannot be annihilated. A robust associated feature of the nontrivial patch Euler class of the stable double node degeneracy $\Gamma_5$ (at $\Gamma$), is its quadratic dispersion in all directions, as opposed to the linear dispersion (``Dirac cone'') around the double degeneracy $K_3$ (at $K$), see Fig.~1C. We emphasize that the pair of nodes at $\Gamma$ are stable upon the breaking of point group $D_{6h}$ of the kagome lattice (L80) as long as the $C_{2}T$ symmetry is preserved as predicted by the Euler class (frame charge) obstruction.

\noindent {\bf Symmetry-constrained braiding of band nodes}\\
We come to a main feature that concerns the transfer of the double degeneracies from one gap to the other. We focus here on the $K$ point and discuss the analog for the $\Gamma$ point in the Supplementary Note~\ref{SM:fullbraiding} (see Fig.\ref{BS_SIs_G}). These processes thrive on the Kagome symmetries that necessitate triple degeneracies at $K$, $M$ (see below), and $\Gamma$ in the charge conversion processes. Namely, we show that the braiding of nodes effectively acts as a pump of simple nodes (at $K$) and double nodes (at $\Gamma$) from one gap to the other other, mediated by triply-degenerate points. 

The process is illustrated through Figs.~2E-K, while the irreducible representation (IRREP) content of the band structures is detailed in Fig.~\ref{BS_SIs} of Supplementary Note~\ref{SM:fullbraiding}. First, two sextets of nodes in gap II (red circles) are generated from $\Gamma$, i.e., there are two nodes on each of the three $M$-$\Gamma$-$M$ and $K$-$\Gamma$-$K^\prime$ lines (Fig.~2F). These extra nodes must appear together upon a band inversion at $\Gamma$ as a result of the crystalline symmetries ($C_{6v} \subset D_{6h}$) of the system, see Figs.~\ref{BS_SIs}{a-b} that show the reordering of IRREPs and the resulting symmetry protected nodes inside the Brillouin zone. Importantly, the charge configuration precisely captures the fact that the nodes can be annihilated in the reversed process. In particular, the Euler class (frame charge) of the nodes at $\Gamma$ is preserved. Then, the $K$-$\Gamma$-$K^\prime$-nodes reach the $K$ points first, thus realizing two triply-degenerate points (represented by the green  stars in Fig.~2G) which is the necessary step upon a band inversion at $K$ ($K^\prime$) (see Figs.~\ref{BS_SIs}{\bf a-c}). Each triply-degenerate point then gives birth to one simple node in gap II at the $K$ ($K^\prime$) point, and three simple nodes in gap I (blue triangles) on the $K$-M-$K^\prime$ lines (Fig.~2H). This follows by symmetry from the band inversion at $K$ ($K^\prime$) (see Fig.~\ref{BS_SIs}{\bf d}). We note that the step from Fig.~2G to H requires the recombination of Dirac strings (cf.~rule $(iii)$). Also, defining a loop around the $K$ ($K^\prime$) encircling the $K$-$\Gamma$-$K^\prime$-nodes in Fig.~2F, the frame charge of $\pm k$ is conserved through Fig.~2G and Fig.~2H. Then, the $\Gamma$-$M$-$\Gamma$-nodes in gap II and the $K$-$M$-$K^\prime$-nodes in gap I join at the $M$ points, realizing three triply-degenerate points (Fig.~2I). Contrary to the previous case, these triple points are not imposed by symmetry (Figs.~\ref{BS_SIs}{\bf d-e}) but are rather model specific (see below). Each triply-degenerate point then gives birth to two $\Gamma$-$M$-$\Gamma$-nodes in gap I and two $K$-$M$-$K^\prime$-nodes in gap II (Fig.~2J). This results by symmetry from the band inversion at $M$ (Figs.~\ref{BS_SIs}{\bf f}). Also, the transition of charges through Fig.~2H-I-J requires the flip of the charges of the nodes in gap II after crossing the DSI (rule ($ii$)). The full band structure at this step of the process (Fig.2J) is shown in Fig.~2B. The $K$-$M$-$K^\prime$-nodes in gap II then reach the $K$ and $K^{\prime}$ points, where two double nodes are formed, each with a nontrivial patch Euler class (frame charge $-1$) (Fig.~2K). We note that the step Fig.~2J-K requires the recombination of DSI (rule ($iii$)).



\noindent {\bf Key bulk features}\\
Given the complexity of the conversion processes, we recapitulate some of the essential features lying at the core of the symmetry-constrained braiding for which symmetry and topology must be married in a new way. In our framework, we already mentioned the quadratic dispersion in all directions of the double nodes, e.g. at $\Gamma$ (Fig.~\ref{BS_SIs}{\bf a}) and at $K^{(\prime)}$ (Fig.~\ref{BS_SIs}{\bf g}). This is a robust feature of the non-trivial patch Euler class (frame charge $-1$) of two stable nodes (see Fig.~1C) that sit on top of each-other as a result of a crystalline symmetry (i.e.~$C_6$ at $\Gamma$, and $C_3$ at $K^{(\prime)}$) \cite{lenggenhager2020multiband}. 

It is known that three-dimensional semimetals with $\mathcal{PT}$ symmetry and fourfold or sixfold rotational symmetry can host a double degeneracy at $\Gamma$ that gives rise to a triply degenerate node along the $k_z$-rotation axis \cite{PhysRevX.6.031003,Vanderbilt_triple,lenggenhager2020multiband}. This is also another important aspect of the bulk topology here for the $K^{(\prime)}$ (and $\Gamma$, see SM \ref{SM:fullbraiding}) point, that are unavoidable by symmetry upon the pumping of nodes from one gap to the other. On the contrary, the triply-degenerate point at the $M$, $M^\prime$ and $M^{\prime\prime}$ points only exist as long as $\boldsymbol{k}$-dependent same-sub-lattice-site hopping terms are discarded (i.e., $\sim c_{i,\boldsymbol{k}}^{\dagger}c_{i,\boldsymbol{k}}$, $i=A,B,C$). In other words, by discarding intra-sublattice-site hoppings the kagome model cannot be gapped, i.e., the three bands must remain connected through simple, double, and triple nodes.

Very interestingly, the frame charge associated to one triply-degenerate node reflects again the dispersion of the bands. A frame charge of $\pm k$ implies a triple node with two linear bands (at $K$ in Fig.~\ref{BS_SIs}{\bf c}), while a frame charge of $-1$ implies a parabolically-dispersed triply-degenerate point (at $\Gamma$ in Fig.~\ref{BS_SIs_G}{\bf b}), as shown in Fig.~1C.

\noindent {\bf Edge projections}\\ 
The above determination of bulk charges (i.e., non-Abelian frame charges, Dirac strings, and patch Euler class) also allows us to systematically obtain the Zak phases for different edge projections and predict the existence of edge branches in the different gaps. Indeed, whenever the Zak phase's integration path crosses an odd (even) number of Dirac strings, we get a $\pi$-($0$-)Zak phase. Zak has shown that in one-dimensional quantum systems (here the 1D slices of the 2D phase) the Zak phase, obtained from the projection to a group of bands isolated by an energy gap, predicts the location of the center of charge of the Bloch eigenstates (i.e.~the expectation value of the position operator) \cite{Zak2}. The bulk-edge correspondence linking the value of the Zak phase with the presence of an edge state follows from the conservation of charges and the incommensurability between the number of atomic orbitals, located at the sites that compose the finite 1D system, and the number of band-projected orbitals originating from the different gap-separated blocks of bands. Incommensurability, also often called \textit{charge anomaly} \cite{SSH}, typically happens when there is a mismatch between the locations of the atomic and the band orbitals. 

We consider the zigzag (see below) and armchair (in Supplementary Note~\ref{SM:edge}) edges, which are defined with respect to the honeycomb structure underlying the kagome lattice (i.e.~connecting the centers of every small triangle forming the kagome lattice), see the green lines in Fig.~3C and F. The integration path in momentum space for a given edge geometry is dictated by the redefinition of the Brillouin zone within the coordinates $(k_{\parallel},k_{\perp})$ dual to the edge axes $(x_{\parallel},x_{\perp})$, where $x_{\perp}$ is perpendicular to the edge and $k_{\parallel}$ remains a good quantum number thanks to the translational invariance along the edge. Then, $k_{\perp}$ is the integration axis for the Zak phase. The edge-oriented Brillouin zones are given as the insets of Fig.~\ref{fig:armchair_edges} for the zigzag edges, and of Fig.~\ref{fig:figs6} for the armchair edges, where $k_{\perp}$ is the vertical axis in both cases. Crucially, we find that there is an edge state, given an edge projection and an energy gap, whenever the Zak phase is zero. This matches with the fact that the atomic orbitals are located on the boundary of the unit cell (i.e.~Wyckoff position $3c$). In other words, for the given geometry, the $0$-Zak phase indicates a charge anomaly and hence the existence of a topological edge state.  

From the known bulk charge configuration, we then expect observable conversions of the edge states through the braiding process. Inversely, the conversion of the edge states through the different steps of the phase diagram provides an {\it indirect} manifestation of the bulk braiding of nodes. We note in this regard recent work on one-dimensional gapped real three-band Hamiltonian realized in electric circuits \cite{guo2020experimental}, where the relation between Zak phases and non-Abelian charges has been observed. Within this work we consider genuine Euler class semimetallic materials, a topological phase that has no one-dimensional equivalent, and the according braiding process (which similarly requires two spatial dimensions) leading to these phases. Moreover, the discussed unavoidable bulk features, such as stability of nodes, the pumping effect and triple degeneracies upon braiding of nodes, are universal and applicable outside the tunable realm of metamaterials. Indeed, we emphasize that, while relations between Zak phases and edge states are well-defined, a general bulk-boundary correspondence in terms of the non-Abelian frame charges is yet to be formally understood. 

\begin{figure*}
	\centering\includegraphics[width=1\linewidth]{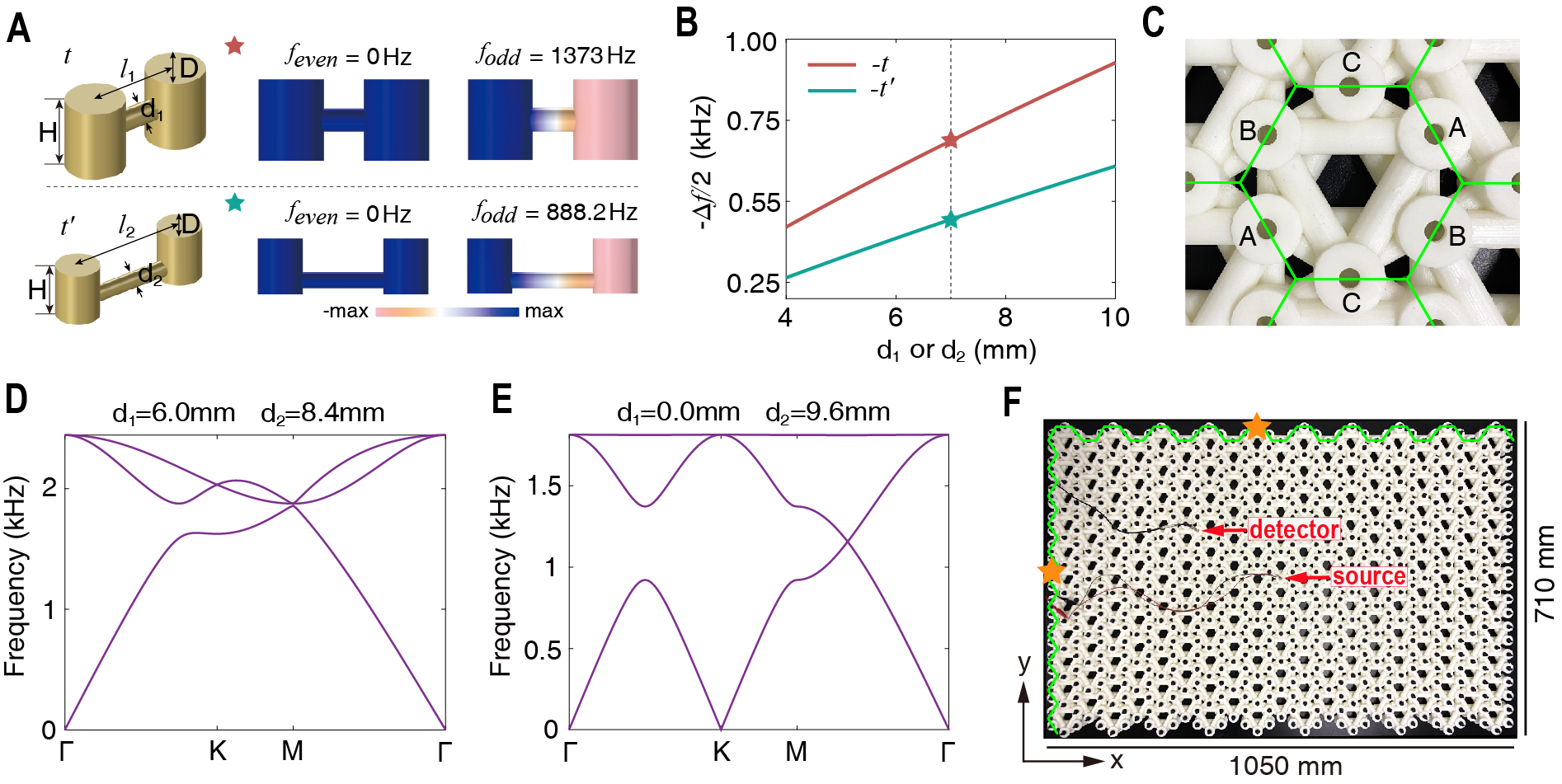}
	\caption{{\bf Acoustic kagome metamaterials and experimental set-ups.} {\bf A}, Cylindrical acoustic resonators coupled by a tube connecting them. Upper/lower panel: Illustration of the coupled resonators (left) and the resultant hybridized even and odd modes (right) for the NN/NNN coupling. $l_1=18\sqrt{3}$~mm, $l_2=54$~mm, $D=19.2$~mm, $H=24$~mm, and $d_1=d_2=7$~mm. {\bf B}, Coupling coefficients $t$ and $t^\prime$ vs. the diameters $d_1$ and $d_2$. The dashed line and the stars denote the case in A. {\bf C}, Photo of a fabricated acoustic metamaterial. Green lines depict the unit-cell boundaries (lattice constant $36\sqrt{3}$~mm). {\bf D} and {\bf E}, The acoustic band structures of the mematerials with ({\bf D}) $d_1=6.0$~mm and $d_2=8.4$~mm and ({\bf E}) $d_1=0.0$~mm and $d_2=9.6$~mm, respectively. {\bf F}, Photo of the experimental set-up for the measurements of the bulk and edge band structures. The upper and lower edges are armchair boundaries, while the left and right edges are zigzag boundaries. The acoustic source and the detector (a microphone) are connected to an Agilent network analyzer. The yellow stars label the positions of the source for the measurement of the zigzag and armchair edge dispersions.} 
\end{figure*}

\noindent {\bf Acoustic metamaterials}\\ 
To verify the above evolution and braiding of the band nodes, we created a physical realization of the kagome model using acoustic metametarials based on coupled acoustic resonators. Such acoustic metamaterials have been demonstrated as a highly-tunable and versatile platform for the study of topological phenomena~\cite{am1,am2,am3,am4,am5,am6,am7,am8,am9}. In our design, each acoustic resonator is a hollow cylinder with an air region of height $H$ and diameter $D$ encapsulated by a shell made of resin which confines the acoustic waves. These acoustic resonators play the roles of the sites in the kagome model. The nearest-neighbor (NN) and next-nearest-neighbor (NNN) couplings between the acoustic resonators are then achieved through the air tubes connecting the resonators with diameters $d_1$ and $d_2$, respectively. The acoustic metamaterial, with a lattice constant of $36\sqrt{3}$~mm, is fabricated using the 3D printing technology based on photosensitive resin (see more details in Supplementary Note~\ref{SM:construction}. As illustrated in Fig.~3A, two acoustic resonators connected with a tube support two hybridized modes (even and odd modes) with a splitting $\Delta f =f_{even}-f_{odd}$. The coupling between the two resonators is equal to $\Delta f/2$. We find that for both the NN and NNN couplings, their strength can be tuned at will through the diameters $d_1$ and $d_2$ (Fig.~3B). A photo of the unit-cell of the fabricated acoustic metamaterial is shown in Fig.~3C for $d_1=6$~mm and $d_2=8.4$~mm. The corresponding acoustic band structure is presented in Fig.~3D, showing the anticipated triple degeneracy at the M point (similar to the case in Fig.~2I). By tuning the diameters to $d_1=0$ and $d_2=9.6$~mm, the acoustic band structure with quadratic band touching at the $\Gamma$ and M points is realized (see Fig.~3E; similar to the case in Fig.~2K). Although the acoustic setup does not exactly match the tight-binding model, the acoustic metamaterial effectively reproduces all topological band features of the model, as shown below.

\begin{figure*}
	\centering\includegraphics[width=1\linewidth]{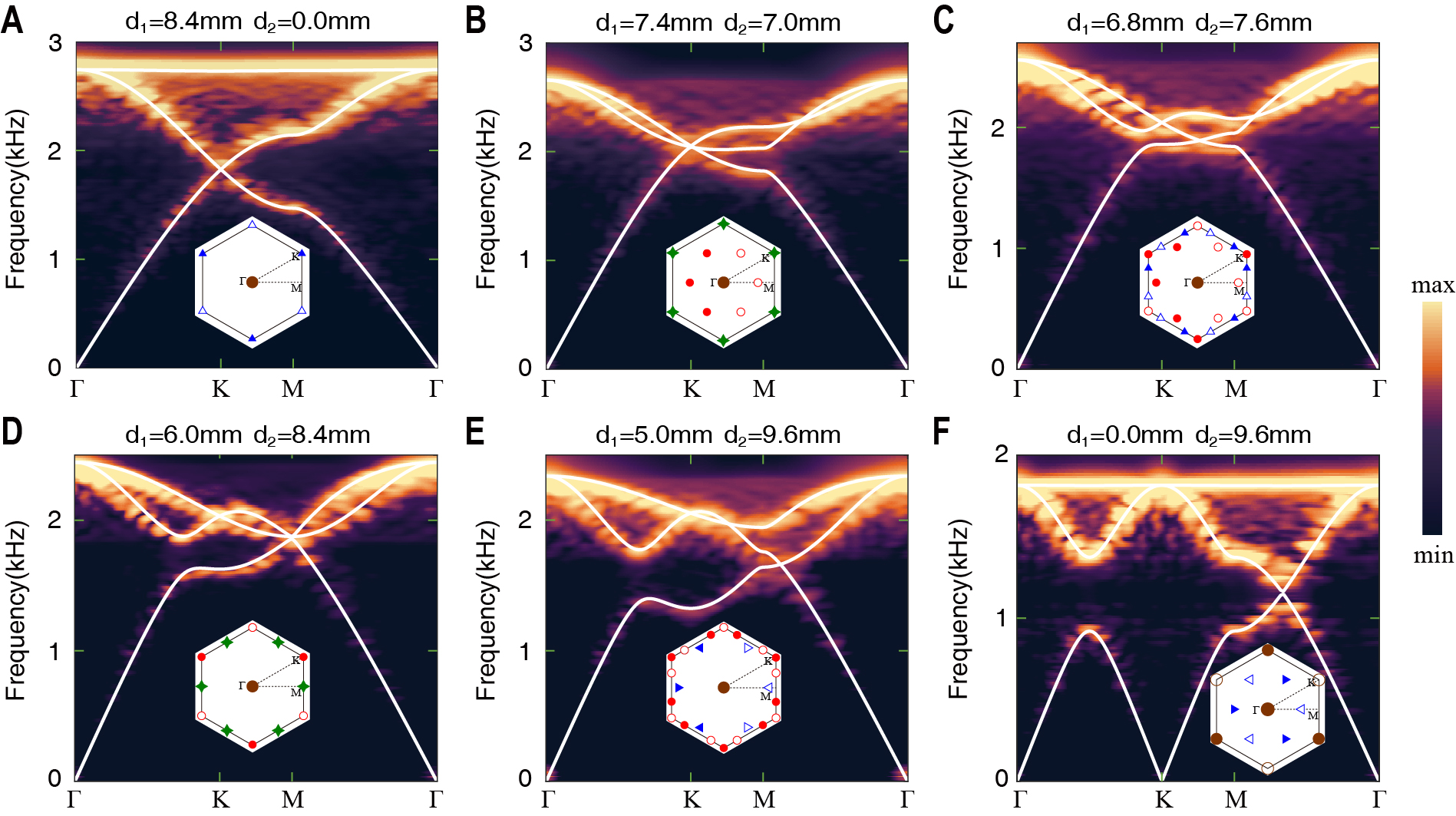}
	\caption{{\bf Acoustic multi-gap topology of the Euler class.} {\bf A}-{\bf F}, The acoustic bulk band structures measured for six kagome metamaterials along the evolution path illustrated in Figs.~2E, 2G-2K. The white lines represent the calculated bulk band structures for the six acoustic metamaterials using numerical methods. Insets show the distributions of the band nodes in gap I and II over the Brillouin zone for each case. Geometry parameters $d_1$ and $d_2$ are shown at the top of each figure. The convention of labeling the band nodes is the same as in Fig.~2. 
} 
\end{figure*}

\noindent {\bf Characterization of the bulk bands}\\
The experimental set-up for the measurement of the acoustic band structures is shown in Fig.~3F. The fabricated acoustic metamaterial has 200 unit-cells, making a rectangular geometry of $1050$~mm$\times 710$mm. The upper and left edge boundaries are closed (except at the corners) while the lower and right edge boundaries are open. Such a structure is optimized for the measurements of the bulk and edge bands (see Supplementary Note~\ref{SM:methods}). For the measurement of the bulk bands, an acoustic source is placed into a resonator at the center of the sample to excite the bulk Bloch waves at various frequencies from 5Hz to 3kHz. A probe microphone scans the acoustic pressure at all other resonators. By Fourier-transforming the detected acoustic pressure distributions at each excitation frequency, we obtain the acoustic bulk band structure for each of the six fabricated acoustic metamaterials (see Supplementary Note~\ref{SM:bulk} for details). Similarly, when measuring the edge bands, the acoustic source is placed at the center of the armchair or zigzag edge boundary (labeled by the yellow stars), while the acoustic pressure distributions along the edge boundaries are scanned at various excitation frequencies.

We first measure the bulk bands for the six acoustic metamaterials, similar (but not exactly identical) to the cases in Figs.~2E, 2G-2K along the braiding trajectory. As presented in Fig.~4, the measured acoustic band structures agree excellently with the calculated acoustic band structures. These band structures faithfully regenerate the band nodes evolution in Fig.~2. Specifically, in Fig.~4A ($d_1=8.4$~mm and $d_2=0$), the acoustic metamaterial exhibits the usual dispersion of the kagome lattice with two Dirac nodes at the K and K$^\prime$ points in gap I as well as a quadratic band touching at the $\Gamma$ point in gap II. Another known feature of the familiar kagome lattice is that the third band emerges as a flat band. In Fig.~4B, by tuning the diameters of the connecting tubes to $d_1=7.4$~mm and $d_2=7$~mm, the triple degeneracy emerges at the K and K$^\prime$ points. Meanwhile, the Dirac nodes are transferred to gap II in the K-$\Gamma$-K$^\prime$ lines. In Fig.~4C ($d_1=6.8$~mm and $d_2=7.6$~mm), the triple degenerate points split into three Dirac nodes: one at the K (or K$^\prime$) point in gap II and two on the K-M-K$^\prime$ lines in gap I. In Fig.~4D, by tuning the parameters to $d_1=6$~mm and $d_2=8.4$, the Dirac nodes in the K-$\Gamma$-K$^\prime$ lines in gap II and those in the K-M-K$^\prime$ lines in gap I coalesce to form the triple degeneracy at the three M points. Meanwhile, the Dirac nodes in gap II remains at the K and K$^\prime$ points. In Fig.~4E ($d_1=5$~mm and $d_2=9.6$~mm), the triple degeneracy splits into Dirac nodes again. But now, the Dirac nodes lying on the K-$\Gamma$-K$^\prime$ lines are transferred into gap I, while the Dirac nodes on the K-M-K$^\prime$ lines are transferred into gap II. The latter Dirac nodes eventually coalesce at the K and K$^\prime$ points to form quadratic band touching, as shown in Fig.~4F by setting $d_1=0$~mm and $d_2=9.6$~mm. Several remarks are in place: First, the whole process indicates that the degeneracy at the K point in gap I switches with the degeneracy on the $\Gamma$-$M$  and $\Gamma$-$K$ lines in gap II. By changing the charges of the nodes in gap II, the reciprocal braiding processes lead to a double node at the $K$ ($K^\prime$) point in gap II with nontrivial patch Euler class. However, since such braiding appears six times due to the $C_6$ symmetry, the total Euler class remains nontrivial, $\xi=1$. Second, from material science aspects, except the band structure in Fig.~3A, all other band structures, particularly the triple degeneracy in Figs.~3B and 3D and the multiple quadratic degeneracy in Fig.~3F, are unprecedented dispersion features. These exotic band structures reveal that kagome materials embrace rich physical properties that are attractive for fundamental science (e.g., kagome Mott insulators) and applications (e.g., Dirac materials with tunable bands).

\begin{figure*}
	\centering\includegraphics[width=1\linewidth]{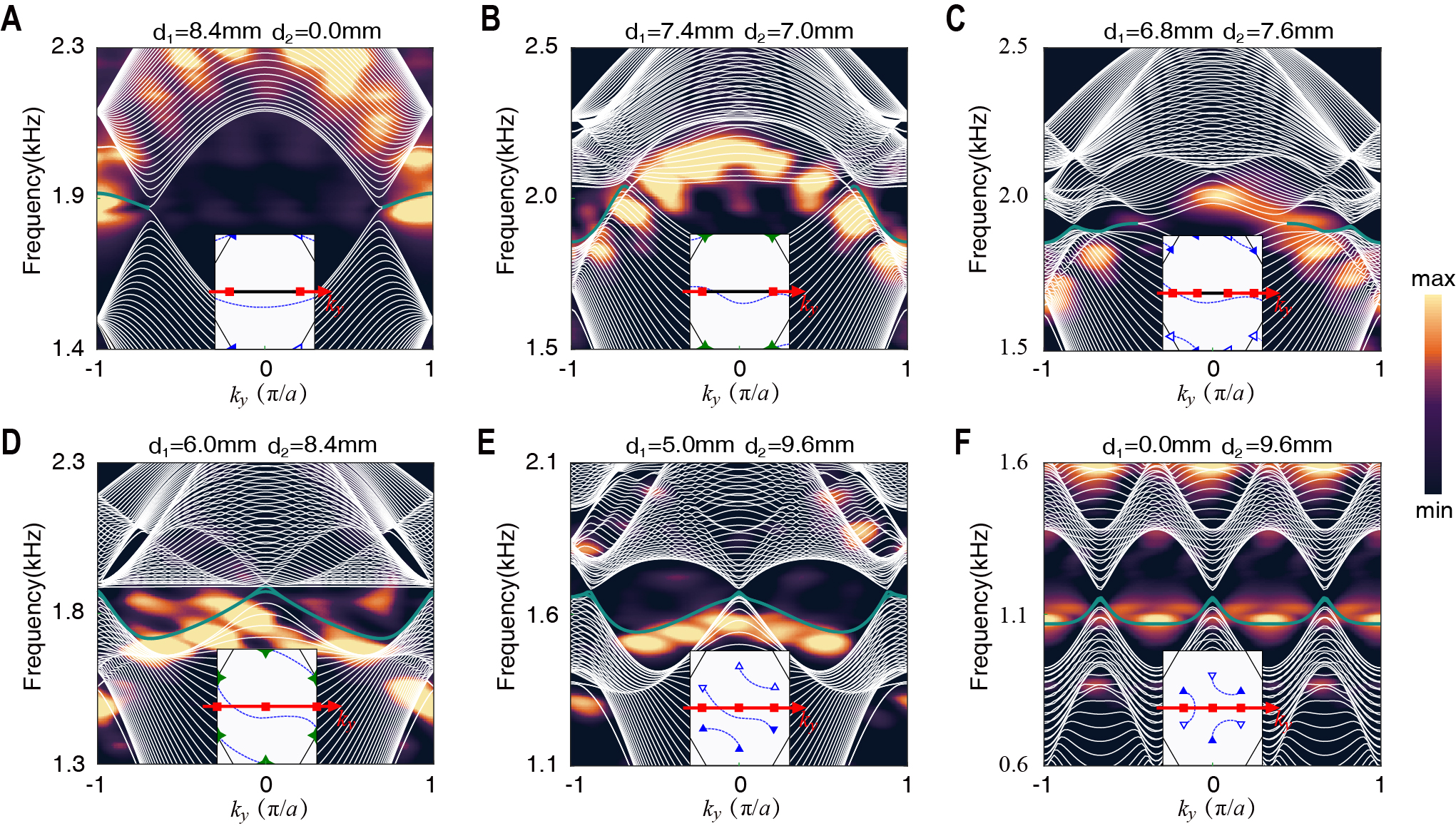}
	\caption{{\bf Bulk-edge correspondence in the kagome metamaterials.} {\bf A}-{\bf F}, Acoustic dispersions obtained from zigzag edge measurements. The lines represent the calculated band diagrams for metamaterial structures that finite in the $x$ direction but periodic in the $y$ direction. The inset of each figure depicts the location of the band nodes in the gap I as well as the Dirac strings. The projection of the band nodes on the $k_y$ axis is shown by the red squares. The red segments on the $k_y$ axis denote the regions with nontrivial Zak phase. Geometry parameters $d_1$ and $d_2$ are shown at the top of each figure. The convention of labeling the bulk band nodes is the same as in Fig.~2. 
} 
\label{fig:armchair_edges}
\end{figure*}

\noindent {\bf Measurement of the edge dispersions}\\
The dispersions of the edge states at the armchair and zigzag boundaries of the six acoustic metamaterials are measured for each acoustic metamaterial. In Fig.~5, we present the measured results for the zigzag boundaries, while the experimental data for the armchair boundaries are presented in the Supplementary Note~\ref{SM:edge}. Since gap II is often much smaller than gap I which hinders the study of edge states in gap II, we focus on the edge states in gap I in this work. Fig.~5A shows that the edge states emerge only at one side of the projection of a Dirac node. The zigzag edge states reflect the nontrivial Zak phase along the $\Gamma$-$K$ direction for each $k_y$. The results in Fig.~5A is consistent with the fact that each Dirac node changes the Zak phase from $\pi$ to 0 or vise versa. The edge band structure in Fig.~5B reflects that a triple degeneracy point also flips the Zak phase between $\pi$ and 0. The triple degenerate point can be regarded as composed of an odd number of Dirac nodes in both gaps I and II (as seen from Figs.~2F and 2G). Fig.~5C again reflects that the Dirac nodes (and the triple degenerate points) are the monopoles of the Berry gauge fields that quantize the Zak phases and therefore are the terminations of the Dirac strings. From Fig.~5D to Fig.~5F, pairs of the band nodes project into the same $k_y$ position. Therefore, the Zak phase does not change with $k_y$ and the edge states emerge at all $k_y$, as confirmed by consistent results from experiments and simulation. The above edge phenomena can be understood equivalently through the projection of the Dirac strings. The $k_y$ regions with an even number of projected Dirac strings correspond to 0 Zak phase, whereas the regions with an odd number of projected Dirac strings correspond to $\pi$ Zak phase. In our system, because the Wyckoff position is at the boundary of the unit-cell, the 0 Zak phase leads to topological edge states, while the regions with $\pi$ Zak phase has no edge states.

\noindent {\bf Conclusion and outlook}\\
We theoretically predict and experimentally observe  non-Abelian topological semimetals, a new type of topological phases with multi-gap topological invariants that depend on  Euler class and crystalline symmetries, in tunable kagome materials. Through the evolution of band nodes and its realization in acoustic metamaterials, we unveil the general rules for the conversion of band nodes in different gaps, leading to an exceptionally rich topological phase diagram. We find that the band evolution processes involve symmetry-enforced intermediate phases hosting doubly and triply degenerate points with unique dispersion features that are directly linked to their non-Abelian charges. Apart from these universal bulk features, we also reveal how the band evolution process is reflected at the edges through the Zak phase. The interplay between  multi-gap topology, the conversion of non-Abelian topological charges of nodes between adjacent band gaps, and crystalline symmetry sets a new benchmark in the study of topological phases of matter. Moreover, the verification of these unprecedented fundamental phenomena opens a new frontier of non-Abelian topological semimetals.

\noindent {\bf Author contributions}\\
A.~B. and R.-J.~S. performed the theory analysis underpinning the project. B.~J. and J.-H.~J. designed the metamaterials. B.~J., Z.-K.~L., X.~Z., B~.H., F.~L. and J.-H.~J. performed the experiments. A.~B., R.-J.~S. and J.-H.~J. wrote the manuscript with input from all authors.

\noindent {\bf Competing interests}\\
The authors declare no competing interests.

\noindent {\bf Data and materials availability}\\
Relevant data are available in the main text and the supplementary materials. Additional data is available upon reasonable request.

\noindent {\bf Acknowledgements}\\
B.~J., Z.-K.~L. and J.-H.~J. are supported by the National Natural Science Foundations of China (Grant No. 12074281) and the Jiangsu Distinguished Professor Funding. X.~Z. and B.~H. are supported by the National Natural Science Foundation of China (Grant No. 12074279), the Major Program of Natural Science Research of Jiangsu Higher Education Institutions (Grant No. 18KJA140003). The work at Soochow University is also supported by the Priority Academic Program Development (PAPD) of Jiangsu Higher Education Institutions. R.-J.~S. acknowledges funding from the Marie Sk{\l}odowska-Curie programme under EC Grant No. 842901 and the Winton programme as well as Trinity College at the University of Cambridge. F.~L. is supported by the Natural Science Foundation of Guangdong Province (No. 2020A1515010549).

\bibliography{references}

\begin{thebibliography}{45}%
\makeatletter
\providecommand \@ifxundefined [1]{%
 \@ifx{#1\undefined}
}%
\providecommand \@ifnum [1]{%
 \ifnum #1\expandafter \@firstoftwo
 \else \expandafter \@secondoftwo
 \fi
}%
\providecommand \@ifx [1]{%
 \ifx #1\expandafter \@firstoftwo
 \else \expandafter \@secondoftwo
 \fi
}%
\providecommand \natexlab [1]{#1}%
\providecommand \enquote  [1]{``#1''}%
\providecommand \bibnamefont  [1]{#1}%
\providecommand \bibfnamefont [1]{#1}%
\providecommand \citenamefont [1]{#1}%
\providecommand \href@noop [0]{\@secondoftwo}%
\providecommand \href [0]{\begingroup \@sanitize@url \@href}%
\providecommand \@href[1]{\@@startlink{#1}\@@href}%
\providecommand \@@href[1]{\endgroup#1\@@endlink}%
\providecommand \@sanitize@url [0]{\catcode `\\12\catcode `\$12\catcode
  `\&12\catcode `\#12\catcode `\^12\catcode `\_12\catcode `\%12\relax}%
\providecommand \@@startlink[1]{}%
\providecommand \@@endlink[0]{}%
\providecommand \url  [0]{\begingroup\@sanitize@url \@url }%
\providecommand \@url [1]{\endgroup\@href {#1}{\urlprefix }}%
\providecommand \urlprefix  [0]{URL }%
\providecommand \Eprint [0]{\href }%
\providecommand \doibase [0]{http://dx.doi.org/}%
\providecommand \selectlanguage [0]{\@gobble}%
\providecommand \bibinfo  [0]{\@secondoftwo}%
\providecommand \bibfield  [0]{\@secondoftwo}%
\providecommand \translation [1]{[#1]}%
\providecommand \BibitemOpen [0]{}%
\providecommand \bibitemStop [0]{}%
\providecommand \bibitemNoStop [0]{.\EOS\space}%
\providecommand \EOS [0]{\spacefactor3000\relax}%
\providecommand \BibitemShut  [1]{\csname bibitem#1\endcsname}%
\let\auto@bib@innerbib\@empty
\bibitem [{\citenamefont {Kruthoff}\ \emph {et~al.}(2017)\citenamefont
  {Kruthoff}, \citenamefont {de~Boer}, \citenamefont {van Wezel}, \citenamefont
  {Kane},\ and\ \citenamefont {Slager}}]{Clas3}%
  \BibitemOpen
  \bibfield  {author} {\bibinfo {author} {\bibfnamefont {Jorrit}\ \bibnamefont
  {Kruthoff}}, \bibinfo {author} {\bibfnamefont {Jan}\ \bibnamefont {de~Boer}},
  \bibinfo {author} {\bibfnamefont {Jasper}\ \bibnamefont {van Wezel}},
  \bibinfo {author} {\bibfnamefont {Charles~L.}\ \bibnamefont {Kane}}, \ and\
  \bibinfo {author} {\bibfnamefont {Robert-Jan}\ \bibnamefont {Slager}},\
  }\bibfield  {title} {\enquote {\bibinfo {title} {Topological classification
  of crystalline insulators through band structure combinatorics},}\ }\href
  {\doibase 10.1103/PhysRevX.7.041069} {\bibfield  {journal} {\bibinfo
  {journal} {Phys. Rev. X}\ }\textbf {\bibinfo {volume} {7}},\ \bibinfo {pages}
  {041069} (\bibinfo {year} {2017})}\BibitemShut {NoStop}%
\bibitem [{\citenamefont {Po}\ \emph {et~al.}(2017)\citenamefont {Po},
  \citenamefont {Vishwanath},\ and\ \citenamefont {Watanabe}}]{Clas4}%
  \BibitemOpen
  \bibfield  {author} {\bibinfo {author} {\bibfnamefont {Hoi~Chun}\
  \bibnamefont {Po}}, \bibinfo {author} {\bibfnamefont {Ashvin}\ \bibnamefont
  {Vishwanath}}, \ and\ \bibinfo {author} {\bibfnamefont {Haruki}\ \bibnamefont
  {Watanabe}},\ }\bibfield  {title} {\enquote {\bibinfo {title} {Symmetry-based
  indicators of band topology in the 230 space groups},}\ }\href {\doibase
  10.1038/s41467-017-00133-2} {\bibfield  {journal} {\bibinfo  {journal} {Nat.
  Commun.}\ }\textbf {\bibinfo {volume} {8}},\ \bibinfo {pages} {50} (\bibinfo
  {year} {2017})}\BibitemShut {NoStop}%
\bibitem [{\citenamefont {Bradlyn}\ \emph {et~al.}(2017)\citenamefont
  {Bradlyn}, \citenamefont {Elcoro}, \citenamefont {Cano}, \citenamefont
  {Vergniory}, \citenamefont {Wang}, \citenamefont {Felser}, \citenamefont
  {Aroyo},\ and\ \citenamefont {Bernevig}}]{Clas5}%
  \BibitemOpen
  \bibfield  {author} {\bibinfo {author} {\bibfnamefont {Barry}\ \bibnamefont
  {Bradlyn}}, \bibinfo {author} {\bibfnamefont {L.}~\bibnamefont {Elcoro}},
  \bibinfo {author} {\bibfnamefont {Jennifer}\ \bibnamefont {Cano}}, \bibinfo
  {author} {\bibfnamefont {M.~G.}\ \bibnamefont {Vergniory}}, \bibinfo {author}
  {\bibfnamefont {Zhijun}\ \bibnamefont {Wang}}, \bibinfo {author}
  {\bibfnamefont {C.}~\bibnamefont {Felser}}, \bibinfo {author} {\bibfnamefont
  {M.~I.}\ \bibnamefont {Aroyo}}, \ and\ \bibinfo {author} {\bibfnamefont
  {B.~Andrei}\ \bibnamefont {Bernevig}},\ }\bibfield  {title} {\enquote
  {\bibinfo {title} {Topological quantum chemistry},}\ }\href
  {http://dx.doi.org/10.1038/nature23268} {\bibfield  {journal} {\bibinfo
  {journal} {Nature}\ }\textbf {\bibinfo {volume} {547}},\ \bibinfo {pages}
  {298} (\bibinfo {year} {2017})}\BibitemShut {NoStop}%
\bibitem [{\citenamefont {Wu}\ \emph {et~al.}(2019)\citenamefont {Wu},
  \citenamefont {Soluyanov},\ and\ \citenamefont {Bzdu{\v s}ek}}]{Wu1273}%
  \BibitemOpen
  \bibfield  {author} {\bibinfo {author} {\bibfnamefont {QuanSheng}\
  \bibnamefont {Wu}}, \bibinfo {author} {\bibfnamefont {Alexey~A.}\
  \bibnamefont {Soluyanov}}, \ and\ \bibinfo {author} {\bibfnamefont
  {Tom{\'a}{\v s}}\ \bibnamefont {Bzdu{\v s}ek}},\ }\bibfield  {title}
  {\enquote {\bibinfo {title} {Non-abelian band topology in noninteracting
  metals},}\ }\href {\doibase 10.1126/science.aau8740} {\bibfield  {journal}
  {\bibinfo  {journal} {Science}\ }\textbf {\bibinfo {volume} {365}},\ \bibinfo
  {pages} {1273--1277} (\bibinfo {year} {2019})}\BibitemShut {NoStop}%
\bibitem [{\citenamefont {Ahn}\ \emph {et~al.}(2019)\citenamefont {Ahn},
  \citenamefont {Park},\ and\ \citenamefont {Yang}}]{BJY_nielsen}%
  \BibitemOpen
  \bibfield  {author} {\bibinfo {author} {\bibfnamefont {Junyeong}\
  \bibnamefont {Ahn}}, \bibinfo {author} {\bibfnamefont {Sungjoon}\
  \bibnamefont {Park}}, \ and\ \bibinfo {author} {\bibfnamefont {Bohm-Jung}\
  \bibnamefont {Yang}},\ }\bibfield  {title} {\enquote {\bibinfo {title}
  {Failure of nielsen-ninomiya theorem and fragile topology in two-dimensional
  systems with space-time inversion symmetry: Application to twisted bilayer
  graphene at magic angle},}\ }\href {\doibase 10.1103/PhysRevX.9.021013}
  {\bibfield  {journal} {\bibinfo  {journal} {Phys. Rev. X}\ }\textbf {\bibinfo
  {volume} {9}},\ \bibinfo {pages} {021013} (\bibinfo {year}
  {2019})}\BibitemShut {NoStop}%
\bibitem [{\citenamefont {Bouhon}\ \emph
  {et~al.}(2020{\natexlab{a}})\citenamefont {Bouhon}, \citenamefont {Wu},
  \citenamefont {Slager}, \citenamefont {Weng}, \citenamefont {Yazyev},\ and\
  \citenamefont {Bzdu{\v s}ek}}]{bouhon2019nonabelian}%
  \BibitemOpen
  \bibfield  {author} {\bibinfo {author} {\bibfnamefont {Adrien}\ \bibnamefont
  {Bouhon}}, \bibinfo {author} {\bibfnamefont {QuanSheng}\ \bibnamefont {Wu}},
  \bibinfo {author} {\bibfnamefont {Robert-Jan}\ \bibnamefont {Slager}},
  \bibinfo {author} {\bibfnamefont {Hongming}\ \bibnamefont {Weng}}, \bibinfo
  {author} {\bibfnamefont {Oleg~V.}\ \bibnamefont {Yazyev}}, \ and\ \bibinfo
  {author} {\bibfnamefont {Tom{\'a}{\v s}}\ \bibnamefont {Bzdu{\v s}ek}},\
  }\bibfield  {title} {\enquote {\bibinfo {title} {Non-abelian reciprocal
  braiding of weyl points and its manifestation in zrte},}\ }\href {\doibase
  10.1038/s41567-020-0967-9} {\bibfield  {journal} {\bibinfo  {journal} {Nature
  Physics}\ }\textbf {\bibinfo {volume} {16}},\ \bibinfo {pages} {1137--1143}
  (\bibinfo {year} {2020}{\natexlab{a}})}\BibitemShut {NoStop}%
\bibitem [{\citenamefont {Fu}(2011)}]{Clas1}%
  \BibitemOpen
  \bibfield  {author} {\bibinfo {author} {\bibfnamefont {Liang}\ \bibnamefont
  {Fu}},\ }\bibfield  {title} {\enquote {\bibinfo {title} {Topological
  crystalline insulators},}\ }\href {\doibase 10.1103/PhysRevLett.106.106802}
  {\bibfield  {journal} {\bibinfo  {journal} {Phys. Rev. Lett.}\ }\textbf
  {\bibinfo {volume} {106}},\ \bibinfo {pages} {106802} (\bibinfo {year}
  {2011})}\BibitemShut {NoStop}%
\bibitem [{\citenamefont {Slager}\ \emph {et~al.}(2012)\citenamefont {Slager},
  \citenamefont {Mesaros}, \citenamefont {Juri{\v c}i{\'c}},\ and\
  \citenamefont {Zaanen}}]{Clas2}%
  \BibitemOpen
  \bibfield  {author} {\bibinfo {author} {\bibfnamefont {Robert-Jan}\
  \bibnamefont {Slager}}, \bibinfo {author} {\bibfnamefont {Andrej}\
  \bibnamefont {Mesaros}}, \bibinfo {author} {\bibfnamefont {Vladimir}\
  \bibnamefont {Juri{\v c}i{\'c}}}, \ and\ \bibinfo {author} {\bibfnamefont
  {Jan}\ \bibnamefont {Zaanen}},\ }\bibfield  {title} {\enquote {\bibinfo
  {title} {The space group classification of topological band-insulators},}\
  }\href {http://dx.doi.org/10.1038/nphys2513} {\bibfield  {journal} {\bibinfo
  {journal} {Nat. Phys.}\ }\textbf {\bibinfo {volume} {9}},\ \bibinfo {pages}
  {98} (\bibinfo {year} {2012})}\BibitemShut {NoStop}%
\bibitem [{\citenamefont {Bouhon}\ and\ \citenamefont
  {Black-Schaffer}(2017)}]{Wi2}%
  \BibitemOpen
  \bibfield  {author} {\bibinfo {author} {\bibfnamefont {Adrien}\ \bibnamefont
  {Bouhon}}\ and\ \bibinfo {author} {\bibfnamefont {Annica~M.}\ \bibnamefont
  {Black-Schaffer}},\ }\bibfield  {title} {\enquote {\bibinfo {title} {Global
  band topology of simple and double dirac-point semimetals},}\ }\href
  {\doibase 10.1103/PhysRevB.95.241101} {\bibfield  {journal} {\bibinfo
  {journal} {Phys. Rev. B}\ }\textbf {\bibinfo {volume} {95}},\ \bibinfo
  {pages} {241101} (\bibinfo {year} {2017})}\BibitemShut {NoStop}%
\bibitem [{\citenamefont {Fang}\ \emph {et~al.}(2012)\citenamefont {Fang},
  \citenamefont {Gilbert},\ and\ \citenamefont {Bernevig}}]{Chenprb2012}%
  \BibitemOpen
  \bibfield  {author} {\bibinfo {author} {\bibfnamefont {Chen}\ \bibnamefont
  {Fang}}, \bibinfo {author} {\bibfnamefont {Matthew~J.}\ \bibnamefont
  {Gilbert}}, \ and\ \bibinfo {author} {\bibfnamefont {B.~Andrei}\ \bibnamefont
  {Bernevig}},\ }\bibfield  {title} {\enquote {\bibinfo {title} {Bulk
  topological invariants in noninteracting point group symmetric insulators},}\
  }\href {\doibase 10.1103/PhysRevB.86.115112} {\bibfield  {journal} {\bibinfo
  {journal} {Phys. Rev. B}\ }\textbf {\bibinfo {volume} {86}},\ \bibinfo
  {pages} {115112} (\bibinfo {year} {2012})}\BibitemShut {NoStop}%
\bibitem [{\citenamefont {Slager}(2019)}]{Codefects2}%
  \BibitemOpen
  \bibfield  {author} {\bibinfo {author} {\bibfnamefont {Robert-Jan}\
  \bibnamefont {Slager}},\ }\bibfield  {title} {\enquote {\bibinfo {title} {The
  translational side of topological band insulators},}\ }\href {\doibase
  https://doi.org/10.1016/j.jpcs.2018.01.023} {\bibfield  {journal} {\bibinfo
  {journal} {Journal of Physics and Chemistry of Solids}\ }\textbf {\bibinfo
  {volume} {128}},\ \bibinfo {pages} {24 -- 38} (\bibinfo {year} {2019})},\
  \bibinfo {note} {spin-Orbit Coupled Materials}\BibitemShut {NoStop}%
\bibitem [{\citenamefont {Cornfeld}\ and\ \citenamefont
  {Carmeli}(2021)}]{Cornfeld_2021}%
  \BibitemOpen
  \bibfield  {author} {\bibinfo {author} {\bibfnamefont {Eyal}\ \bibnamefont
  {Cornfeld}}\ and\ \bibinfo {author} {\bibfnamefont {Shachar}\ \bibnamefont
  {Carmeli}},\ }\bibfield  {title} {\enquote {\bibinfo {title} {Tenfold
  topology of crystals: Unified classification of crystalline topological
  insulators and superconductors},}\ }\href {\doibase
  10.1103/PhysRevResearch.3.013052} {\bibfield  {journal} {\bibinfo  {journal}
  {Phys. Rev. Research}\ }\textbf {\bibinfo {volume} {3}},\ \bibinfo {pages}
  {013052} (\bibinfo {year} {2021})}\BibitemShut {NoStop}%
\bibitem [{\citenamefont {Yang}\ and\ \citenamefont
  {Nagaosa}(2014)}]{yang2014classification}%
  \BibitemOpen
  \bibfield  {author} {\bibinfo {author} {\bibfnamefont {Bohm-Jung}\
  \bibnamefont {Yang}}\ and\ \bibinfo {author} {\bibfnamefont {Naoto}\
  \bibnamefont {Nagaosa}},\ }\bibfield  {title} {\enquote {\bibinfo {title}
  {Classification of stable three-dimensional dirac semimetals with nontrivial
  topology},}\ }\href@noop {} {\bibfield  {journal} {\bibinfo  {journal}
  {Nature communications}\ }\textbf {\bibinfo {volume} {5}},\ \bibinfo {pages}
  {1--10} (\bibinfo {year} {2014})}\BibitemShut {NoStop}%
\bibitem [{\citenamefont {Bradlyn}\ \emph {et~al.}(2016)\citenamefont
  {Bradlyn}, \citenamefont {Cano}, \citenamefont {Wang}, \citenamefont
  {Vergniory}, \citenamefont {Felser}, \citenamefont {Cava},\ and\
  \citenamefont {Bernevig}}]{Bradlynaaf5037}%
  \BibitemOpen
  \bibfield  {author} {\bibinfo {author} {\bibfnamefont {Barry}\ \bibnamefont
  {Bradlyn}}, \bibinfo {author} {\bibfnamefont {Jennifer}\ \bibnamefont
  {Cano}}, \bibinfo {author} {\bibfnamefont {Zhijun}\ \bibnamefont {Wang}},
  \bibinfo {author} {\bibfnamefont {M.~G.}\ \bibnamefont {Vergniory}}, \bibinfo
  {author} {\bibfnamefont {C.}~\bibnamefont {Felser}}, \bibinfo {author}
  {\bibfnamefont {R.~J.}\ \bibnamefont {Cava}}, \ and\ \bibinfo {author}
  {\bibfnamefont {B.~Andrei}\ \bibnamefont {Bernevig}},\ }\bibfield  {title}
  {\enquote {\bibinfo {title} {Beyond dirac and weyl fermions: Unconventional
  quasiparticles in conventional crystals},}\ }\href {\doibase
  10.1126/science.aaf5037} {\bibfield  {journal} {\bibinfo  {journal}
  {Science}\ }\textbf {\bibinfo {volume} {353}} (\bibinfo {year} {2016}),\
  10.1126/science.aaf5037}\BibitemShut {NoStop}%
\bibitem [{\citenamefont {Bouhon}\ \emph
  {et~al.}(2020{\natexlab{b}})\citenamefont {Bouhon}, \citenamefont
  {Bzdu\v{s}ek},\ and\ \citenamefont {Slager}}]{bouhonGeometric2020}%
  \BibitemOpen
  \bibfield  {author} {\bibinfo {author} {\bibfnamefont {Adrien}\ \bibnamefont
  {Bouhon}}, \bibinfo {author} {\bibfnamefont {Tom\'a\v{s}}\ \bibnamefont
  {Bzdu\v{s}ek}}, \ and\ \bibinfo {author} {\bibfnamefont {Robert-Jan}\
  \bibnamefont {Slager}},\ }\bibfield  {title} {\enquote {\bibinfo {title}
  {Geometric approach to fragile topology beyond symmetry indicators},}\ }\href
  {\doibase 10.1103/PhysRevB.102.115135} {\bibfield  {journal} {\bibinfo
  {journal} {Phys. Rev. B}\ }\textbf {\bibinfo {volume} {102}},\ \bibinfo
  {pages} {115135} (\bibinfo {year} {2020}{\natexlab{b}})}\BibitemShut
  {NoStop}%
\bibitem [{\citenamefont {Johansson}\ and\ \citenamefont
  {Sj\"oqvist}(2004)}]{Sjoeqvist_2004}%
  \BibitemOpen
  \bibfield  {author} {\bibinfo {author} {\bibfnamefont {Niklas}\ \bibnamefont
  {Johansson}}\ and\ \bibinfo {author} {\bibfnamefont {Erik}\ \bibnamefont
  {Sj\"oqvist}},\ }\bibfield  {title} {\enquote {\bibinfo {title} {Optimal
  {T}opological {T}est for {D}egeneracies of {R}eal {H}amiltonians},}\ }\href
  {\doibase 10.1103/PhysRevLett.92.060406} {\bibfield  {journal} {\bibinfo
  {journal} {Phys. Rev. Lett.}\ }\textbf {\bibinfo {volume} {92}},\ \bibinfo
  {pages} {060406} (\bibinfo {year} {2004})}\BibitemShut {NoStop}%
\bibitem [{\citenamefont {Alexander}\ \emph {et~al.}(2012)\citenamefont
  {Alexander}, \citenamefont {Chen}, \citenamefont {Matsumoto},\ and\
  \citenamefont {Kamien}}]{Kamienrmp}%
  \BibitemOpen
  \bibfield  {author} {\bibinfo {author} {\bibfnamefont {Gareth~P.}\
  \bibnamefont {Alexander}}, \bibinfo {author} {\bibfnamefont {Bryan Gin-ge}\
  \bibnamefont {Chen}}, \bibinfo {author} {\bibfnamefont {Elisabetta~A.}\
  \bibnamefont {Matsumoto}}, \ and\ \bibinfo {author} {\bibfnamefont
  {Randall~D.}\ \bibnamefont {Kamien}},\ }\bibfield  {title} {\enquote
  {\bibinfo {title} {Colloquium: Disclination loops, point defects, and all
  that in nematic liquid crystals},}\ }\href {\doibase
  10.1103/RevModPhys.84.497} {\bibfield  {journal} {\bibinfo  {journal} {Rev.
  Mod. Phys.}\ }\textbf {\bibinfo {volume} {84}},\ \bibinfo {pages} {497--514}
  (\bibinfo {year} {2012})}\BibitemShut {NoStop}%
\bibitem [{\citenamefont {Liu}\ \emph {et~al.}(2016)\citenamefont {Liu},
  \citenamefont {Nissinen}, \citenamefont {Slager}, \citenamefont {Wu},\ and\
  \citenamefont {Zaanen}}]{Prx2016}%
  \BibitemOpen
  \bibfield  {author} {\bibinfo {author} {\bibfnamefont {Ke}~\bibnamefont
  {Liu}}, \bibinfo {author} {\bibfnamefont {Jaakko}\ \bibnamefont {Nissinen}},
  \bibinfo {author} {\bibfnamefont {Robert-Jan}\ \bibnamefont {Slager}},
  \bibinfo {author} {\bibfnamefont {Kai}\ \bibnamefont {Wu}}, \ and\ \bibinfo
  {author} {\bibfnamefont {Jan}\ \bibnamefont {Zaanen}},\ }\bibfield  {title}
  {\enquote {\bibinfo {title} {Generalized liquid crystals: Giant fluctuations
  and the vestigial chiral order of $i$, $o$, and $t$ matter},}\ }\href
  {\doibase 10.1103/PhysRevX.6.041025} {\bibfield  {journal} {\bibinfo
  {journal} {Phys. Rev. X}\ }\textbf {\bibinfo {volume} {6}},\ \bibinfo {pages}
  {041025} (\bibinfo {year} {2016})}\BibitemShut {NoStop}%
\bibitem [{\citenamefont {Volovik}\ and\ \citenamefont
  {Mineev}(2018)}]{volovik2018investigation}%
  \BibitemOpen
  \bibfield  {author} {\bibinfo {author} {\bibfnamefont {GE}~\bibnamefont
  {Volovik}}\ and\ \bibinfo {author} {\bibfnamefont {VP}~\bibnamefont
  {Mineev}},\ }\bibfield  {title} {\enquote {\bibinfo {title} {Investigation of
  singularities in superfluid he3 in liquid crystals by the homotopic topology
  methods},}\ }in\ \href@noop {} {\emph {\bibinfo {booktitle} {Basic Notions Of
  Condensed Matter Physics}}}\ (\bibinfo  {publisher} {CRC Press},\ \bibinfo
  {year} {2018})\ pp.\ \bibinfo {pages} {392--401}\BibitemShut {NoStop}%
\bibitem [{\citenamefont {Tiwari}\ and\ \citenamefont
  {Bzdu\ifmmode~\check{s}\else \v{s}\fi{}ek}(2020)}]{Tiwari}%
  \BibitemOpen
  \bibfield  {author} {\bibinfo {author} {\bibfnamefont {Apoorv}\ \bibnamefont
  {Tiwari}}\ and\ \bibinfo {author} {\bibfnamefont {Tom\'a\ifmmode
  \check{s}\else~\v{s}\fi{}}\ \bibnamefont {Bzdu\ifmmode~\check{s}\else
  \v{s}\fi{}ek}},\ }\bibfield  {title} {\enquote {\bibinfo {title} {Non-abelian
  topology of nodal-line rings in $\mathcal{PT}$-symmetric systems},}\ }\href
  {\doibase 10.1103/PhysRevB.101.195130} {\bibfield  {journal} {\bibinfo
  {journal} {Phys. Rev. B}\ }\textbf {\bibinfo {volume} {101}},\ \bibinfo
  {pages} {195130} (\bibinfo {year} {2020})}\BibitemShut {NoStop}%
\bibitem [{\citenamefont {Po}\ \emph {et~al.}(2018)\citenamefont {Po},
  \citenamefont {Watanabe},\ and\ \citenamefont {Vishwanath}}]{Ft1}%
  \BibitemOpen
  \bibfield  {author} {\bibinfo {author} {\bibfnamefont {Hoi~Chun}\
  \bibnamefont {Po}}, \bibinfo {author} {\bibfnamefont {Haruki}\ \bibnamefont
  {Watanabe}}, \ and\ \bibinfo {author} {\bibfnamefont {Ashvin}\ \bibnamefont
  {Vishwanath}},\ }\bibfield  {title} {\enquote {\bibinfo {title} {Fragile
  topology and wannier obstructions},}\ }\href {\doibase
  10.1103/PhysRevLett.121.126402} {\bibfield  {journal} {\bibinfo  {journal}
  {Phys. Rev. Lett.}\ }\textbf {\bibinfo {volume} {121}},\ \bibinfo {pages}
  {126402} (\bibinfo {year} {2018})}\BibitemShut {NoStop}%
\bibitem [{\citenamefont {Bouhon}\ \emph {et~al.}(2019)\citenamefont {Bouhon},
  \citenamefont {Black-Schaffer},\ and\ \citenamefont
  {Slager}}]{bouhon2019wilson}%
  \BibitemOpen
  \bibfield  {author} {\bibinfo {author} {\bibfnamefont {Adrien}\ \bibnamefont
  {Bouhon}}, \bibinfo {author} {\bibfnamefont {Annica~M.}\ \bibnamefont
  {Black-Schaffer}}, \ and\ \bibinfo {author} {\bibfnamefont {Robert-Jan}\
  \bibnamefont {Slager}},\ }\bibfield  {title} {\enquote {\bibinfo {title}
  {Wilson loop approach to fragile topology of split elementary band
  representations and topological crystalline insulators with time-reversal
  symmetry},}\ }\href {\doibase 10.1103/PhysRevB.100.195135} {\bibfield
  {journal} {\bibinfo  {journal} {Phys. Rev. B}\ }\textbf {\bibinfo {volume}
  {100}},\ \bibinfo {pages} {195135} (\bibinfo {year} {2019})}\BibitemShut
  {NoStop}%
\bibitem [{\citenamefont {Song}\ \emph {et~al.}(2019)\citenamefont {Song},
  \citenamefont {Elcoro}, \citenamefont {Regnault},\ and\ \citenamefont
  {Bernevig}}]{song2019fragile}%
  \BibitemOpen
  \bibfield  {author} {\bibinfo {author} {\bibfnamefont {Zhida}\ \bibnamefont
  {Song}}, \bibinfo {author} {\bibfnamefont {L.}~\bibnamefont {Elcoro}},
  \bibinfo {author} {\bibfnamefont {Nicolas}\ \bibnamefont {Regnault}}, \ and\
  \bibinfo {author} {\bibfnamefont {B.~Andrei}\ \bibnamefont {Bernevig}},\
  }\href@noop {} {\enquote {\bibinfo {title} {Fragile phases as affine monoids:
  Full classification and material examples},}\ } (\bibinfo {year} {2019}),\
  \Eprint {http://arxiv.org/abs/1905.03262} {arXiv:1905.03262
  [cond-mat.mes-hall]} \BibitemShut {NoStop}%
\bibitem [{\citenamefont {Peri}\ \emph {et~al.}(2020)\citenamefont {Peri},
  \citenamefont {Song}, \citenamefont {Serra-Garcia}, \citenamefont {Engeler},
  \citenamefont {Queiroz}, \citenamefont {Huang}, \citenamefont {Deng},
  \citenamefont {Liu}, \citenamefont {Bernevig},\ and\ \citenamefont
  {Huber}}]{Peri797}%
  \BibitemOpen
  \bibfield  {author} {\bibinfo {author} {\bibfnamefont {Valerio}\ \bibnamefont
  {Peri}}, \bibinfo {author} {\bibfnamefont {Zhi-Da}\ \bibnamefont {Song}},
  \bibinfo {author} {\bibfnamefont {Marc}\ \bibnamefont {Serra-Garcia}},
  \bibinfo {author} {\bibfnamefont {Pascal}\ \bibnamefont {Engeler}}, \bibinfo
  {author} {\bibfnamefont {Raquel}\ \bibnamefont {Queiroz}}, \bibinfo {author}
  {\bibfnamefont {Xueqin}\ \bibnamefont {Huang}}, \bibinfo {author}
  {\bibfnamefont {Weiyin}\ \bibnamefont {Deng}}, \bibinfo {author}
  {\bibfnamefont {Zhengyou}\ \bibnamefont {Liu}}, \bibinfo {author}
  {\bibfnamefont {B.~Andrei}\ \bibnamefont {Bernevig}}, \ and\ \bibinfo
  {author} {\bibfnamefont {Sebastian~D.}\ \bibnamefont {Huber}},\ }\bibfield
  {title} {\enquote {\bibinfo {title} {Experimental characterization of fragile
  topology in an acoustic metamaterial},}\ }\href {\doibase
  10.1126/science.aaz7654} {\bibfield  {journal} {\bibinfo  {journal}
  {Science}\ }\textbf {\bibinfo {volume} {367}},\ \bibinfo {pages} {797--800}
  (\bibinfo {year} {2020})}\BibitemShut {NoStop}%
\bibitem [{\citenamefont {Song}\ \emph {et~al.}(2020)\citenamefont {Song},
  \citenamefont {Elcoro},\ and\ \citenamefont {Bernevig}}]{Song794}%
  \BibitemOpen
  \bibfield  {author} {\bibinfo {author} {\bibfnamefont {Zhi-Da}\ \bibnamefont
  {Song}}, \bibinfo {author} {\bibfnamefont {Luis}\ \bibnamefont {Elcoro}}, \
  and\ \bibinfo {author} {\bibfnamefont {B.~Andrei}\ \bibnamefont {Bernevig}},\
  }\bibfield  {title} {\enquote {\bibinfo {title} {Twisted bulk-boundary
  correspondence of fragile topology},}\ }\href {\doibase
  10.1126/science.aaz7650} {\bibfield  {journal} {\bibinfo  {journal}
  {Science}\ }\textbf {\bibinfo {volume} {367}},\ \bibinfo {pages} {794--797}
  (\bibinfo {year} {2020})}\BibitemShut {NoStop}%
\bibitem [{\citenamefont {\"Unal}\ \emph {et~al.}(2020)\citenamefont {\"Unal},
  \citenamefont {Bouhon},\ and\ \citenamefont {Slager}}]{Euler_drive}%
  \BibitemOpen
  \bibfield  {author} {\bibinfo {author} {\bibfnamefont {F.~Nur}\ \bibnamefont
  {\"Unal}}, \bibinfo {author} {\bibfnamefont {Adrien}\ \bibnamefont {Bouhon}},
  \ and\ \bibinfo {author} {\bibfnamefont {Robert-Jan}\ \bibnamefont
  {Slager}},\ }\bibfield  {title} {\enquote {\bibinfo {title} {Topological
  euler class as a dynamical observable in optical lattices},}\ }\href
  {\doibase 10.1103/PhysRevLett.125.053601} {\bibfield  {journal} {\bibinfo
  {journal} {Phys. Rev. Lett.}\ }\textbf {\bibinfo {volume} {125}},\ \bibinfo
  {pages} {053601} (\bibinfo {year} {2020})}\BibitemShut {NoStop}%
\bibitem [{\citenamefont {Zak}(1989)}]{Zak2}%
  \BibitemOpen
  \bibfield  {author} {\bibinfo {author} {\bibfnamefont {J.}~\bibnamefont
  {Zak}},\ }\bibfield  {title} {\enquote {\bibinfo {title} {Berry's phase for
  energy bands in solids},}\ }\href {\doibase 10.1103/PhysRevLett.62.2747}
  {\bibfield  {journal} {\bibinfo  {journal} {Phys. Rev. Lett.}\ }\textbf
  {\bibinfo {volume} {62}},\ \bibinfo {pages} {2747--2750} (\bibinfo {year}
  {1989})}\BibitemShut {NoStop}%
\bibitem [{\citenamefont {Ahn}\ \emph {et~al.}(2018)\citenamefont {Ahn},
  \citenamefont {Kim}, \citenamefont {Kim},\ and\ \citenamefont
  {Yang}}]{Ahn2018b}%
  \BibitemOpen
  \bibfield  {author} {\bibinfo {author} {\bibfnamefont {Junyeong}\
  \bibnamefont {Ahn}}, \bibinfo {author} {\bibfnamefont {Dongwook}\
  \bibnamefont {Kim}}, \bibinfo {author} {\bibfnamefont {Youngkuk}\
  \bibnamefont {Kim}}, \ and\ \bibinfo {author} {\bibfnamefont {Bohm-Jung}\
  \bibnamefont {Yang}},\ }\bibfield  {title} {\enquote {\bibinfo {title} {Band
  topology and linking structure of nodal line semimetals with ${Z}_{2}$
  monopole charges},}\ }\href {\doibase 10.1103/PhysRevLett.121.106403}
  {\bibfield  {journal} {\bibinfo  {journal} {Phys. Rev. Lett.}\ }\textbf
  {\bibinfo {volume} {121}},\ \bibinfo {pages} {106403} (\bibinfo {year}
  {2018})}\BibitemShut {NoStop}%
\bibitem [{ITC(2006)}]{ITCA}%
  \BibitemOpen
  \href {http://it.iucr.org/Ac/} {\emph {\bibinfo {title} {International Tables
  for Crystallography. Volume A, Space-group symmetry}}},\ \bibinfo {edition}
  {online}\ ed.\ (\bibinfo {year} {2006})\ \bibinfo {note} {edited by T.
  Hahn}\BibitemShut {NoStop}%
\bibitem [{ITC(2010)}]{ITCE}%
  \BibitemOpen
  \href {https://it.iucr.org/E/} {\emph {\bibinfo {title} {International Tables
  for Crystallography. Volume E, Subperiodic groups}}},\ \bibinfo {edition}
  {online}\ ed.\ (\bibinfo {year} {2010})\ \bibinfo {note} {edited by T.
  Hahn}\BibitemShut {NoStop}%
\bibitem [{\citenamefont {Lenggenhager}\ \emph {et~al.}(2020)\citenamefont
  {Lenggenhager}, \citenamefont {Liu}, \citenamefont {Tsirkin}, \citenamefont
  {Neupert},\ and\ \citenamefont {Bzdušek}}]{lenggenhager2020multiband}%
  \BibitemOpen
  \bibfield  {author} {\bibinfo {author} {\bibfnamefont {Patrick~M.}\
  \bibnamefont {Lenggenhager}}, \bibinfo {author} {\bibfnamefont {Xiaoxiong}\
  \bibnamefont {Liu}}, \bibinfo {author} {\bibfnamefont {Stepan~S.}\
  \bibnamefont {Tsirkin}}, \bibinfo {author} {\bibfnamefont {Titus}\
  \bibnamefont {Neupert}}, \ and\ \bibinfo {author} {\bibfnamefont {Tomáš}\
  \bibnamefont {Bzdušek}},\ }\href@noop {} {\enquote {\bibinfo {title}
  {Multi-band nodal links in triple-point materials},}\ } (\bibinfo {year}
  {2020}),\ \Eprint {http://arxiv.org/abs/2008.02807} {arXiv:2008.02807
  [cond-mat.mes-hall]} \BibitemShut {NoStop}%
\bibitem [{\citenamefont {Zhu}\ \emph {et~al.}(2016)\citenamefont {Zhu},
  \citenamefont {Winkler}, \citenamefont {Wu}, \citenamefont {Li},\ and\
  \citenamefont {Soluyanov}}]{PhysRevX.6.031003}%
  \BibitemOpen
  \bibfield  {author} {\bibinfo {author} {\bibfnamefont {Ziming}\ \bibnamefont
  {Zhu}}, \bibinfo {author} {\bibfnamefont {Georg~W.}\ \bibnamefont {Winkler}},
  \bibinfo {author} {\bibfnamefont {QuanSheng}\ \bibnamefont {Wu}}, \bibinfo
  {author} {\bibfnamefont {Ju}~\bibnamefont {Li}}, \ and\ \bibinfo {author}
  {\bibfnamefont {Alexey~A.}\ \bibnamefont {Soluyanov}},\ }\bibfield  {title}
  {\enquote {\bibinfo {title} {Triple point topological metals},}\ }\href
  {\doibase 10.1103/PhysRevX.6.031003} {\bibfield  {journal} {\bibinfo
  {journal} {Phys. Rev. X}\ }\textbf {\bibinfo {volume} {6}},\ \bibinfo {pages}
  {031003} (\bibinfo {year} {2016})}\BibitemShut {NoStop}%
\bibitem [{\citenamefont {Kim}\ \emph {et~al.}(2018)\citenamefont {Kim},
  \citenamefont {Kim},\ and\ \citenamefont {Vanderbilt}}]{Vanderbilt_triple}%
  \BibitemOpen
  \bibfield  {author} {\bibinfo {author} {\bibfnamefont {Jinwoong}\
  \bibnamefont {Kim}}, \bibinfo {author} {\bibfnamefont {Heung-Sik}\
  \bibnamefont {Kim}}, \ and\ \bibinfo {author} {\bibfnamefont {David}\
  \bibnamefont {Vanderbilt}},\ }\bibfield  {title} {\enquote {\bibinfo {title}
  {Nearly triple nodal point topological phase in half-metallic gdn},}\ }\href
  {\doibase 10.1103/PhysRevB.98.155122} {\bibfield  {journal} {\bibinfo
  {journal} {Phys. Rev. B}\ }\textbf {\bibinfo {volume} {98}},\ \bibinfo
  {pages} {155122} (\bibinfo {year} {2018})}\BibitemShut {NoStop}%
\bibitem [{\citenamefont {Su}\ \emph {et~al.}(1979)\citenamefont {Su},
  \citenamefont {Schrieffer},\ and\ \citenamefont {Heeger}}]{SSH}%
  \BibitemOpen
  \bibfield  {author} {\bibinfo {author} {\bibfnamefont {W.~P.}\ \bibnamefont
  {Su}}, \bibinfo {author} {\bibfnamefont {J.~R.}\ \bibnamefont {Schrieffer}},
  \ and\ \bibinfo {author} {\bibfnamefont {A.~J.}\ \bibnamefont {Heeger}},\
  }\bibfield  {title} {\enquote {\bibinfo {title} {Solitons in
  polyacetylene},}\ }\href {\doibase 10.1103/PhysRevLett.42.1698} {\bibfield
  {journal} {\bibinfo  {journal} {Phys. Rev. Lett.}\ }\textbf {\bibinfo
  {volume} {42}},\ \bibinfo {pages} {1698--1701} (\bibinfo {year}
  {1979})}\BibitemShut {NoStop}%
\bibitem [{\citenamefont {Guo}\ \emph {et~al.}(2020)\citenamefont {Guo},
  \citenamefont {Jiang}, \citenamefont {Zhang}, \citenamefont {Zhang},
  \citenamefont {Zhang}, \citenamefont {Yang}, \citenamefont {Zhang},\ and\
  \citenamefont {Chan}}]{guo2020experimental}%
  \BibitemOpen
  \bibfield  {author} {\bibinfo {author} {\bibfnamefont {Qinghua}\ \bibnamefont
  {Guo}}, \bibinfo {author} {\bibfnamefont {Tianshu}\ \bibnamefont {Jiang}},
  \bibinfo {author} {\bibfnamefont {Ruo-Yang}\ \bibnamefont {Zhang}}, \bibinfo
  {author} {\bibfnamefont {Lei}\ \bibnamefont {Zhang}}, \bibinfo {author}
  {\bibfnamefont {Zhao-Qing}\ \bibnamefont {Zhang}}, \bibinfo {author}
  {\bibfnamefont {Biao}\ \bibnamefont {Yang}}, \bibinfo {author} {\bibfnamefont
  {Shuang}\ \bibnamefont {Zhang}}, \ and\ \bibinfo {author} {\bibfnamefont
  {C.~T.}\ \bibnamefont {Chan}},\ }\href@noop {} {\enquote {\bibinfo {title}
  {Experimental observation of non-abelian topological charges and bulk-edge
  correspondence},}\ } (\bibinfo {year} {2020}),\ \Eprint
  {http://arxiv.org/abs/2008.06100} {arXiv:2008.06100 [cond-mat.mes-hall]}
  \BibitemShut {NoStop}%
\bibitem [{\citenamefont {Yang}\ \emph {et~al.}(2015)\citenamefont {Yang},
  \citenamefont {Gao}, \citenamefont {Shi}, \citenamefont {Lin}, \citenamefont
  {Gao}, \citenamefont {Chong},\ and\ \citenamefont {Zhang}}]{am1}%
  \BibitemOpen
  \bibfield  {author} {\bibinfo {author} {\bibfnamefont {Zhaoju}\ \bibnamefont
  {Yang}}, \bibinfo {author} {\bibfnamefont {Fei}\ \bibnamefont {Gao}},
  \bibinfo {author} {\bibfnamefont {Xihang}\ \bibnamefont {Shi}}, \bibinfo
  {author} {\bibfnamefont {Xiao}\ \bibnamefont {Lin}}, \bibinfo {author}
  {\bibfnamefont {Zhen}\ \bibnamefont {Gao}}, \bibinfo {author} {\bibfnamefont
  {Yidong}\ \bibnamefont {Chong}}, \ and\ \bibinfo {author} {\bibfnamefont
  {Baile}\ \bibnamefont {Zhang}},\ }\bibfield  {title} {\enquote {\bibinfo
  {title} {Topological acoustics},}\ }\href
  {https://link.aps.org/doi/10.1103/PhysRevLett.114.114301} {\bibfield
  {journal} {\bibinfo  {journal} {Phys. Rev. Lett.}\ }\textbf {\bibinfo
  {volume} {114}},\ \bibinfo {pages} {114301} (\bibinfo {year}
  {2015})}\BibitemShut {NoStop}%
\bibitem [{\citenamefont {Xiao}\ \emph {et~al.}(2015)\citenamefont {Xiao},
  \citenamefont {Chen}, \citenamefont {He},\ and\ \citenamefont {Chan}}]{am2}%
  \BibitemOpen
  \bibfield  {author} {\bibinfo {author} {\bibfnamefont {Meng}\ \bibnamefont
  {Xiao}}, \bibinfo {author} {\bibfnamefont {Wen-Jie}\ \bibnamefont {Chen}},
  \bibinfo {author} {\bibfnamefont {Wen-Yu}\ \bibnamefont {He}}, \ and\
  \bibinfo {author} {\bibfnamefont {C.~T.}\ \bibnamefont {Chan}},\ }\bibfield
  {title} {\enquote {\bibinfo {title} {Synthetic gauge flux and weyl points in
  acoustic systems},}\ }\href@noop {} {\bibfield  {journal} {\bibinfo
  {journal} {Nat. Phys.}\ }\textbf {\bibinfo {volume} {11}},\ \bibinfo {pages}
  {920--924} (\bibinfo {year} {2015})}\BibitemShut {NoStop}%
\bibitem [{\citenamefont {He}\ \emph {et~al.}(2016)\citenamefont {He},
  \citenamefont {Ni}, \citenamefont {Ge}, \citenamefont {Sun}, \citenamefont
  {Chen}, \citenamefont {Lu}, \citenamefont {Liu},\ and\ \citenamefont
  {Chen}}]{am3}%
  \BibitemOpen
  \bibfield  {author} {\bibinfo {author} {\bibfnamefont {C.}~\bibnamefont
  {He}}, \bibinfo {author} {\bibfnamefont {X.}~\bibnamefont {Ni}}, \bibinfo
  {author} {\bibfnamefont {H.}~\bibnamefont {Ge}}, \bibinfo {author}
  {\bibfnamefont {X.~C.}\ \bibnamefont {Sun}}, \bibinfo {author} {\bibfnamefont
  {Y.-B.}\ \bibnamefont {Chen}}, \bibinfo {author} {\bibfnamefont {M.-H.}\
  \bibnamefont {Lu}}, \bibinfo {author} {\bibfnamefont {X.-P.}\ \bibnamefont
  {Liu}}, \ and\ \bibinfo {author} {\bibfnamefont {Y.-F.}\ \bibnamefont
  {Chen}},\ }\bibfield  {title} {\enquote {\bibinfo {title} {Acoustic
  topological insulator and robust one-way sound transport},}\ }\href@noop {}
  {\bibfield  {journal} {\bibinfo  {journal} {Nat. Phys.}\ }\textbf {\bibinfo
  {volume} {12}},\ \bibinfo {pages} {1124--1129} (\bibinfo {year}
  {2016})}\BibitemShut {NoStop}%
\bibitem [{\citenamefont {Lu}\ \emph {et~al.}(2017)\citenamefont {Lu},
  \citenamefont {Qiu}, \citenamefont {Ye}, \citenamefont {Fan}, \citenamefont
  {Ke}, \citenamefont {Zhang},\ and\ \citenamefont {Liu}}]{am4}%
  \BibitemOpen
  \bibfield  {author} {\bibinfo {author} {\bibfnamefont {J.}~\bibnamefont
  {Lu}}, \bibinfo {author} {\bibfnamefont {C.}~\bibnamefont {Qiu}}, \bibinfo
  {author} {\bibfnamefont {L.}~\bibnamefont {Ye}}, \bibinfo {author}
  {\bibfnamefont {X.}~\bibnamefont {Fan}}, \bibinfo {author} {\bibfnamefont
  {M.}~\bibnamefont {Ke}}, \bibinfo {author} {\bibfnamefont {F.}~\bibnamefont
  {Zhang}}, \ and\ \bibinfo {author} {\bibfnamefont {Z.}~\bibnamefont {Liu}},\
  }\bibfield  {title} {\enquote {\bibinfo {title} {Observation of topological
  valley transport of sound in sonic crystals},}\ }\href@noop {} {\bibfield
  {journal} {\bibinfo  {journal} {Nat. Phys.}\ }\textbf {\bibinfo {volume}
  {13}},\ \bibinfo {pages} {369--374} (\bibinfo {year} {2017})}\BibitemShut
  {NoStop}%
\bibitem [{\citenamefont {Li}\ \emph {et~al.}(2017)\citenamefont {Li},
  \citenamefont {Huang}, \citenamefont {Lu}, \citenamefont {Ma},\ and\
  \citenamefont {Liu}}]{am5}%
  \BibitemOpen
  \bibfield  {author} {\bibinfo {author} {\bibfnamefont {F.}~\bibnamefont
  {Li}}, \bibinfo {author} {\bibfnamefont {X.}~\bibnamefont {Huang}}, \bibinfo
  {author} {\bibfnamefont {J.}~\bibnamefont {Lu}}, \bibinfo {author}
  {\bibfnamefont {J.}~\bibnamefont {Ma}}, \ and\ \bibinfo {author}
  {\bibfnamefont {Z.}~\bibnamefont {Liu}},\ }\bibfield  {title} {\enquote
  {\bibinfo {title} {Weyl points and fermi arcs in a chiral phononic
  crystal},}\ }\href@noop {} {\bibfield  {journal} {\bibinfo  {journal} {Nat.
  Phys.}\ }\textbf {\bibinfo {volume} {14}},\ \bibinfo {pages} {30--34}
  (\bibinfo {year} {2017})}\BibitemShut {NoStop}%
\bibitem [{\citenamefont {He}\ \emph {et~al.}(2018)\citenamefont {He},
  \citenamefont {Qiu}, \citenamefont {Ye}, \citenamefont {Cai}, \citenamefont
  {Fan}, \citenamefont {Ke}, \citenamefont {Zhang},\ and\ \citenamefont
  {Liu}}]{am6}%
  \BibitemOpen
  \bibfield  {author} {\bibinfo {author} {\bibfnamefont {Hailong}\ \bibnamefont
  {He}}, \bibinfo {author} {\bibfnamefont {Chunyin}\ \bibnamefont {Qiu}},
  \bibinfo {author} {\bibfnamefont {Liping}\ \bibnamefont {Ye}}, \bibinfo
  {author} {\bibfnamefont {Xiangxi}\ \bibnamefont {Cai}}, \bibinfo {author}
  {\bibfnamefont {Xiying}\ \bibnamefont {Fan}}, \bibinfo {author}
  {\bibfnamefont {Manzhu}\ \bibnamefont {Ke}}, \bibinfo {author} {\bibfnamefont
  {Fan}\ \bibnamefont {Zhang}}, \ and\ \bibinfo {author} {\bibfnamefont
  {Zhengyou}\ \bibnamefont {Liu}},\ }\bibfield  {title} {\enquote {\bibinfo
  {title} {Topological negative refraction of surface acoustic waves in a weyl
  phononic crystal},}\ }\href@noop {} {\bibfield  {journal} {\bibinfo
  {journal} {Nature}\ }\textbf {\bibinfo {volume} {560}},\ \bibinfo {pages}
  {61--64} (\bibinfo {year} {2018})}\BibitemShut {NoStop}%
\bibitem [{\citenamefont {Xue}\ \emph {et~al.}(2019)\citenamefont {Xue},
  \citenamefont {Yang}, \citenamefont {Gao}, \citenamefont {Chong},\ and\
  \citenamefont {Zhang}}]{am7}%
  \BibitemOpen
  \bibfield  {author} {\bibinfo {author} {\bibfnamefont {H.}~\bibnamefont
  {Xue}}, \bibinfo {author} {\bibfnamefont {Y.}~\bibnamefont {Yang}}, \bibinfo
  {author} {\bibfnamefont {F.}~\bibnamefont {Gao}}, \bibinfo {author}
  {\bibfnamefont {Y.}~\bibnamefont {Chong}}, \ and\ \bibinfo {author}
  {\bibfnamefont {B.}~\bibnamefont {Zhang}},\ }\bibfield  {title} {\enquote
  {\bibinfo {title} {Acoustic higher-order topological insulator on a kagome
  lattice},}\ }\href@noop {} {\bibfield  {journal} {\bibinfo  {journal} {Nat.
  Mater.}\ }\textbf {\bibinfo {volume} {18}},\ \bibinfo {pages} {108--112}
  (\bibinfo {year} {2019})}\BibitemShut {NoStop}%
\bibitem [{\citenamefont {Ni}\ \emph {et~al.}(2019)\citenamefont {Ni},
  \citenamefont {Weiner}, \citenamefont {Al\'u},\ and\ \citenamefont
  {Khanikaev}}]{am8}%
  \BibitemOpen
  \bibfield  {author} {\bibinfo {author} {\bibfnamefont {X.}~\bibnamefont
  {Ni}}, \bibinfo {author} {\bibfnamefont {M.}~\bibnamefont {Weiner}}, \bibinfo
  {author} {\bibfnamefont {A}~\bibnamefont {Al\'u}}, \ and\ \bibinfo {author}
  {\bibfnamefont {A.~B.}\ \bibnamefont {Khanikaev}},\ }\bibfield  {title}
  {\enquote {\bibinfo {title} {Observation of higher-order topological acoustic
  states protected by generalized chiral symmetry},}\ }\href@noop {} {\bibfield
   {journal} {\bibinfo  {journal} {Nat. Mater.}\ }\textbf {\bibinfo {volume}
  {18}},\ \bibinfo {pages} {113–120} (\bibinfo {year} {2019})}\BibitemShut
  {NoStop}%
\bibitem [{\citenamefont {Zhang}\ \emph {et~al.}(2019)\citenamefont {Zhang},
  \citenamefont {Wang}, \citenamefont {Lin}, \citenamefont {Tian},
  \citenamefont {Xie}, \citenamefont {Lu}, \citenamefont {Chen},\ and\
  \citenamefont {Jiang}}]{am9}%
  \BibitemOpen
  \bibfield  {author} {\bibinfo {author} {\bibfnamefont {X.}~\bibnamefont
  {Zhang}}, \bibinfo {author} {\bibfnamefont {H.-X.}\ \bibnamefont {Wang}},
  \bibinfo {author} {\bibfnamefont {Z.-K.}\ \bibnamefont {Lin}}, \bibinfo
  {author} {\bibfnamefont {Y.}~\bibnamefont {Tian}}, \bibinfo {author}
  {\bibfnamefont {B.}~\bibnamefont {Xie}}, \bibinfo {author} {\bibfnamefont
  {M.-H.}\ \bibnamefont {Lu}}, \bibinfo {author} {\bibfnamefont {Y.-F.}\
  \bibnamefont {Chen}}, \ and\ \bibinfo {author} {\bibfnamefont {J.-H.}\
  \bibnamefont {Jiang}},\ }\bibfield  {title} {\enquote {\bibinfo {title}
  {Second-order topology and multidimensional topological transitions in sonic
  crystals},}\ }\href@noop {} {\bibfield  {journal} {\bibinfo  {journal} {Nat.
  Phys.}\ }\textbf {\bibinfo {volume} {15}},\ \bibinfo {pages} {582–588}
  (\bibinfo {year} {2019})}\BibitemShut {NoStop}%
\bibitem [{\citenamefont {Lidorikis}\ \emph {et~al.}(1998)\citenamefont
  {Lidorikis}, \citenamefont {Sigalas}, \citenamefont {Economou},\ and\
  \citenamefont {Soukoulis}}]{PhysRevLett.81.1405}%
  \BibitemOpen
  \bibfield  {author} {\bibinfo {author} {\bibfnamefont {E.}~\bibnamefont
  {Lidorikis}}, \bibinfo {author} {\bibfnamefont {M.~M.}\ \bibnamefont
  {Sigalas}}, \bibinfo {author} {\bibfnamefont {E.~N.}\ \bibnamefont
  {Economou}}, \ and\ \bibinfo {author} {\bibfnamefont {C.~M.}\ \bibnamefont
  {Soukoulis}},\ }\bibfield  {title} {\enquote {\bibinfo {title} {Tight-binding
  parametrization for photonic band gap materials},}\ }\href {\doibase
  10.1103/PhysRevLett.81.1405} {\bibfield  {journal} {\bibinfo  {journal}
  {Phys. Rev. Lett.}\ }\textbf {\bibinfo {volume} {81}},\ \bibinfo {pages}
  {1405--1408} (\bibinfo {year} {1998})}\BibitemShut {NoStop}%
\end{thebibliography}%

\clearpage


\clearpage

\newpage

\setcounter{equation}{0}
\setcounter{page}{1}
\setcounter{figure}{0}
\renewcommand\thefigure{{S-\arabic{figure}}}
\renewcommand\thesection{\Alph{section}}
\renewcommand\thesubsection{\arabic{subsection}}


\section{More on Dirac strings}\label{SM:DS}
\begin{figure}
 \centering
 \includegraphics[width=0.3\linewidth]{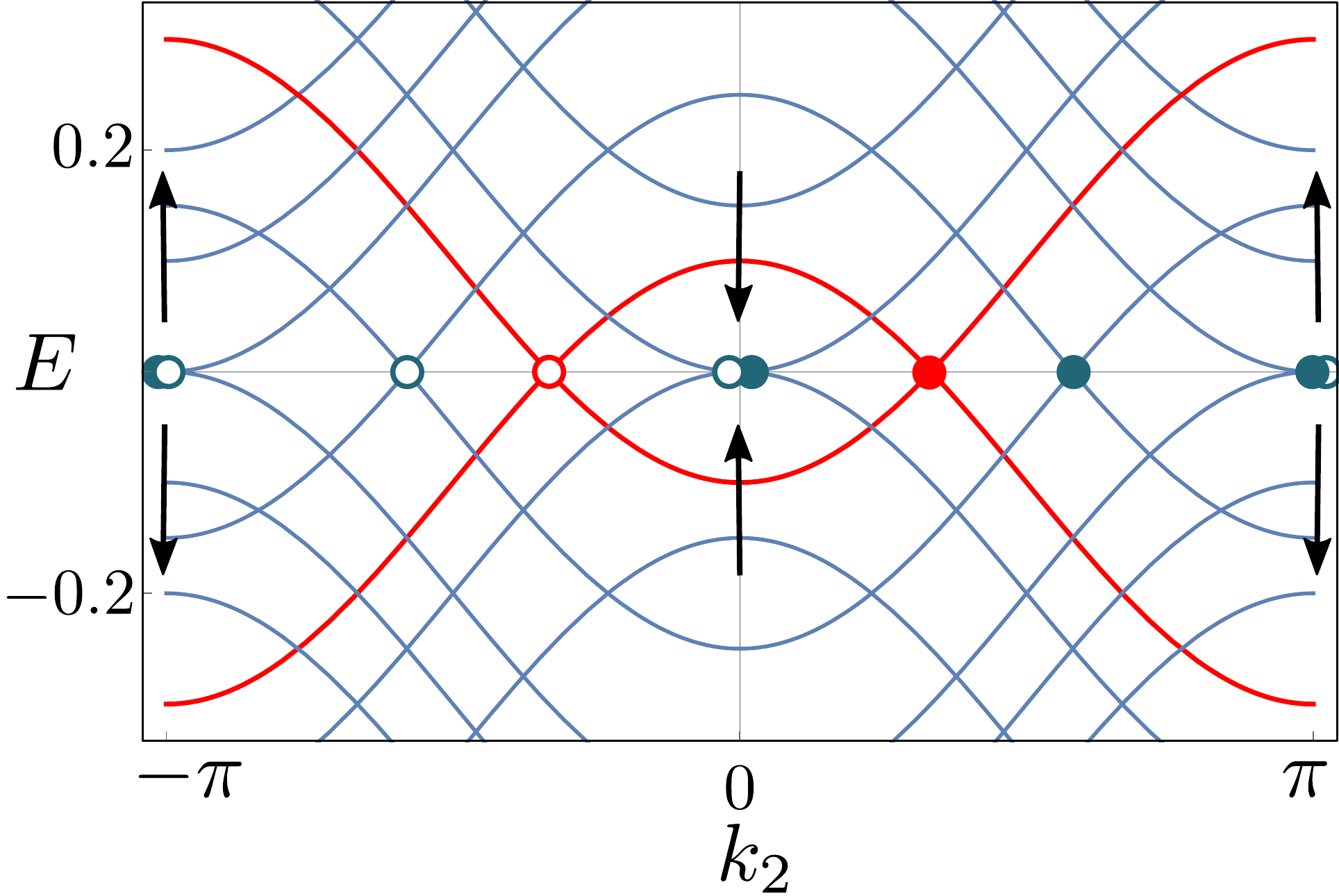}
 \caption{{\it Dirac string creation in the two-band model (see text). Upon increasing the control parameter $t$, a pair of nodes is created at the $\Gamma$-point that move outward and recombine over the edge of the Brillouinz zone. This thus creates a Dirac string over the $k_1$ line for $t>6/5$.}}
    \label{fig:Nodes_pair_1D}
\end{figure}

We here wish to further elaborate on the Dirac strings (DS) as motivated in the main text. As a first step we can make matters more concrete by showing how a DS is formed in the context of a simple model. To this end consider a two-band model $H = h_x \sigma_x + h_z \sigma_z$ with $h_x(\boldsymbol{k}) = \sin k_1$ and $h_z(\boldsymbol{k}) = 1-1/5 \cos k_2 - t \cos k_1$ for $t\in[0,2]$. The two bands are gapped for $0\leq t<4/5$, then a band inversion occurs at $(k_1,k_2)=(0,0)$ for $t\geq 4/5$ from which two simple nodes are produced on the $k_1=0$ line that move outward for increasing $t$, see Fig.~\ref{fig:Nodes_pair_1D}. 
When the nodes annihilate for $t=6/5$ at the Brillouin zone boundary, the system exhibits a non-contractible Dirac string at $k_1=0$ that can be shown to mark a switch in orientation.

\begin{figure}
 \centering
 \includegraphics[width=0.5\linewidth]{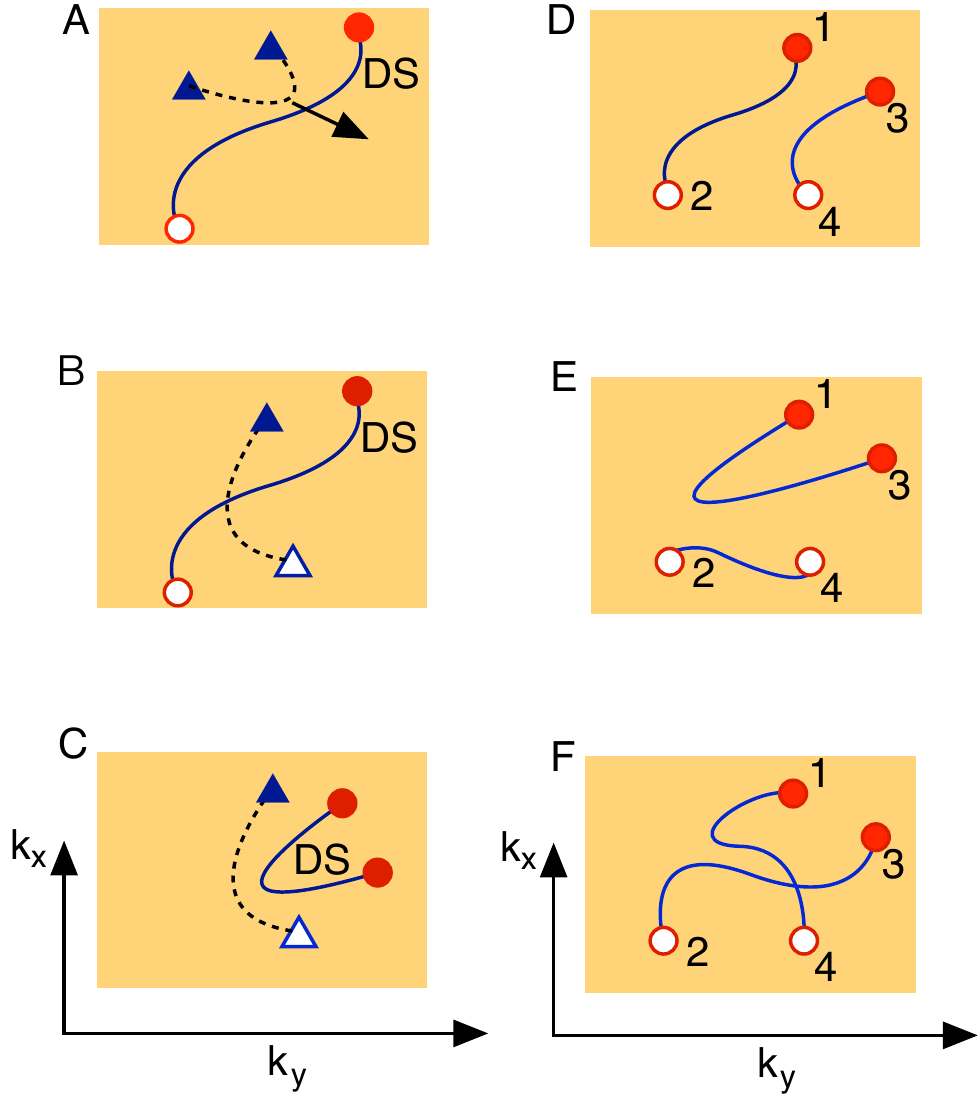}
 \caption{{\it Schematic illustration of action of Dirac string (DS). The left column shown the effect of a DS residing in the second gap on nodes in the first gap. The charge inversion processes of pulling a first-gap node through this DS  or subsequently retracting the string over the other DS that connects the pair in the first gap is illustrated in the bottom panels, respectively. The right column depicts the recombination rules of Dirac strings as outlined in the main text. Every panel is merely a different representation of the same physics.}}
    \label{fig:DS_recombinationrules}
\end{figure}

\begin{figure}
 \centering
 \includegraphics[width=0.5\linewidth]{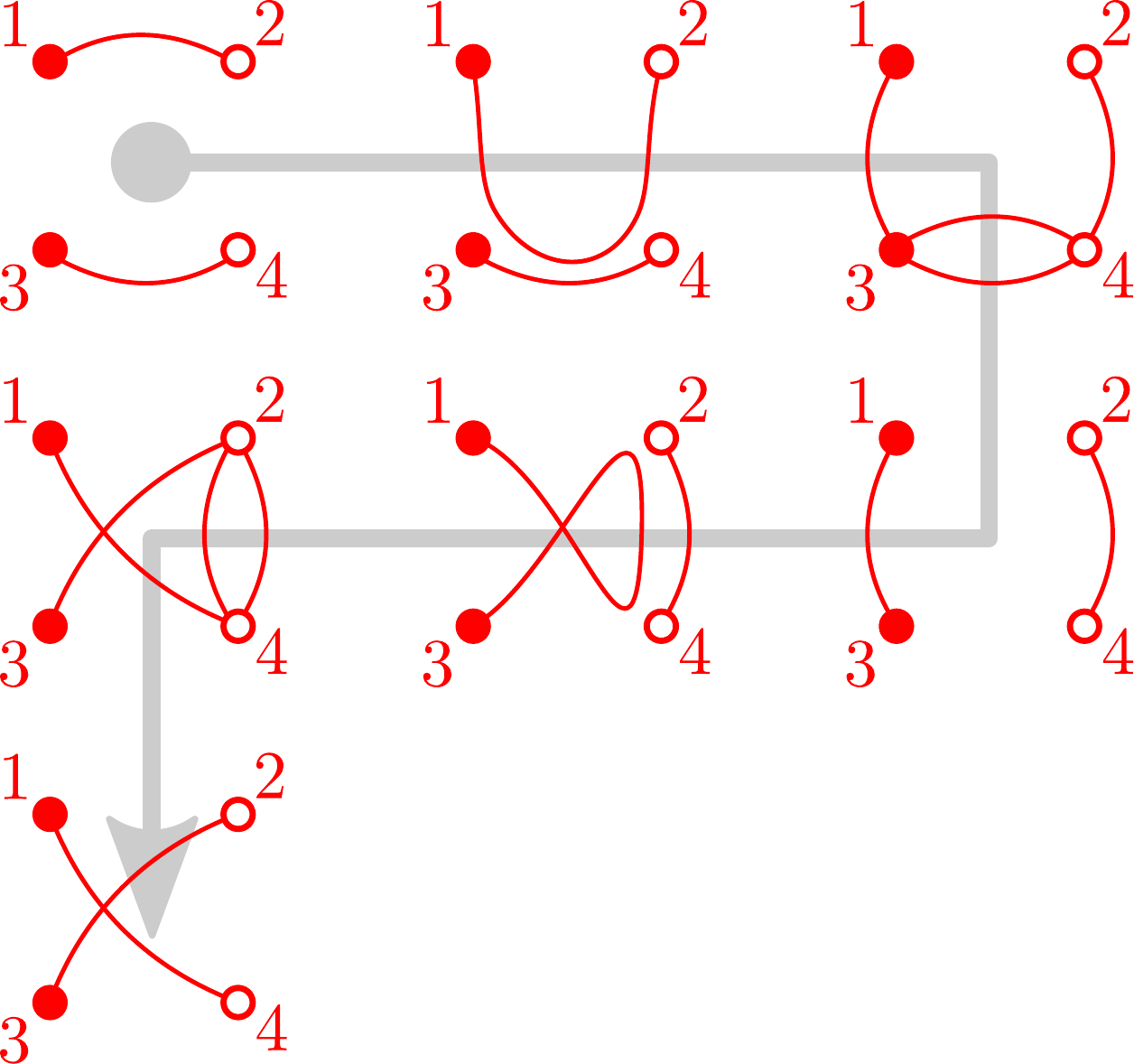}
 \caption{{\it Dirac string recombination. As stated in the main text, the Dirac string (DS) configuration at the first, fourth and final stage are the same. Deforming DS strings and using that crossing a double DS always results in a trivial phase factor ($(-1)^2=1$), we see that these configurations indeed amount to the same physics, underpinning the outlined recombination rules. }}
    \label{fig:DS_recombination}
\end{figure}

As a next step, we make the recombination rules more insightful. In the main text we already motivated the rules concerning charge conversion when nodes of the adjacent gap pass a Dirac string. In addition, we outlined that upon permuting the end point of a DS, we obtain a set of similar configurations that merely form different representations of the same system. For concreteness we recapitulate these charge conversion and recombination rules in Fig.~\ref{fig:DS_recombinationrules}. The consistency of the underlying physics is effectively illustrated in Fig.~\ref{fig:DS_recombination}. Using that passing a double Dirac string always gives a trivial phase factor, owing to the fact that $(-1)^2=1$, we see that the different recombination configurations in the main text can indeed be adiabatically deformed into each other.


\section{The kagome model}
\label{sm:model}
We here elaborate more on the proposed model in the main text. We start from a kagome geometry, see Fig.~2 in the main text. The primitive Bravais lattice vectors are
\begin{align*}
	 \boldsymbol{a}_1 &= a( 3/2   \hat{x} + \sqrt{3}/2 \hat{y} )\\
	 \boldsymbol{a}_2 &= a( -3/2   \hat{x} + \sqrt{3}/2 \hat{y} ), \\
	 \boldsymbol{a}_3 &= \hat{z},  
\end{align*} 
(with the orthonormal Cartesian frame $\langle \hat{x},\hat{y},\hat{z}\rangle $, e.g., $\hat{x} = (1,0,0)$, $\hat{y} = (0,1,0)$, $\hat{z} = (0,0,1)$), and the reciprocal primitive vectors are
\begin{align*}
	 \boldsymbol{b}_1 &= 2\pi \dfrac{\boldsymbol{a}_2 \times  \boldsymbol{a}_3}{ \boldsymbol{a}_1 \cdot (\boldsymbol{a}_2 \times  \boldsymbol{a}_3)} , \\
	 \boldsymbol{b}_2 &= 2\pi \dfrac{\boldsymbol{a}_3 \times  \boldsymbol{a}_1}{ \boldsymbol{a}_1 \cdot (\boldsymbol{a}_2 \times  \boldsymbol{a}_3)} , \\
	 \boldsymbol{b}_3 &= 2\pi \dfrac{\boldsymbol{a}_1 \times  \boldsymbol{a}_2}{ \boldsymbol{a}_1 \cdot (\boldsymbol{a}_2 \times  \boldsymbol{a}_3)} .
\end{align*} 

The Kagome lattice can be obtained from the layer group LG80, with point group $D_{6h} = p6/m mm$, and taking the Wyckoff position (WP) $3c$. We define the position of sub-lattice sites $\{A,B,C\}$  
\begin{align*}
	 \boldsymbol{r}_A &= \bs{a}_1/2, \\
	 \boldsymbol{r}_B &= \bs{a}_2/2, \\
	 \boldsymbol{r}_C &= -(\bs{a}_1+\bs{a}_2)/2,
\end{align*} 

In the following we assume that a single s-wave orbital is occupied on each sub-lattice site. The minimal kagome model is then defined as
\begin{widetext}
\begin{equation}
\label{eq_kagome}
	H_{\mathrm{K}}(\bs{k}) = \left(\begin{array}{ccc} 
		\varepsilon & t c_2(\bs{k}) + t' c'_2(\bs{k}) & t c_1(\bs{k}) + t' c'_1(\bs{k}) \\
		t c_2(\bs{k}) + t' c'_2(\bs{k}) & \varepsilon &  t c_3(\bs{k}) + t' c'_3(\bs{k}) \\
		t c_1(\bs{k}) + t' c'_1(\bs{k})  & t c_3(\bs{k}) + t' c'_3(\bs{k}) & \varepsilon
	\end{array}
	\right),
\end{equation}
\end{widetext}
where we only have included the nearest-neighbor and next-nearest-neighbor hopping terms, where 
\begin{align*}
	c^{(')}_1(\bs{k}) &= \cos( \bs{k}\cdot \bs{\delta}^{(')}_1) , \\
	c^{(')}_2(\bs{k}) &= \cos( \bs{k}\cdot \bs{\delta}^{(')}_2) , \\
	c^{(')}_3(\bs{k}) &= \cos( \bs{k}\cdot \bs{\delta}^{(')}_3) ,
\end{align*} 
with the nearest-neighbor bond vectors 
\begin{align*}
	\bs{\delta}_1 &= -\bs{r}_B , \\
	\bs{\delta}_2 &= -\bs{r}_C , \\
	\bs{\delta}_3 &= -\bs{r}_A , 
\end{align*} 
and the next-nearest-neighbor bond vectors
\begin{align*}
	\bs{\delta}'_1 &= \bs{r}_A - \bs{r}_C , \\
	\bs{\delta}'_2 &= \bs{r}_B - \bs{r}_A , \\
	\bs{\delta}'_3 &= \bs{r}_C -\bs{r}_B .
\end{align*} 

Let us write the momentum in units of the reciprocal lattice vectors as $\bs{k} = k_1 \bs{b}_1/(2\pi) + k_2 \bs{b}_2/(2\pi)$, with $(k_1,k_2) \in [-\pi,\pi]^2$ inside the Brillouin zone. We then have 
\begin{equation}
\begin{aligned}
	c_1(\bs{k}) &= \cos [k_2 / 2 ], \\
	c_2(\bs{k}) &=  \cos [(k_1 +k_2)/2] , \\
	c_3(\bs{k}) &= \cos [k_1/2 ] , 
\end{aligned}
\end{equation}
and 
\begin{equation}
\begin{aligned}
	c'_1(\bs{k}) &= \cos [k_1 + k_2/2] , \\
	c'_2(\bs{k}) &= \cos [(k_1 - k_2)/2] , \\
	c'_3(\bs{k}) &= \cos [k_1/2 + k_2 ],
\end{aligned}
\end{equation}
which we can readily substitute in $H_{\mathrm{K}}$ Eq.~(\ref{eq_kagome}).


\section{Band structures and IRREP ordering through the complete braiding process}
\label{SM:fullbraiding}

In conjunction with the detailed enumeration of the charge conversion processes in the main text (Fig.~2), we here give the band structure corresponding to each step of the braiding process together with all the relevant irreducible representations (IRREPs) over the Brillouin zone. This allows us to relate the movement of the nodes to band inversions at specific points and regions of the Brillouin zone. By doing so, we have a simple prediction of the nodes that must (dis-)appear together by symmetry inside the Brillouin zone, and similarly for the existence of triply degenerate points at $K$ and $\Gamma$.

Starting with Fig.~\ref{BS_SIs}{\bf a)} the set of IRREPs correspond to the induced IRREPs for the elementary band representation of $s$-orbitals maximally localized at the three sub-lattice sites of Wyckoff position $3c$ for the layer group $L80$. Since the basal mirror symmetry ($\sigma_h :  z \rightarrow -z$) does not play any role here, it is convenient to use the IRREPs of point group $C_{6v}$ (instead of $D_{6h}$) \cite{ITCA,ITCE}. 

Fig.~\ref{BS_SIs} provides the data with the IRREP reordering upon the band inversions associated to the symmetry-constrained braiding of nodes, as discussed in detail in the main text. 


\clearpage
\begin{figure*}[t!]
	\centering
	\begin{tabular}{|c|c|c|}
	\hline
    \textbf{a)}~$(t,t')=(-1,0)$~\textbf{[E]}  & 
    \textbf{b)}~$(t,t')=(-0.72,-0.28)~$\textbf{[F]} &
	\textbf{c)}~$(t,t')=(-0.4,-0.2)$~\textbf{[G]} 
\\
	\includegraphics[width=0.22\linewidth]{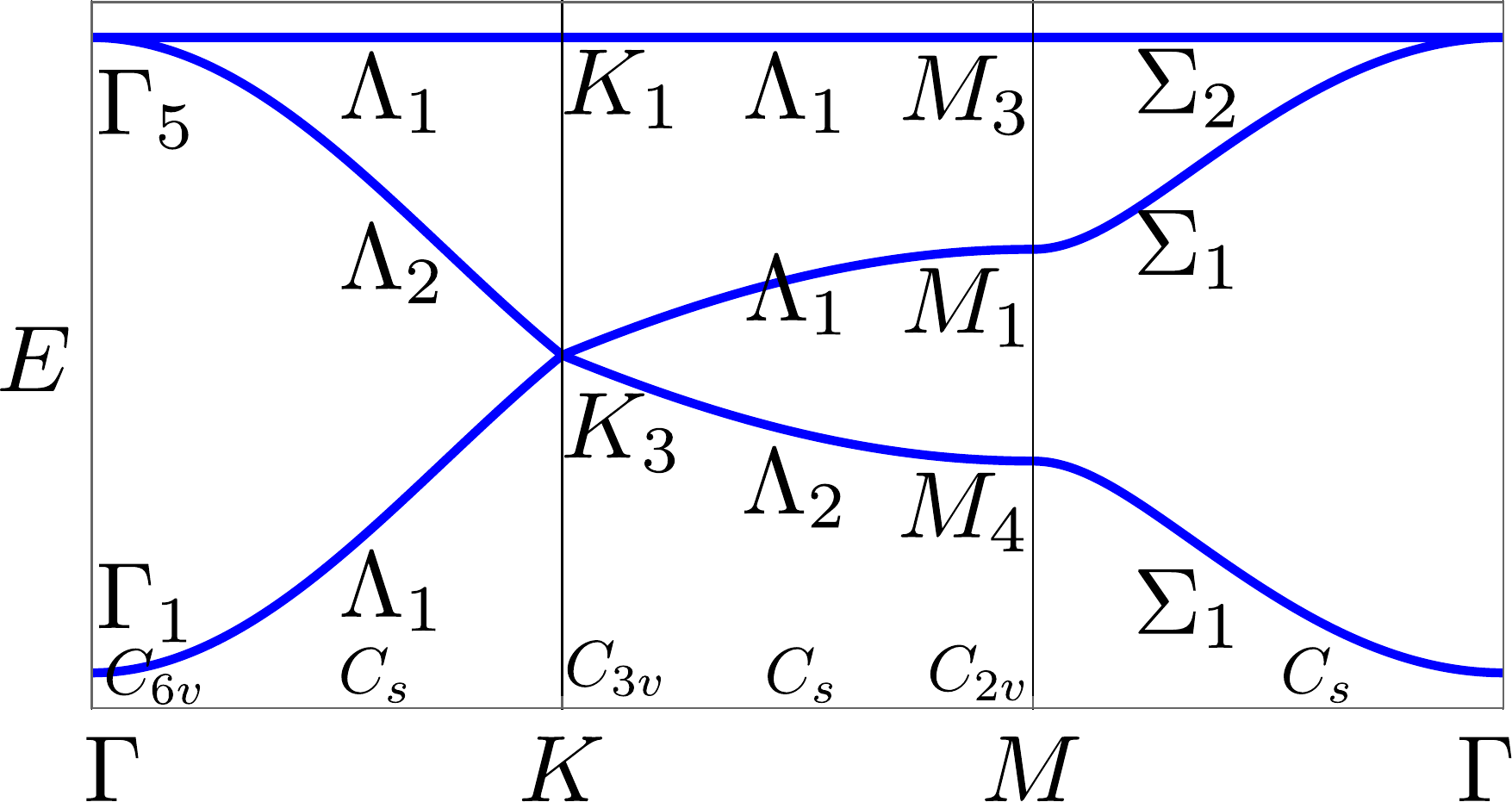} &
	\includegraphics[width=0.22\linewidth]{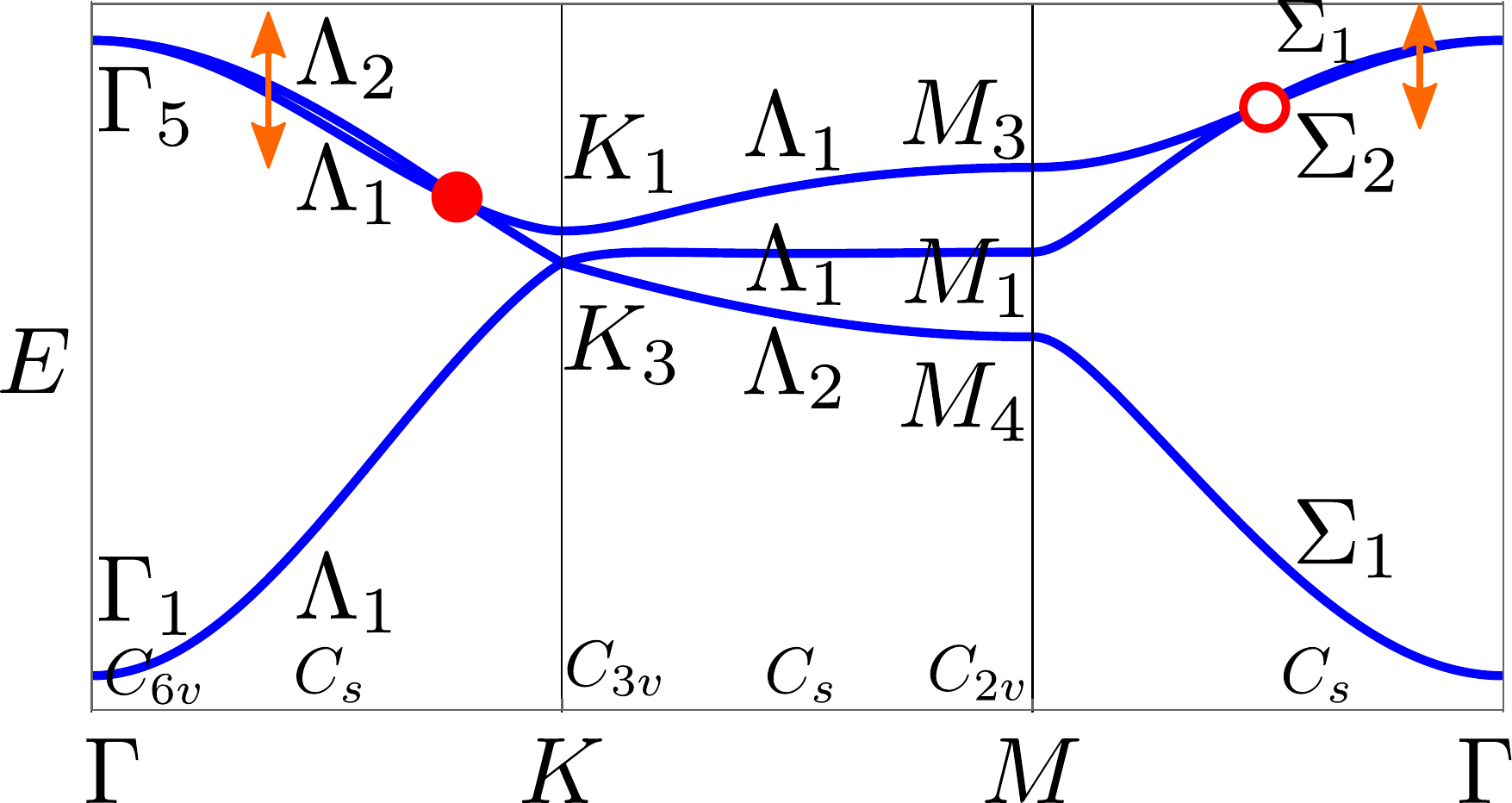} &
	\includegraphics[width=0.22\linewidth]{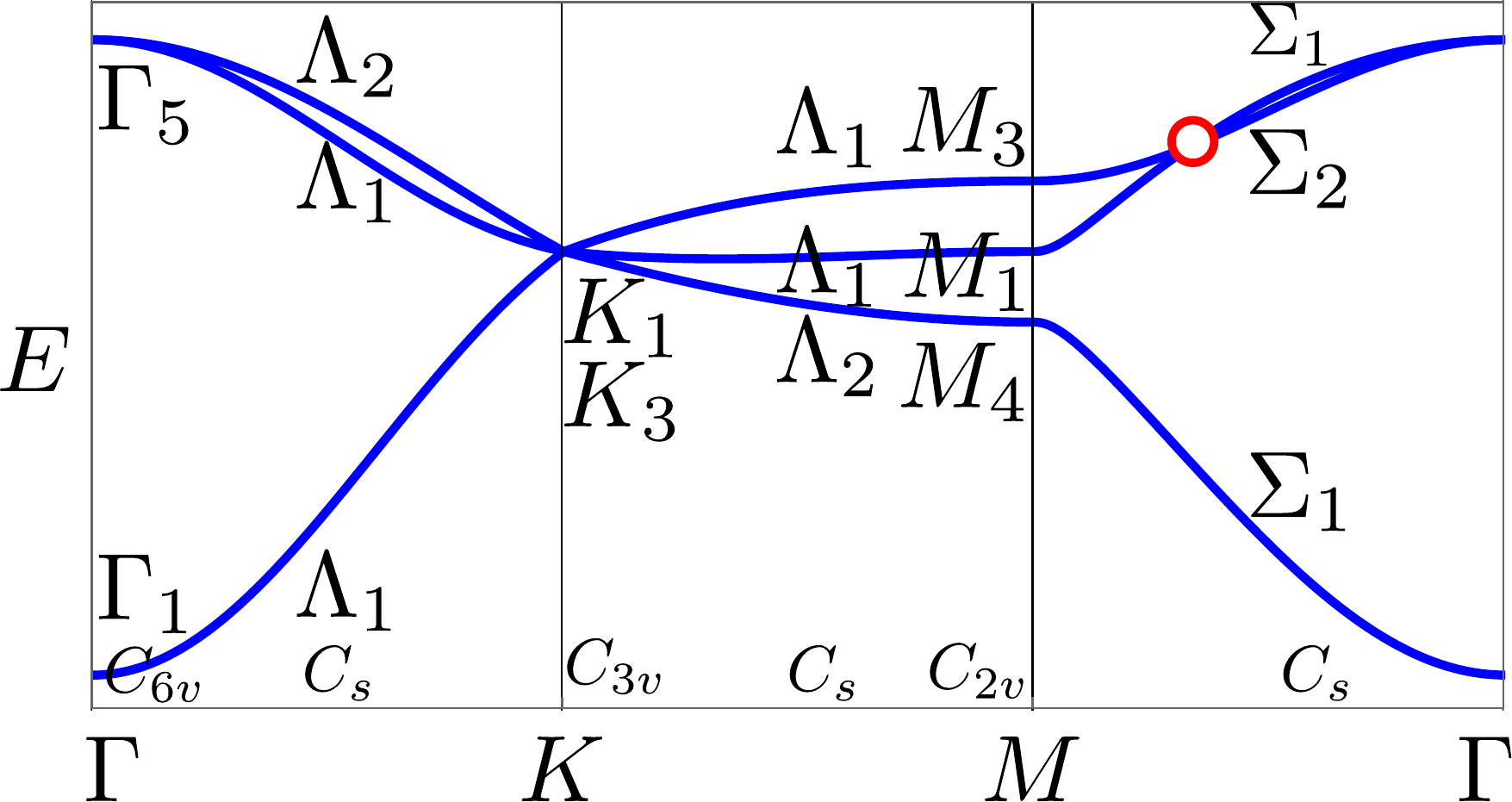}  \\
	\includegraphics[height=0.24\linewidth]{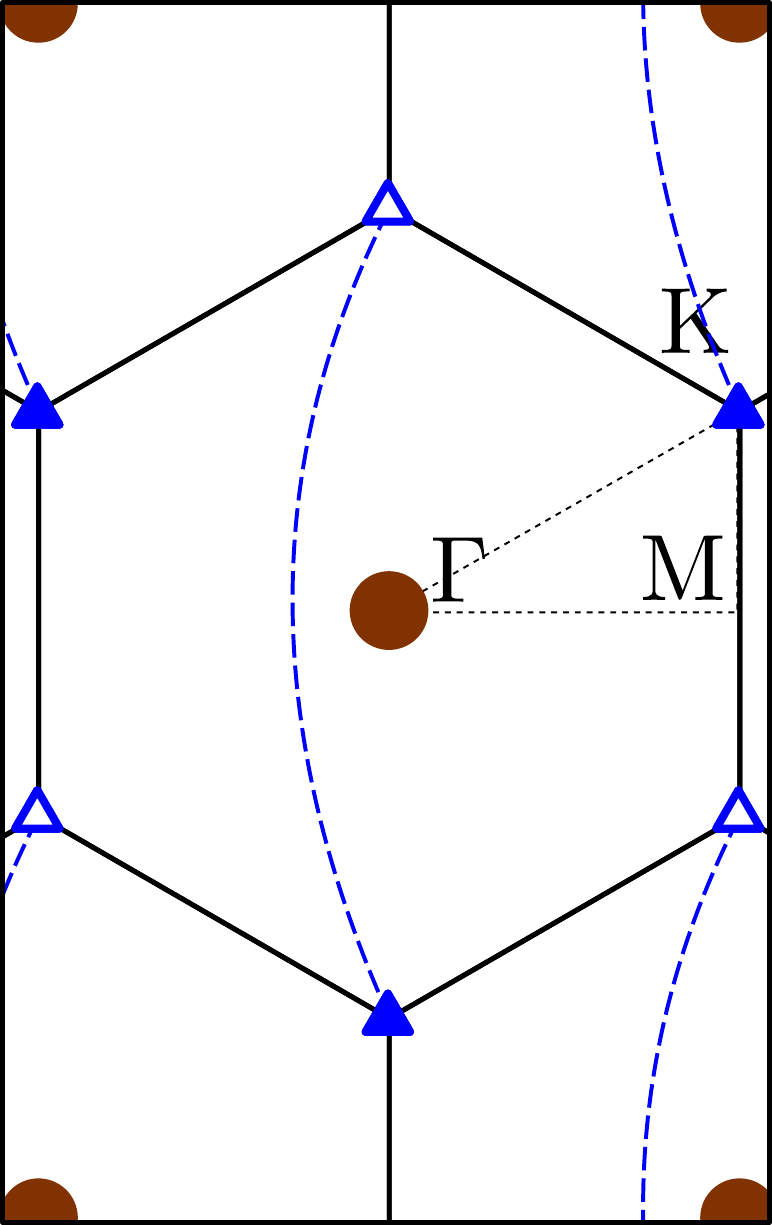} &
	\includegraphics[height=0.24\linewidth]{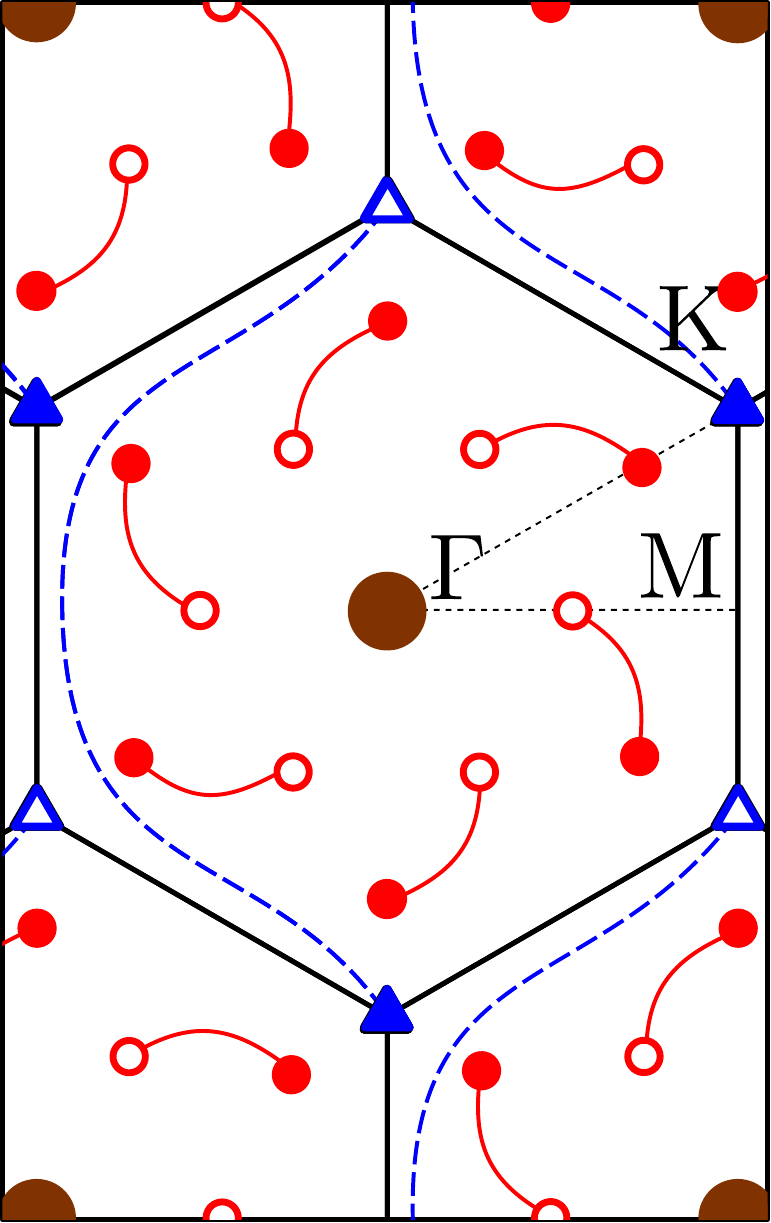} &
	\includegraphics[height=0.24\linewidth]{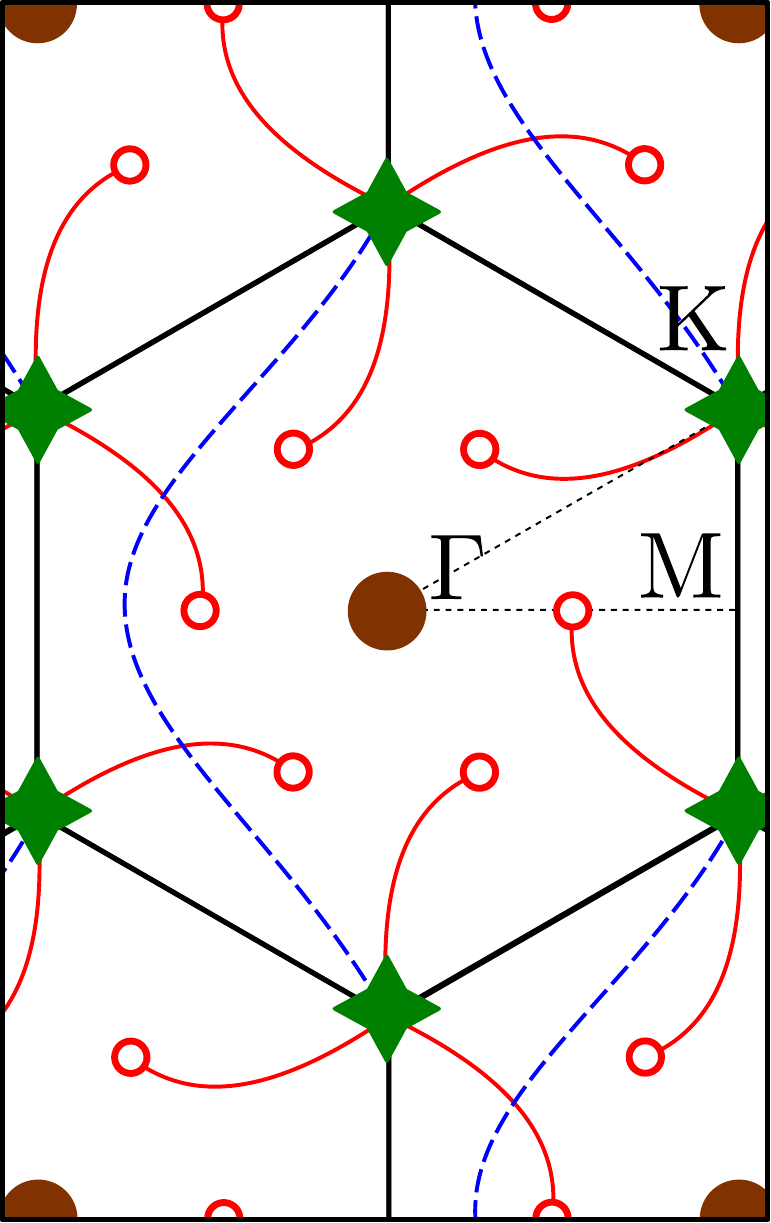} \\
	\hline
        &\textbf{d)}~$(t,t')=(-0.32,-0.23)$~\textbf{[H]}&\\	
        &\includegraphics[width=0.25\linewidth]{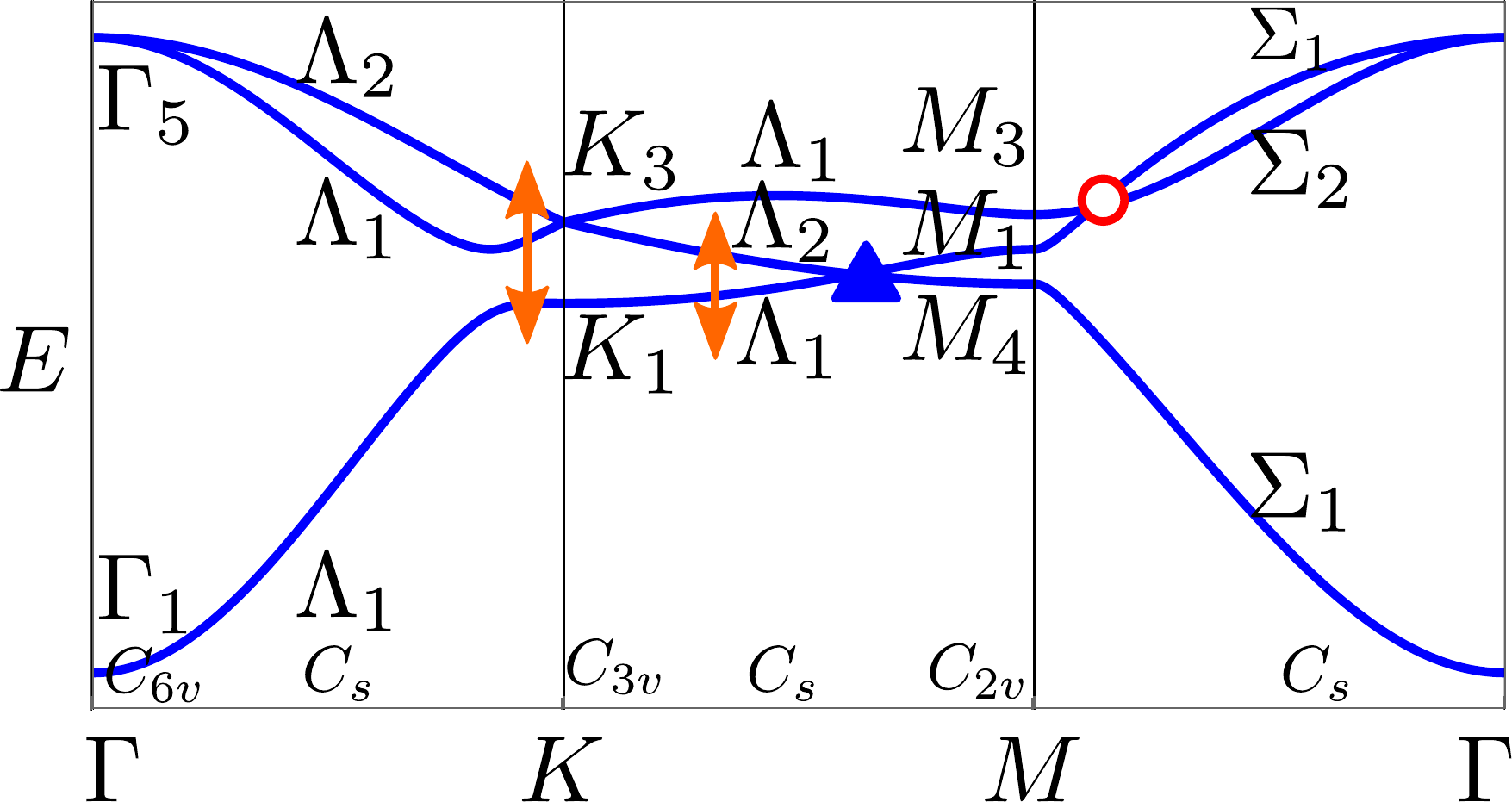}&\\
        &\includegraphics[height=0.24\linewidth]{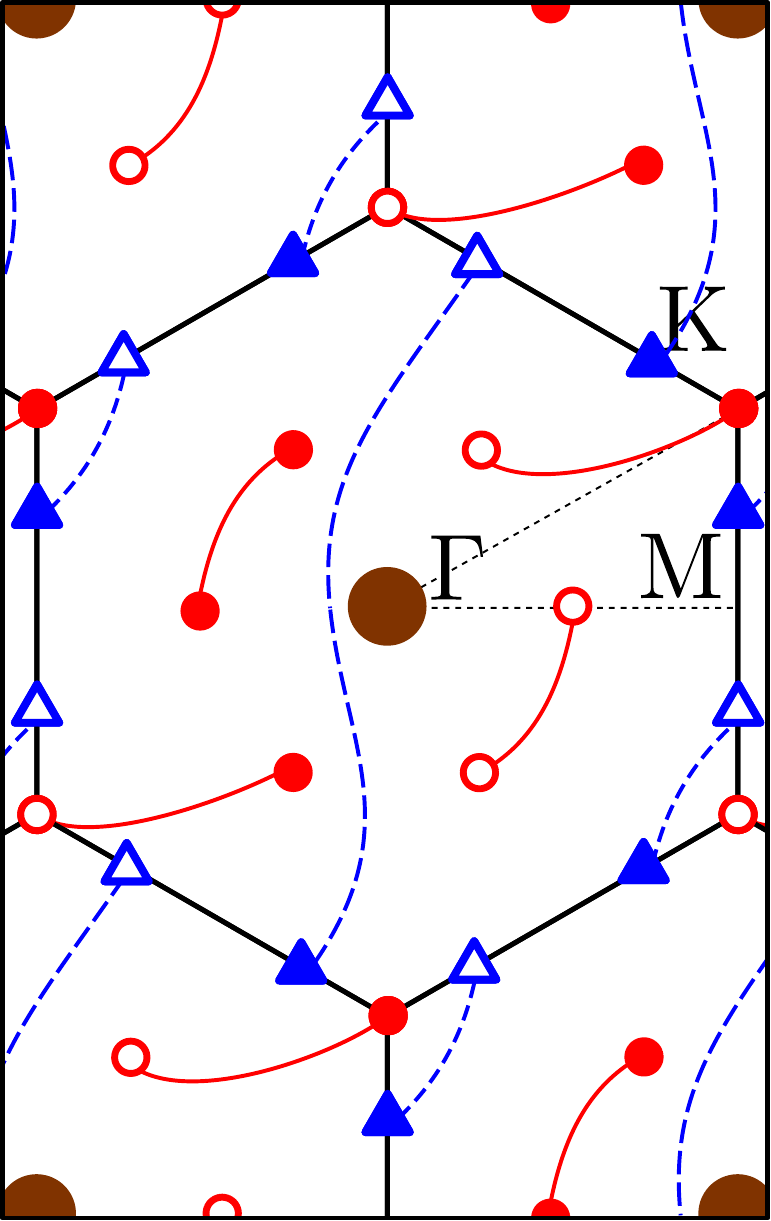}& \\
        \hline
	\textbf{e)}~$(t,t')=(-0.25,-0.25)$~\textbf{[I]} & \textbf{f)}~$(t,t')=(-0.1,-0.3)$~\textbf{[J]} &
	\textbf{g)}~$(t,t')=(t,t')=(0,-1)$~\textbf{[K]}      \\
	\includegraphics[width=0.22\linewidth]{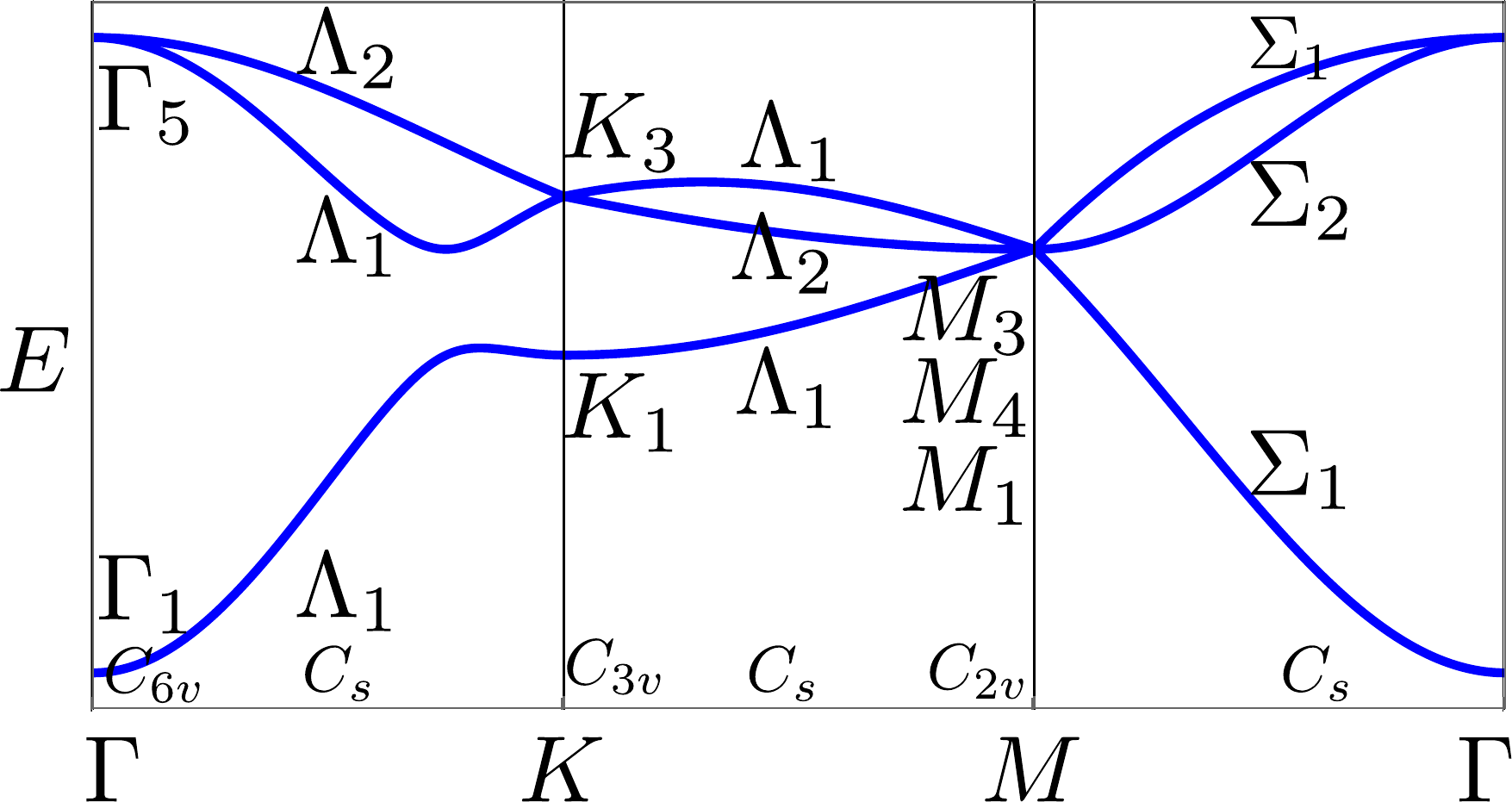} &
	\includegraphics[width=0.22\linewidth]{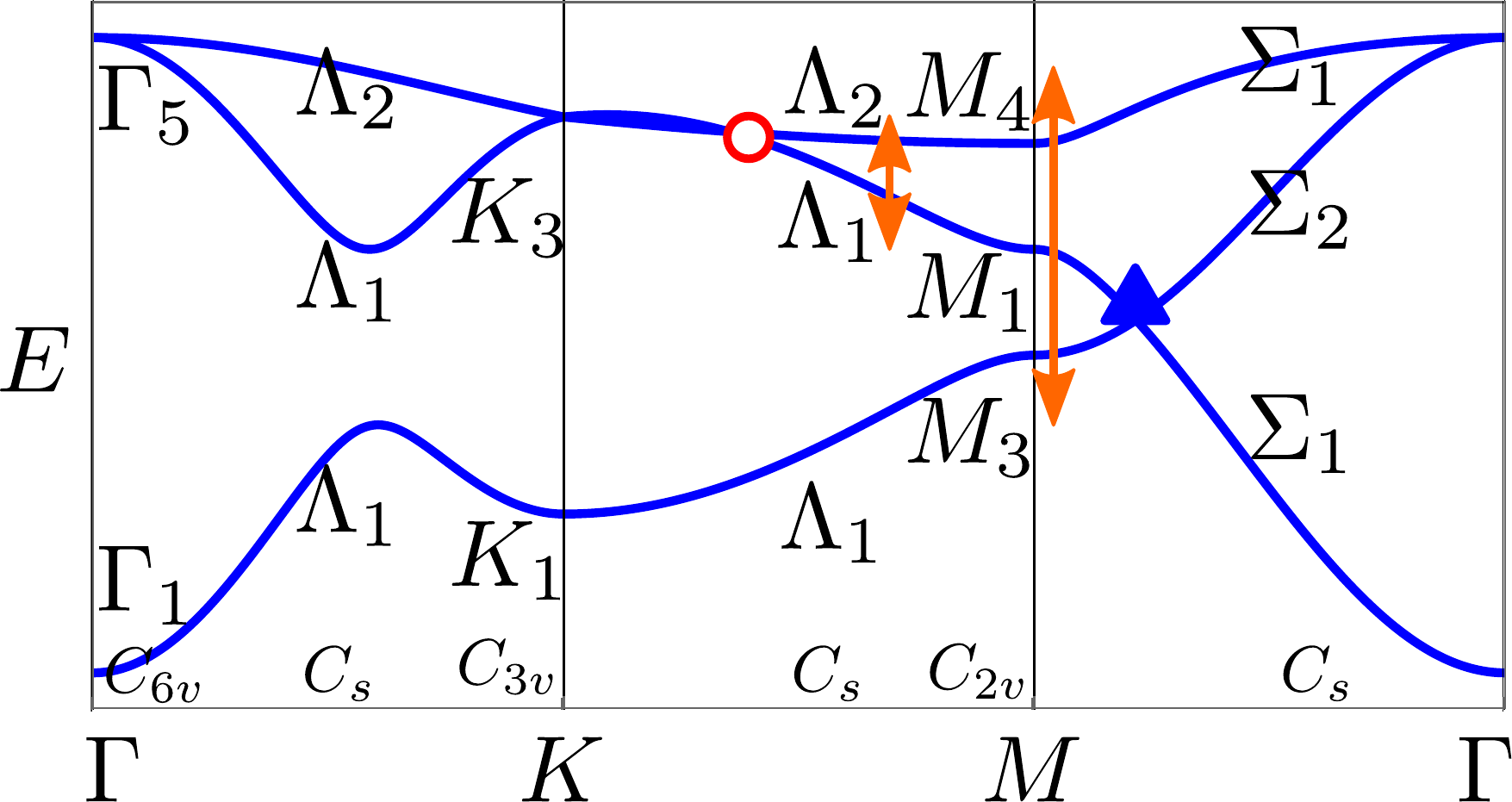} &
	\includegraphics[width=0.22\linewidth]{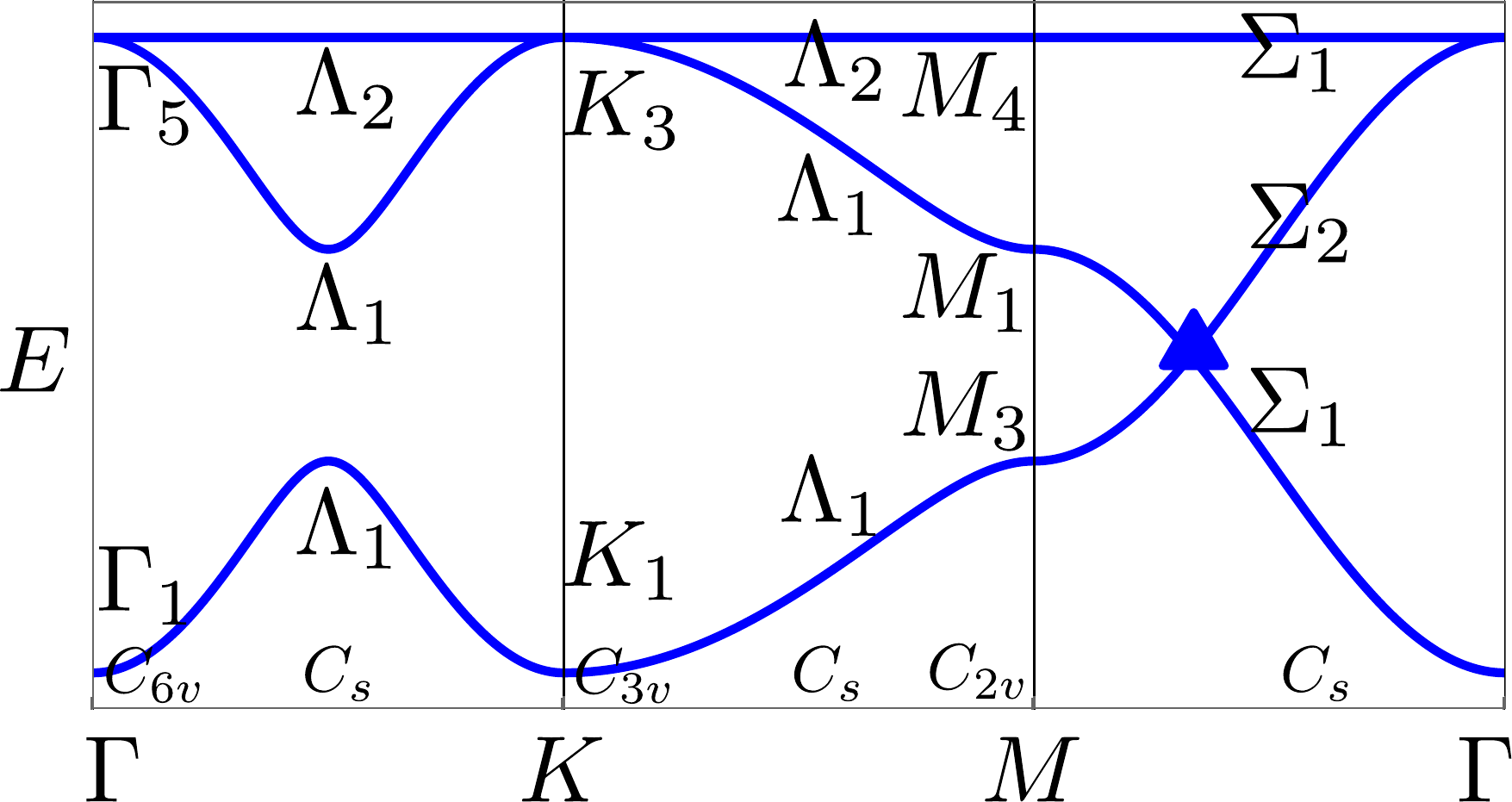}  
	 \\
	\includegraphics[height=0.24\linewidth]{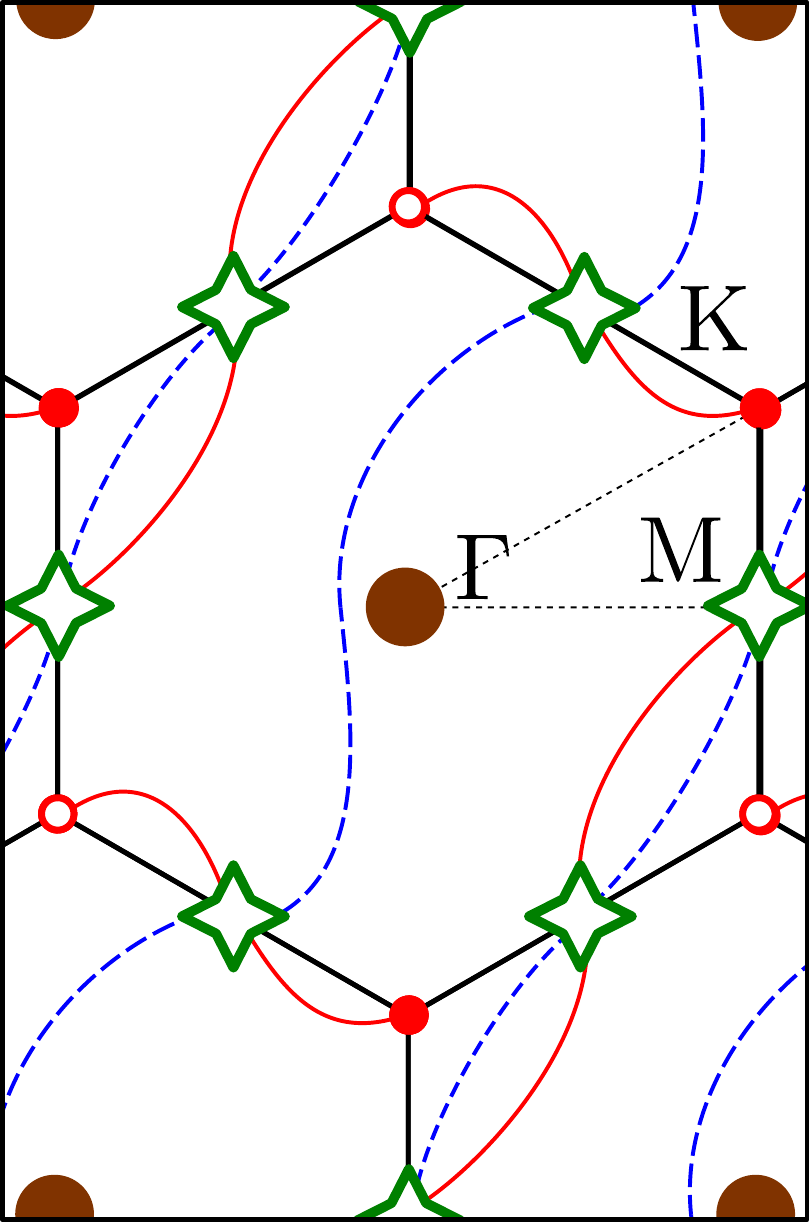} &
	\includegraphics[height=0.24\linewidth]{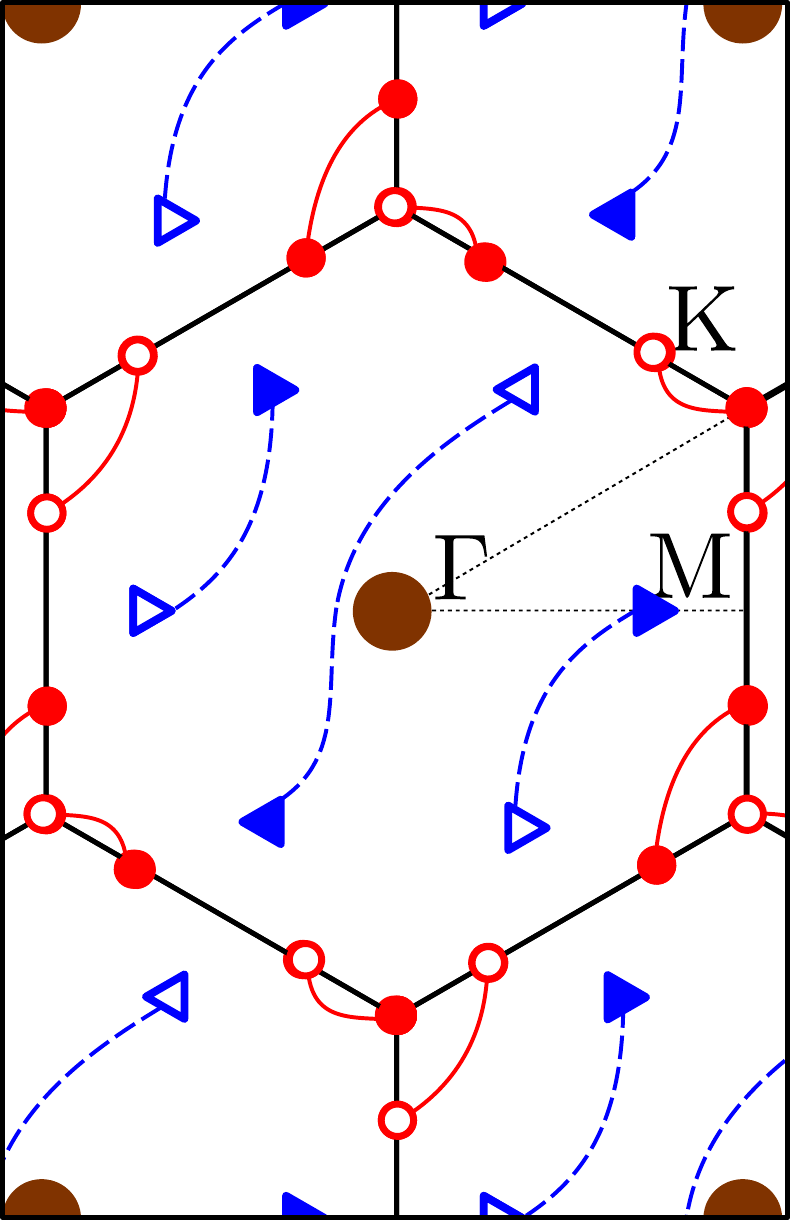} &
	\includegraphics[height=0.24\linewidth]{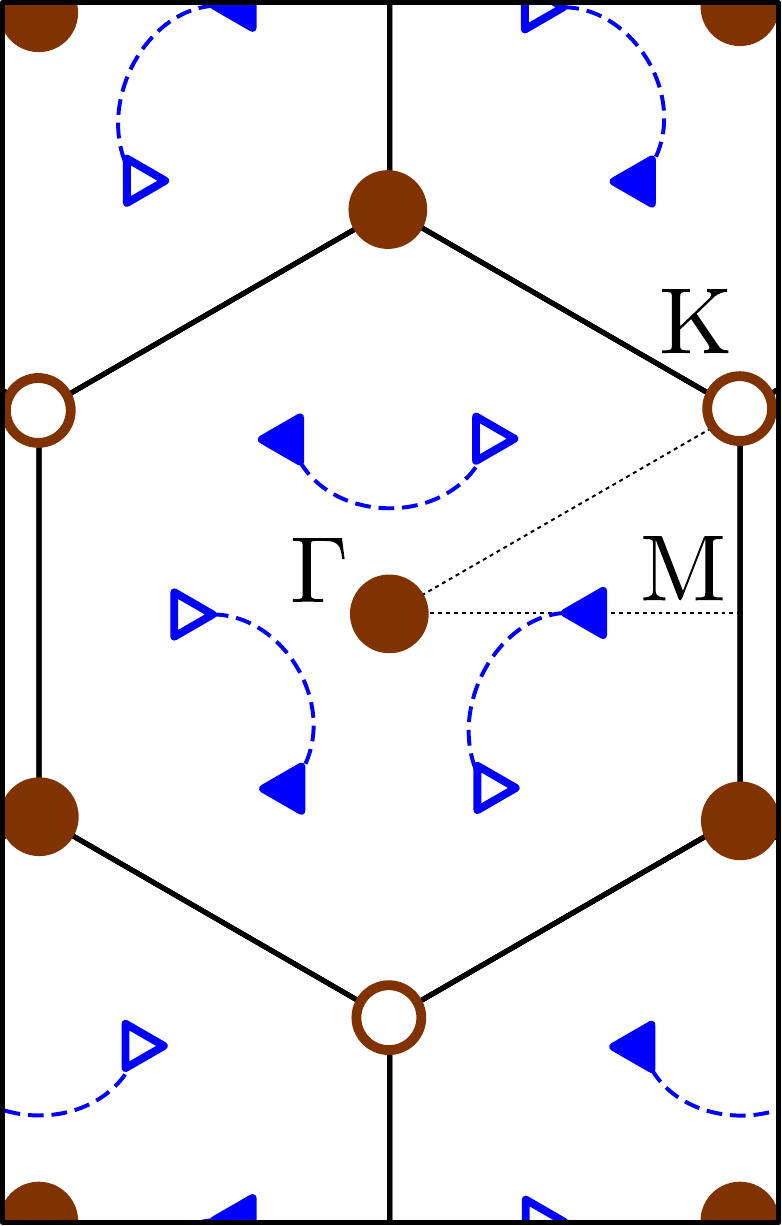} 
	 \\
	\hline
	\end{tabular}
\caption{{\it Symmetry enforced nodes inside the Brillouin zone at band inversions (marked by orange arrows) through the braiding process in the kagome model. The IRREPs refer to the little co-groups of the high-symmetry momenta written below. We show the bulk configuration of charges below each phase for convenience and we give in brackets the label of the corresponding panel in Fig.~\ref{kagome_bulk_charges}. 
} }
\label{BS_SIs}
\end{figure*}

\clearpage

\begin{figure*}[t!]
	\centering
	\begin{tabular}{|c|c|c|}
	\hline
	&&\\
	\textbf{a)}~$(t,t')=(0.2,-0.27)$ &
    \textbf{b)}~$(t,t')=(0.25,-0.25)$ &
    \textbf{c)}~$(t,t')=(0.3,-0.23)$   \\
    &&\\
    \includegraphics[width=0.25\linewidth]{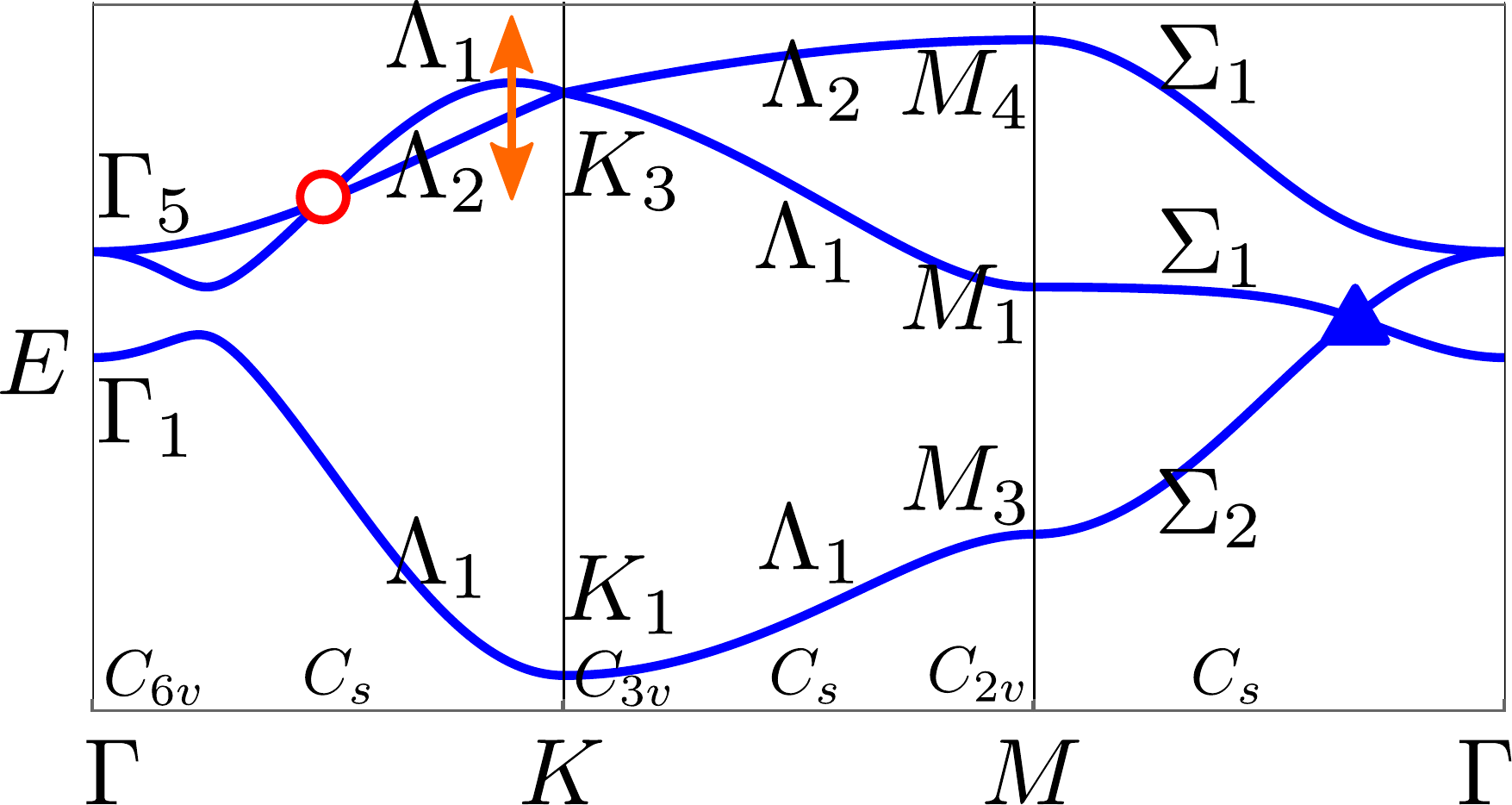} &
	\includegraphics[width=0.25\linewidth]{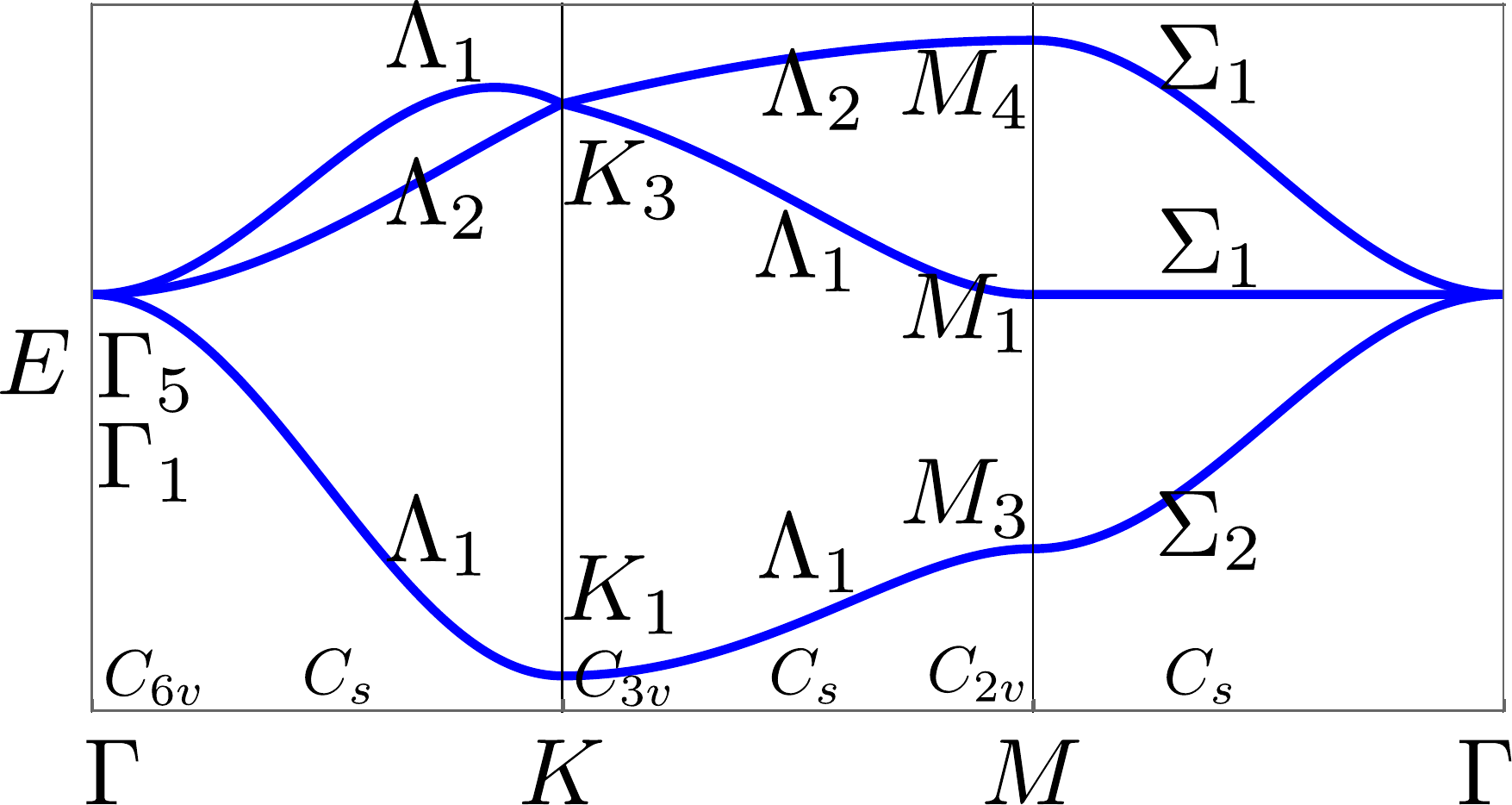} & 
	\includegraphics[width=0.25\linewidth]{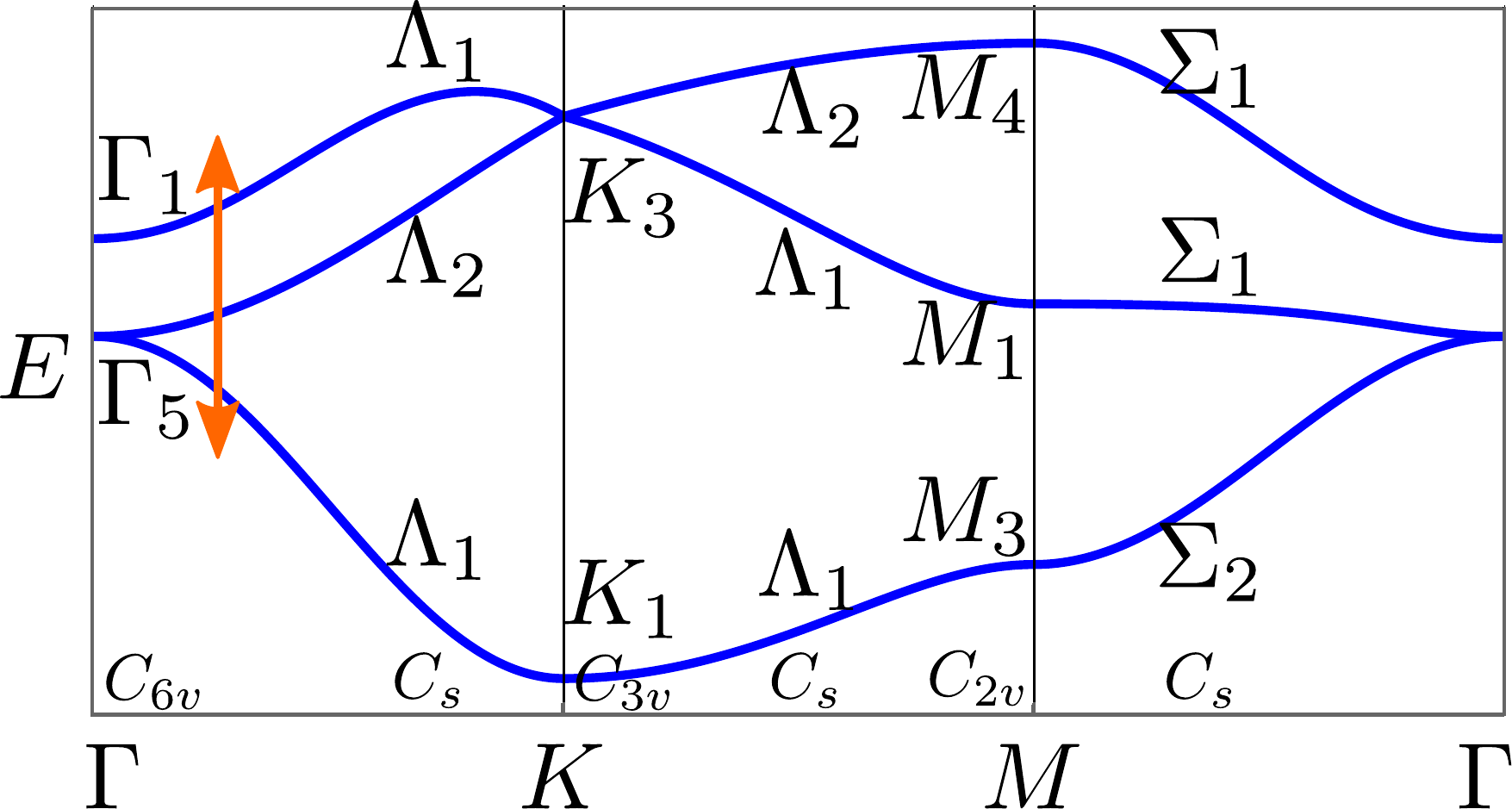}   \\
	\includegraphics[height=0.24\linewidth]{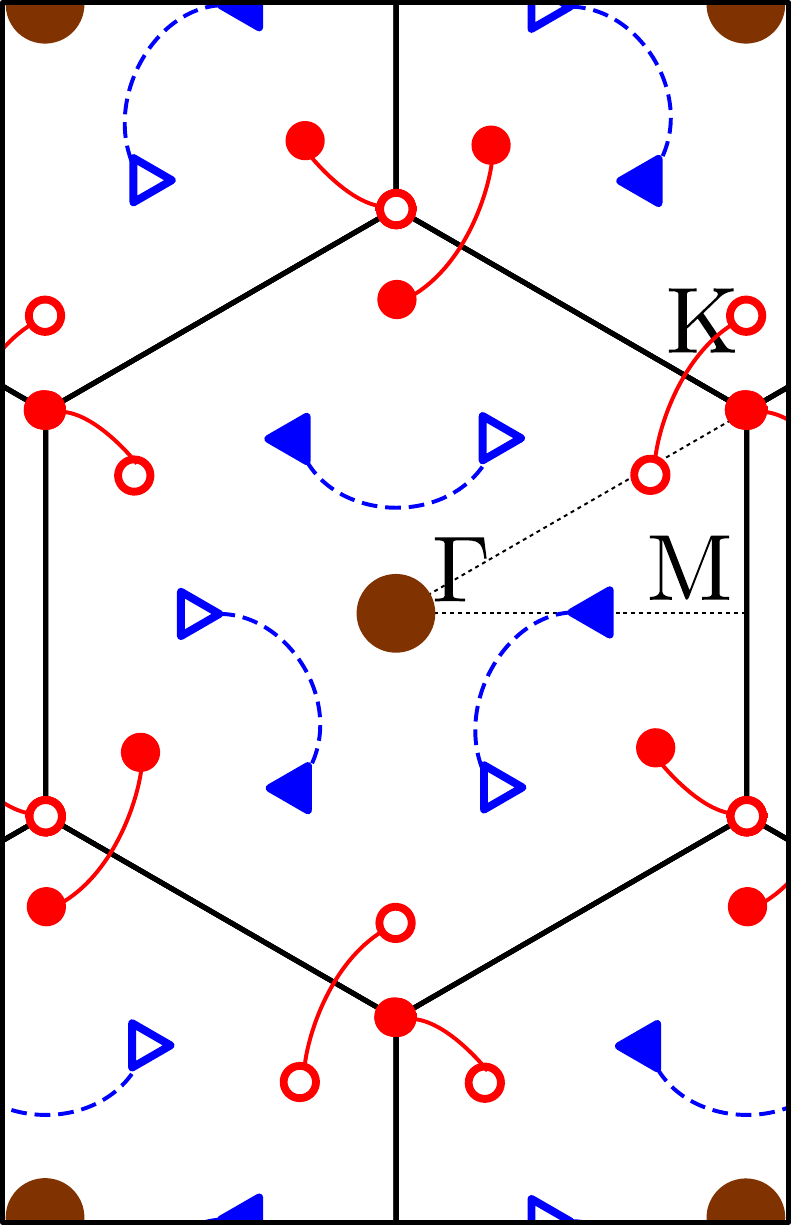} &
	\includegraphics[height=0.24\linewidth]{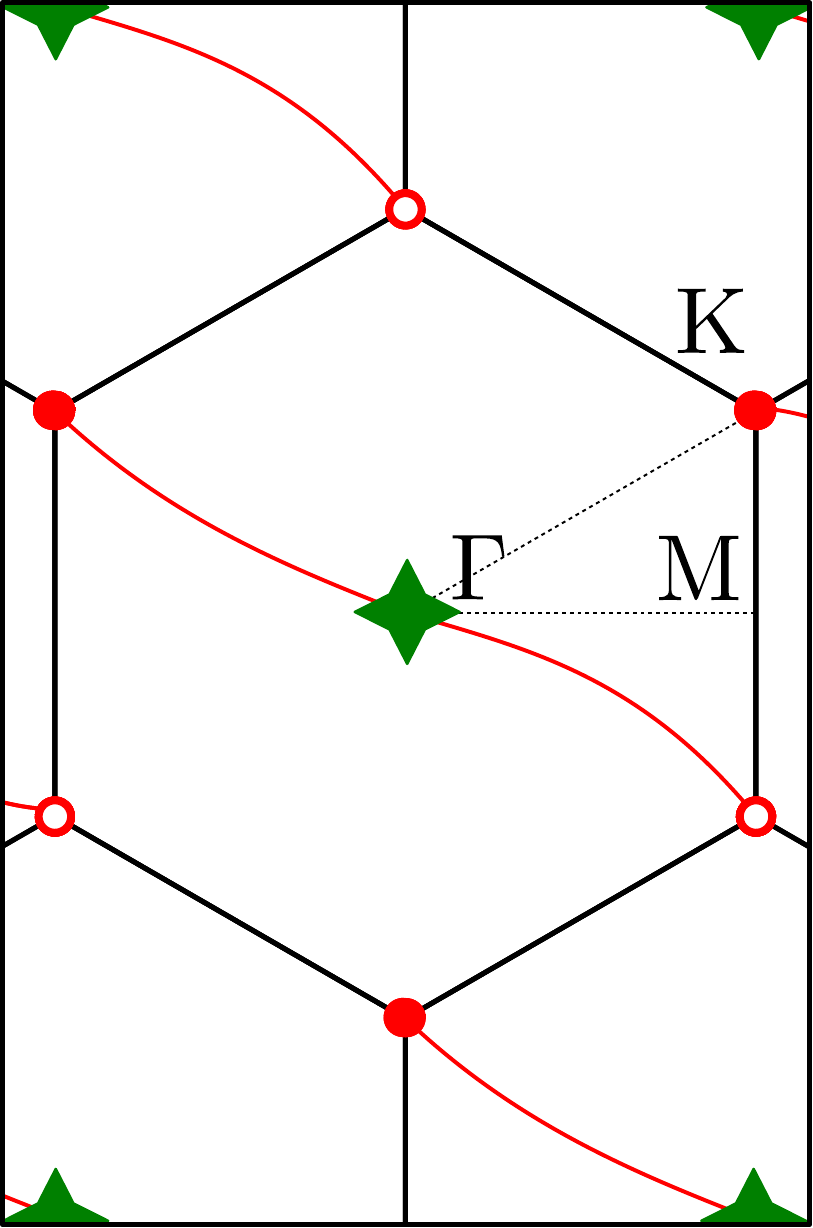} &
	\includegraphics[height=0.24\linewidth]{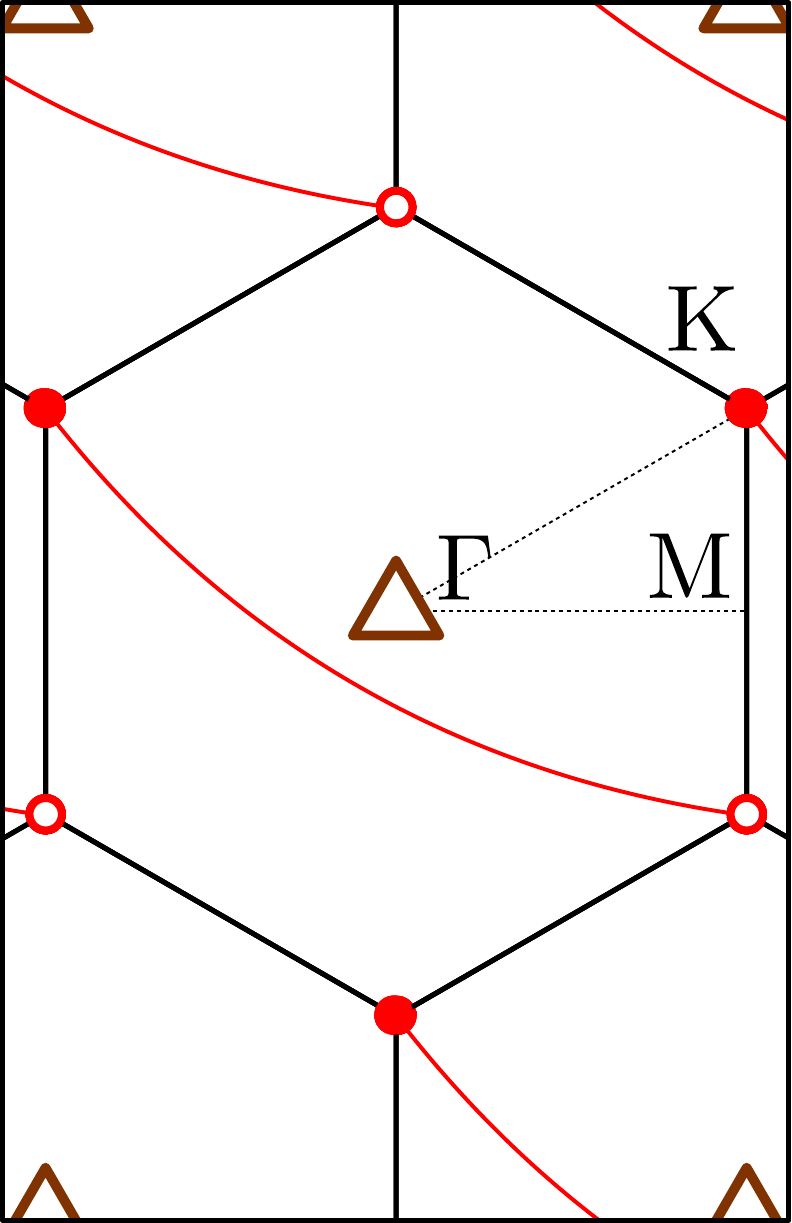}   \\
	\hline
	\end{tabular}
\caption{{\it Continuation of the braiding process of Fig.~\ref{BS_SIs}, here going through the transfer of the double node at $\Gamma$ (with non-trivial Euler class) from gap I to gap II, with the appearance of a triply degenerate point.
} }
\label{BS_SIs_G}
\end{figure*}

\subsection{Pumping of a double node from gap II to gap I and triply degenerate point at $\Gamma$}

In addition to the braiding process of Fig.~2 (Fig.~\ref{BS_SIs}), we consider the continuation of the braiding process with the transfer of the double node at $\Gamma$ from gap II to gap I. As shown in Fig.~\ref{BS_SIs_G} following the process of the main text we may indeed further tune the parameters. Taking opposite signs between the NN and NNN hopping parameters, and upon increasing $t$ and decreasing $t'$, we see that a characteristic triple band node is formed at $\Gamma$. First, the transfer of simple nodes in gap II from $K^{(\prime)}$ to $\Gamma$, and of the simple nodes in gap I to $\Gamma$ as well (Fig.~\ref{BS_SIs_G}{\bf a}) leads to the triple point at $\Gamma$ (Fig.~\ref{BS_SIs_G}{\bf b}). Evolving the parameters further we see the triple node gaps and the double node degeneracy, quantified by a finite patch Euler class, now resides in the {\it first} gap (Fig.~\ref{BS_SIs_G}{\bf c}). This process is readily obtained by symmetry through a band inversion at $\Gamma$ between the IRRPEs $\Gamma_1$ and $\Gamma_5$ (Fig.~\ref{BS_SIs_G}{\bf a-b-c}). Remarkably, we can characterize the triple point at $\Gamma$ (Fig.~\ref{BS_SIs_G}{\bf b}), that must necessarily occur upon the band inversion, in terms of its frame charge. 

Given the nontrivial patch Euler class of the double node at $\Gamma$ before and after the band inversion, the corresponding frame charge is $-1$. Remarkably, the dispersion of the triple point directly reflects the non-trivial frame charge $-1$, i.e.~there must be two bands from the three with a quadratic dispersion (see the band structure of Fig.~\ref{BS_SIs_G}{\bf b}). This must be contrasted with the dispersion of the triple point at $K^{(\prime)}$ in Fig.~\ref{BS_SIs}{\bf c)} with a frame charge of $\pm k$ where two of the bands must be linear instead.   

We thus conclude the intrinsic relationship between the dispersion at a triply degenerate point and the value of its frame charge, in the same way as the patch Euler class dictates the dispersion of a pair of nodes between two bands.


\section{Construction principles of the kagome acoustic metamaterials}
\label{SM:construction}
The kagome acoustic metamaterials are constructed using coupled acoustic resonators. We start from the basic element of the construction, two identical, cylindrical resonators connected with an air tube in between them (see Fig. 3A in the main text). The associated effective Hamiltonian (in unit of frequency) is
\begin{equation}
    H = \left ( \begin{array}{cc}
         \omega_0 & \kappa \\
          \kappa & \omega_0 \\
    \end{array} \right)
\end{equation}
Here, $\omega_0$ is the frequency of the lowest resonant modes (s-like acoustic modes) in the two identical resonators, and $\kappa$ represents a real-valued coupling coefficient between them. The eigen-frequencies of the above Hamiltonian are $\omega_0\pm \kappa$ associated with the two hybridized modes. These are the even and odd modes described by the eigenvectors $1/\sqrt{2}(1, \pm 1)^T$, respectively. We use the signs $+$ and $-$ to represent the even and odd modes, respectively, in analog to the bonding and anti-bonding states in electronic systems. Therefore, a positive (negative) coupling coefficient $\kappa$ indicates that the eigen-frequency of the even mode $f_{\rm even}$ is higher (lower) than that of the odd mode $f_{\rm odd}$. Quantitatively, $\kappa$ is equal to half of the frequency difference, i.e., $\kappa=\frac{1}{2}(f_{\rm even}-f_{\rm odd})$. In our construction, the geometry parameters of the acoustic resonators are $D=19.2$~mm and $H=24$~mm for the inner space, while the thickness of the resin is about 2~mm. Such a design is to ensure that the frequency of the lowest $s$ mode is far away from the other resonant modes, such as the $p_z$ mode (7145.8Hz) and the doubly degenerate $p_x$-$p_y$ modes (10470Hz). Note that the lowest $s$ mode is very special. After hybridization, the even eigen-mode in the coupled two resonators system always has zero frequency and uniform acoustic pressure everywhere, as indicated by Fig. 3A in the main text. Nevertheless, such a zero-frequency even-mode can develop into plane wave like Bloch states in the lowest acoustic band. We find that as long as the $s$ mode is utilized, the coupling coefficient $\kappa$ between the resonators is always negative (as shown in Fig. 3B in the main text).

The kagome lattices are constructed by many coupled acoustic resonators. The kagome model studied in this work consists of coexisting nearest-neighbor (NN) and next-nearest-neighbor (NNN) couplings. We realize the NN couplings with horizontal tubes centered at $z=H/2$ and the NNN couplings with horizontal tubes either centered at $z=H/4$ or at $z=3H/4$. In this way, all the tubes do not cross each other and thus remains independent. It is worth noting that the coupling strengths depend only on the diameters and the lengths of the tubes, and independent of the z positions of the tubes. Once the lattice constant is fixed, the lengths of the tubes are determined. The NN and NNN couplings are then tuned by the diameters of the tubes. The quantitative dependences of the NN and NNN couplings on the diameters of the connecting tubes are shown in Fig. 3B in the man text which are obtained from numerical simulations. Here, we remark that although the acoustic bands cannot be quantitatively explained by a tight-binding picture~\cite{PhysRevLett.81.1405}, the ability to tune the NN and NNN couplings at will through the diameters of the connecting tubes still enable us to reproduce all features of the non-Abelian reciprocal braiding and the associated band evolution for the lowest three acoustic bands.

\begin{figure*}
 \centering
 \includegraphics[width=0.9\linewidth]{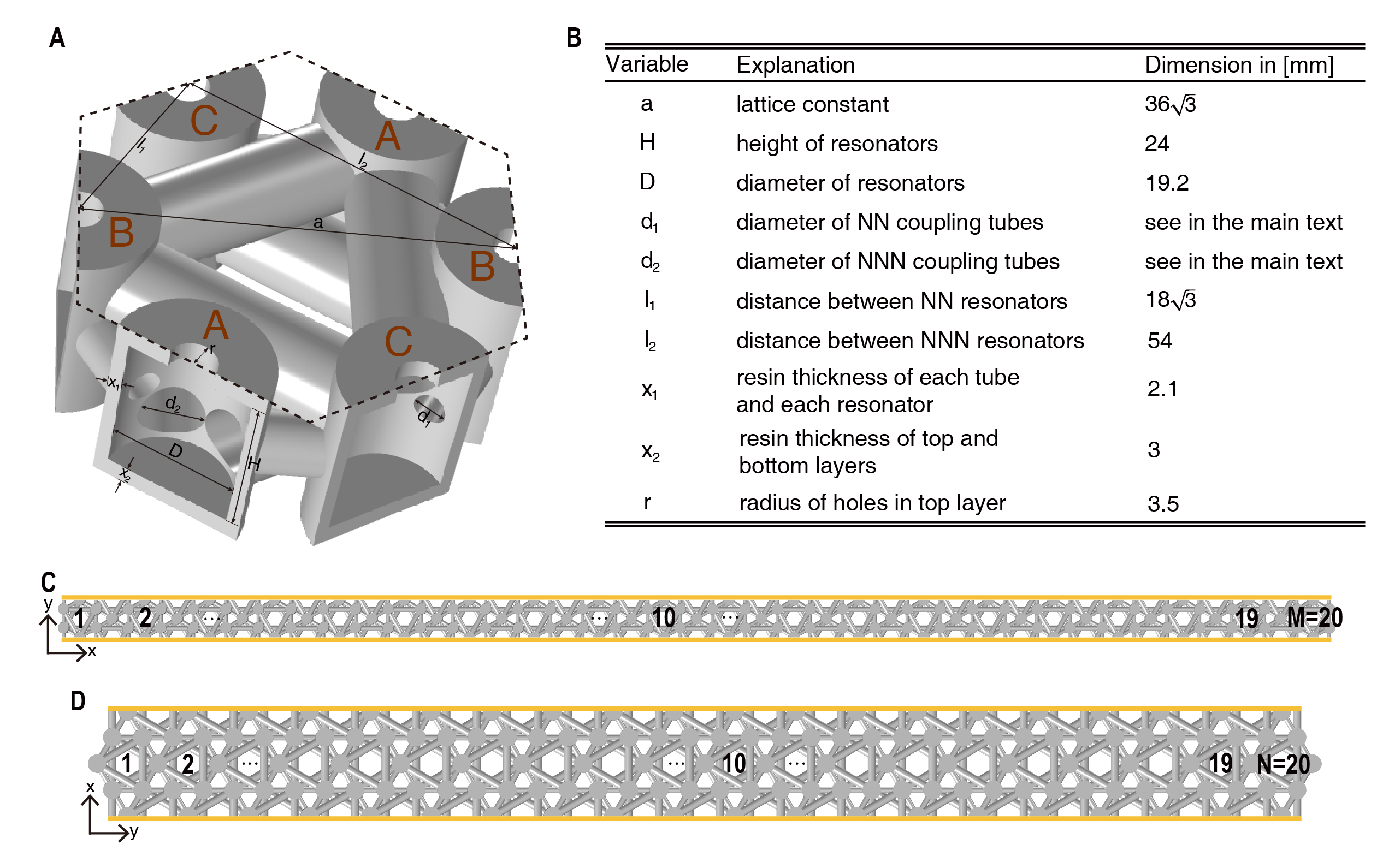}
 \caption{{\it The designed unit-cell structure. A, The unit-cell of one of the designed kagome acoustic metamaterial. Cylindrical acoustic resonators are located at the centers of the unit-cell boundaries (because of the periodicity of the structure, half cylinders are shown in the figure). The gray regions represent the photosensitive resin which is used in the 3D printing technology to fabricate the acoustic metamaterials. Horizontal tubes are used to connect the acoustic resonators in order to realize the NN and NNN couplings. B, The geometry parameters of the acoustic kagome metamaterials. The diameters of the horizontal tubes, $d_1$ and $d_2$, are given in Figs. 4 and 5 in the main text, and in Supplementary Fig.~\ref{fig:figs4} for the six samples. C, Ribbon-like structure with zigzag boundaries at the left and right sides. D, Ribbon-like structure with armchair boundaries at the left and right sides. In C and D, the gray color represents the resin structures. These structures are finite along the horizontal direction, whereas along the vertical direction, periodic boundary conditions are used (denoted by the golden lines above and below these structures). The lengths of these ribbon-like structures are described by the parameters M and N. In Fig. 5 in the main text and in the Supplementary Fig.~\ref{fig:figs6}, we take $M=20$ and $N=20$ in the simulation.}}
    \label{fig:figs1}
\end{figure*}

\section{Methods and samples}
\label{SM:methods}
All simulations of acoustic wave dynamics and calculations of acoustic energy bands were performed using the commercial finite-element solver COMSOL Multiphysics. The photosensitive resin used in 3D printing can be modelled as rigid walls for acoustic waves, considering its huge acoustic impedance contrast to the air background. The systems are filled with air in which the acoustic waves propagate (modeled with a mass density of 1.300 kg/m3 and the speed of sound of 343 m/s at room temperatures). In the metamaterials, the bulk band structures of the acoustic waves are calculated by solving the eigen-mode equation for acoustic waves in a single unit cell with Bloch boundary conditions in two-dimensional x-y plane. In comparison, the edge band structures are calculated by solving the same equation in ribbon-like structures with the zigzag or armchair boundary in one direction while periodic boundary condition in the other direction (see next section). Particularly, in our set-up, the zigzag boundary is along the y direction, while the armchair boundary is along the $x$ direction (see Fig. 3F in the main text).

The six experimental samples are all fabricated by commercial 3D printing technology based on photosensitive resin provided by a company in Shenzhen city. The fabrication precision is literally 0.1 mm. All samples are constructed using the coupled acoustic resonators. Each sample has 200 unit cells and has a geometry of $1050\times710\times 30$mm$^3$ along the $x$, $y$ and $z$ directions, separately. Each acoustic resonator and each connecting tube are encapsulated by photosensitive resin with a thickness of 2.1 mm. The top and bottom resin layers are enhanced to a thickness of 3 mm. A hole is open at the top of each resonator with a diameter of 7 mm for the detection of the acoustic wave amplitude in each resonator, $p_{j,\alpha}$ with $j$ and $\alpha$ labeling the unit cell and the type of sites (A, B or C), respectively. When detecting the acoustic wave amplitude in the j-th unit cell at the  site, we keep the top holes in all other resonators (except the resonators with the source and the detector) as closed by rubber plugs.

\begin{figure}
 \centering
 \includegraphics[width=\linewidth]{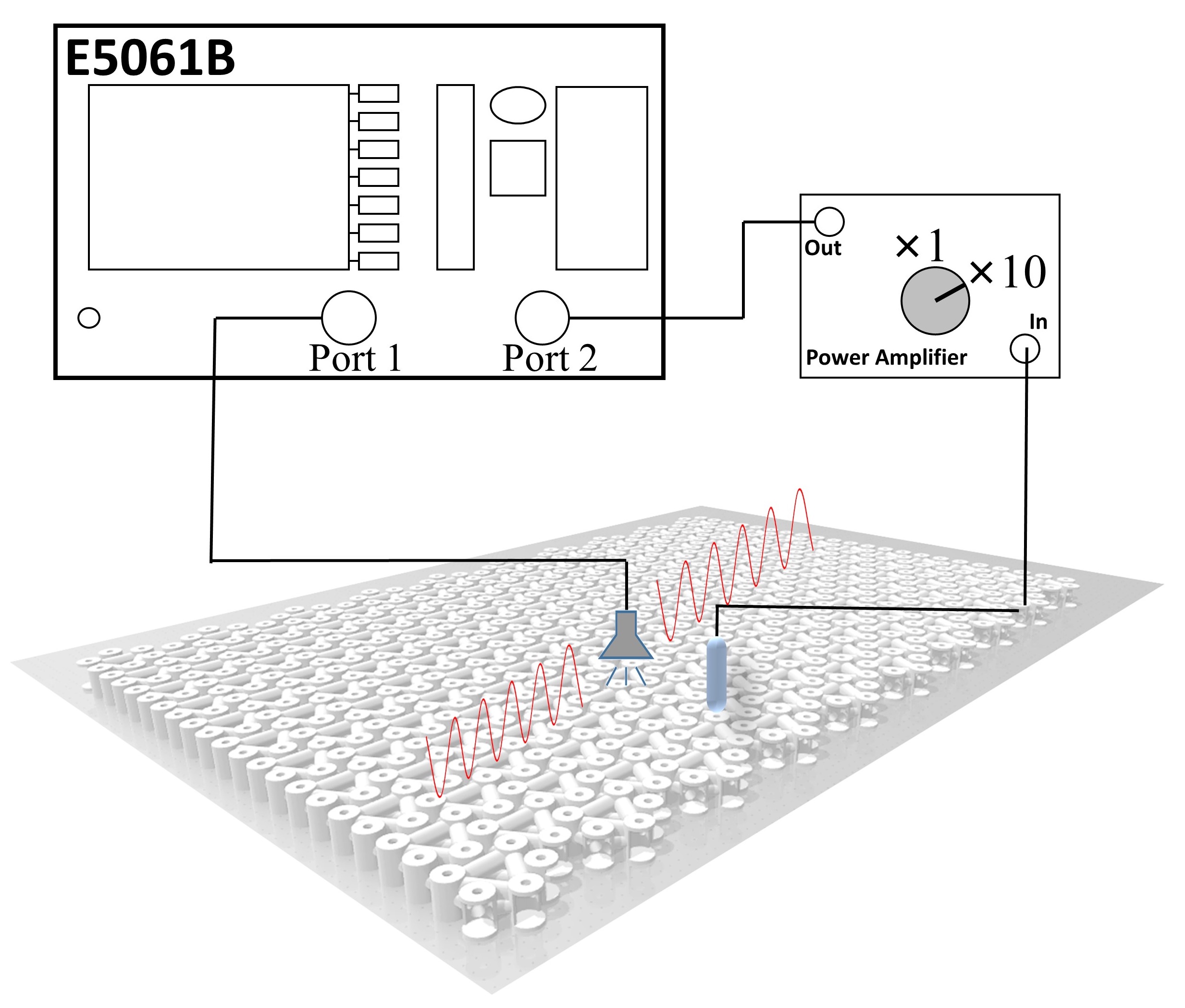}
 \caption{{\it Schematic illustration of the experiment setup. The measurement system includes a speaker as the acoustic source inserted into a resonator to excite the acoustic waves, a microphone as the acoustic detector connected to a power amplifier, and an Agilent network analyzer recording the acoustic signals and analyzing the detected signals at each frequency when the source frequency sweeps the from 5Hz to 3kHz.}
}
    \label{fig:figs2}
\end{figure}

To measure both the bulk and edge dispersions in the same sample, we create a zigzag boundary at the left edge of the sample and an armchair boundary at the upper edge, while other edge boundaries are left open for the bulk modes to leak out of the sample. To disconnect the zigzag and armchair edge boundaries, we also open the resonators at the corner between the two edge boundaries. Here, by ``open", we mean that the side surfaces of the resonators are half open (but the top and the bottom surfaces remain unchanged), so that the acoustic waves are able to propagate into the free space.

An acoustic source is placed into a resonator at the center of the sample for the excitation of the bulk modes. The acoustic wavefunction (technically, the acoustic pressure field) distributions are manually detected for each resonator through a portable probe microphone with a diameter about 7 mm (just fitting the size of the top hole in each resonator). The amplitude and phase distributions of the acoustic wavefunctions (precisely the acoustic pressure fields) are detected and then automatically recorded by an Agilent network analyzer (model: Keysight E5061B). By Fourier transforming the detected acoustic pressure distributions at each excitation frequency, we obtain the acoustic bulk band structure for each sample (the results are presented in Fig. 4 in main text). To measure the edge band structures for the zigzag (armchair) edge boundary, we place the acoustic source into a resonator in the middle of the zigzag (armchair) boundary.

\section{Details of the acoustic bulk band structure measurements}
\label{SM:bulk}
To increase the acoustic signal resolution, the probe microphone is connected to a power amplifier with 10 times amplification. A speaker as an acoustic source is inserted into a resonant cavity through the top hole for the excitation of the bulk states. The excited acoustic wave is then detected manually in each resonator using the probe microphone. Technically, the detected signal is the acoustic pressure in each resonator with both amplitude and phase signals. Therefore, for each frequency, we obtain a profile of the detected acoustic field. Some examples of the detected acoustic field profiles are shown in Supplementary Figure~\ref{fig:figs3}. Note that the measured acoustic field profiles are different when the acoustic source (labeled as the green star in the figures) is placed in the A, B or C sites in the same unit cell, since in kagome lattices, these sites are inequivalent. In particular, for the sixth sample which has $d_1=0$~mm and $d_2=9.6$~mm, the kagome lattice can be divided into two independent parts. The measured acoustic field profiles are quite different when the position of the acoustic source is different, as indicated by Supplementary Figs.~\ref{fig:figs3}D-F.

We then carefully check the experimental results by performing the measurements for each sample with the source in sites A, B and C in the central unit cell, i.e., we perform three independent measurements. We then carry out the Fourier transformation of the detected acoustic pressure profiles on the A, B and C sub-lattice sites and sum the intensity of the three Fourier transformation signals together. That is, from the detected real-space acoustic wavefunctions $p_{j,\alpha}$, we obtain the Fourier transformed signals, $g_\alpha({\bf k})=FFT(p_{j,\alpha})$, using the fast Fourier transformation (FFT) function in Matlab. The quantity presented in Supplementary Fig.~\ref{fig:figs4} is $P({\bf k})=\sum_\alpha |g_\alpha({\bf k})|^2$. Specifically, we present for each sample the obtained $P({\bf k})$ for three different positions of the acoustic source: with the source in sites C (left column), B (middle column) and A (right column) in the central unit cell. The obtained $P({\bf k})$'s are slightly different for the three source positions. For some samples, such differences (essentially due to the fact that the A, B and C sites are inequivalent in kagome lattices) are more noticeable, which may be due to the properties of the bulk eigen-states of the measured energy bands. Nevertheless, the overall results show excellent consistency with each other and with the calculated bulk band structure (denoted by the white curves). For each sample, we adopt the $P({\bf k})$ which has the best spectral resolution and/or the best consistency with the calculated acoustic bulk band structures, since each $P({\bf k})$ is a legitimated experimental representation of the acoustic bulk band structure for one sample. The adopted $P({\bf k})$ in Fig. 4 in the main text are labeled by the dashed boxes in Supplementary Fig.~\ref{fig:figs4} for each of the six fabricated samples (Each sample corresponds to one row in the figure). Finally, we remark that because of the low excitation efficiency of the speaker at low frequencies (limited mainly by the size effect), the detected acoustic signals at low frequencies are much weaker and so are the Fourier transformed signals $P({\bf k})$ at low frequencies, as indicated by Fig. 4 in the main text and Supplementary Fig.\ref{fig:figs4}.

\begin{figure}
 \centering
 \includegraphics[width=\linewidth]{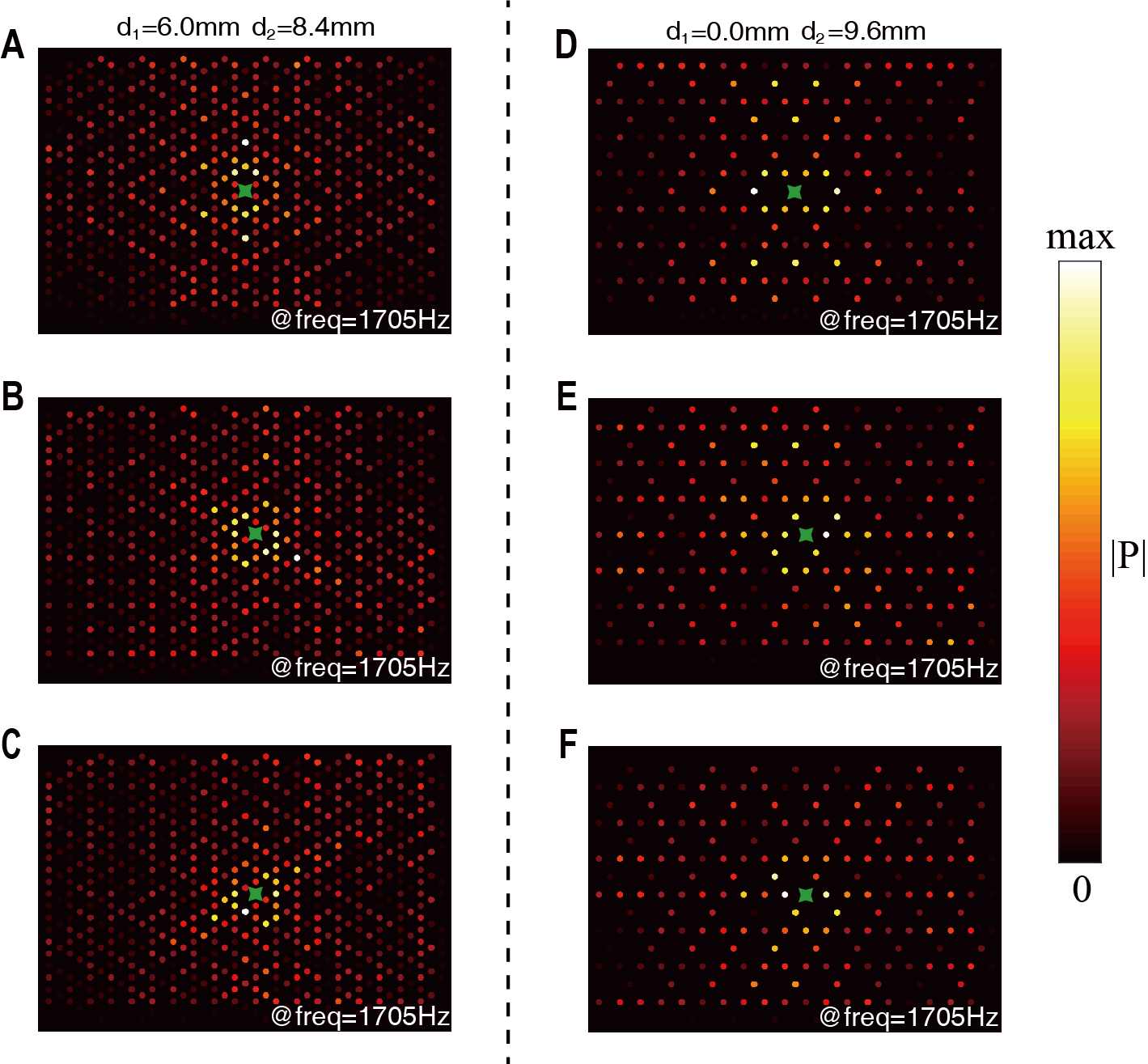}
 \caption{ {\it Examples of the measured acoustic pressure profiles in the bulk. A-C, Measured acoustic pressure profiles of the fifth kagome metamaterial (corresponding to Fig. 4E in the main text). D-F, Measured acoustic pressure profiles of the sixth kagome metamaterial (corresponding to Fig. 4F in the main text). The excitation frequency, 1705 Hz, is labeled in each figure. The sample parameters, $d_1$ and $d_2$, are given at the top of the Figures D and F. The first row (i.e., A and D): the detected acoustic signal profiles when the source is placed at the A site. The second row (i.e., B and E): the detected acoustic signal profiles when the source is placed at the B site. The third row (i.e., C and F): the detected acoustic signal profiles when the source is placed at the C site. The color scale represents the absolution value of the acoustic pressure.}
}
    \label{fig:figs3}
\end{figure}

\begin{figure*}
 \centering
 \includegraphics[width=0.7\linewidth]{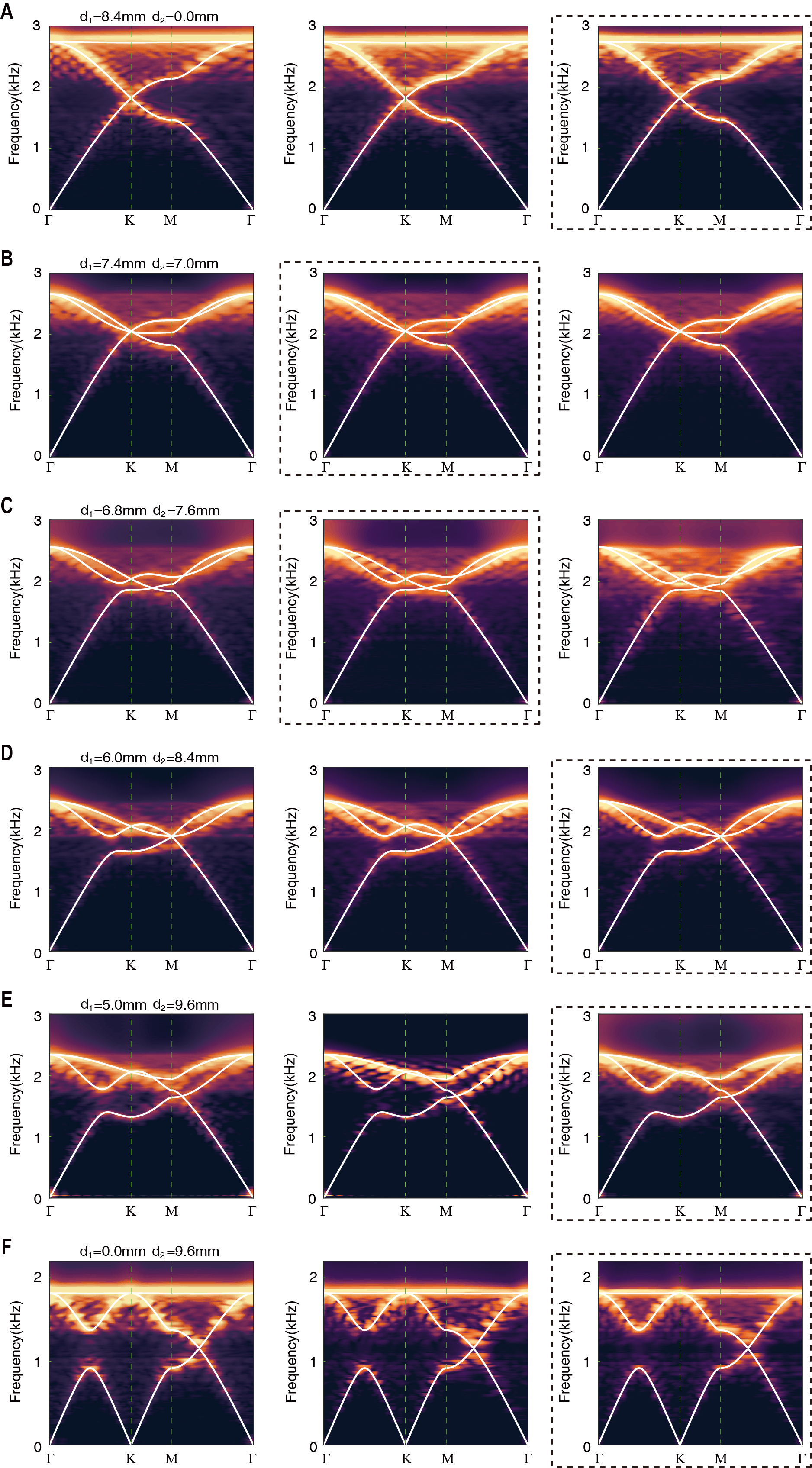}
 \caption{ {\it Experimentally measured acoustic bulk bands. Each row corresponds to a sample. The left, middle and right columns correspond to the cases with the acoustic source placed at the C, B and A sites, respectively. The sample parameters, $d_1$ and $d_2$, are given at the top of the figures in the left column. Figures with a dashed box are selected into Fig. 4 in the main text.}
}
\label{fig:figs4}
\end{figure*}

\section{Details of the acoustic edge dispersion measurements}
\label{SM:edge}
To measure the acoustic edge dispersions, we insert the acoustic source (i.e., the speaker) into a resonator in the middle of the zigzag (armchair) boundary and detect the acoustic pressure profile along the zigzag (armchair) edge. Examples of the measured acoustic pressure profiles along the edges for each sample are presented in Supplementary Fig.~\ref{fig:figs5} for both the armchair edges (Figs. 5A-F) and the zigzag edges (Figs. 5G-L).

\begin{figure*}
 \centering
 \includegraphics[width=\linewidth]{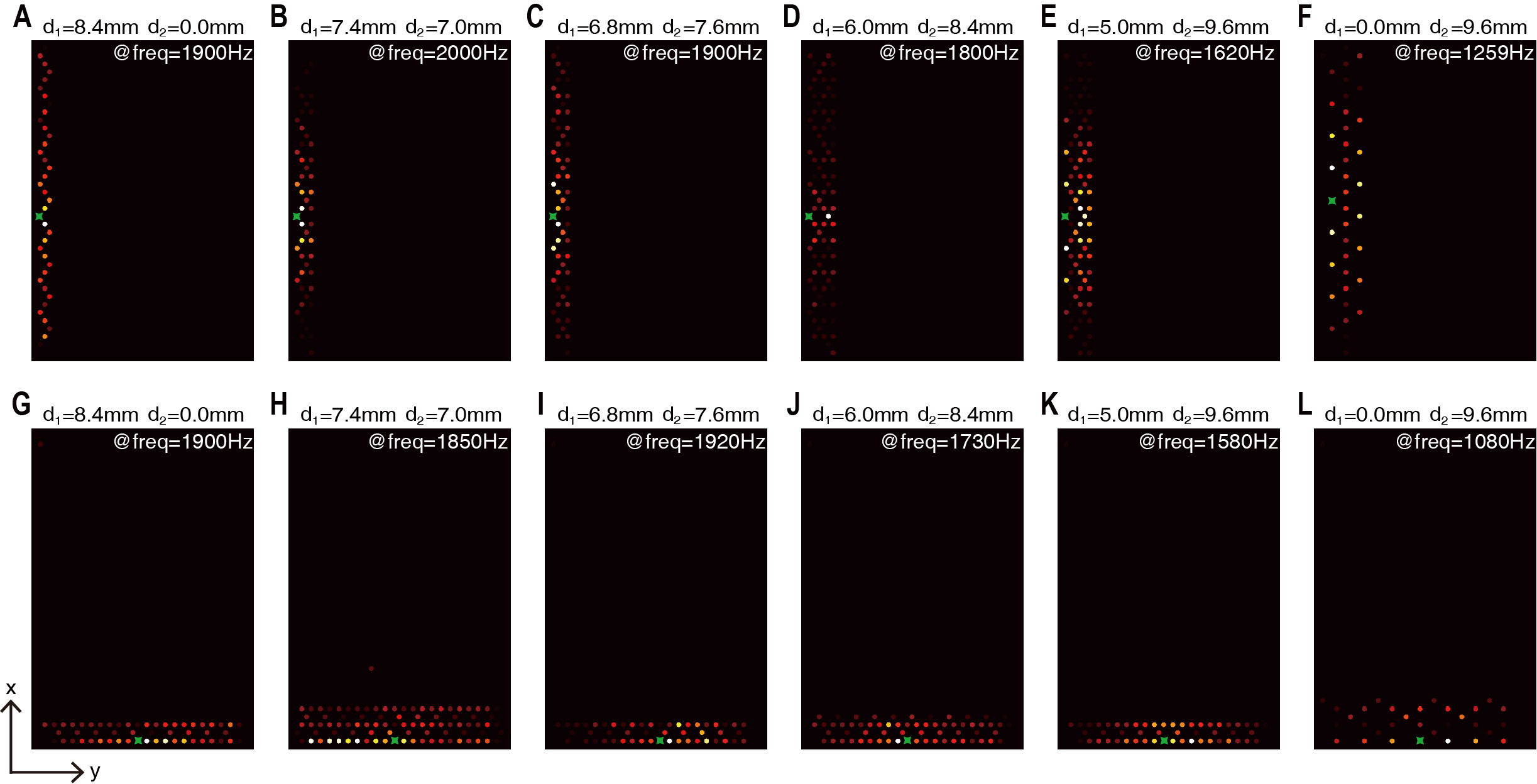}
 \caption{ {\it Observation of topological edge states in finite-sized acoustic kagome metamaterials. A-F: The measured acoustic pressure profiles along the armchair edge when the source (labeled by the green star in each figure) is placed at the middle of the armchair boundary. G-L: The measured acoustic pressure profiles along the armchair edge when the source (labeled by the green star in each figure) is placed at the middle of the zigzag boundary. The geometry parameters of the sample and the excitation frequency of the source are given at the top of each figure. Acoustic pressure profiles at a few layers of resonators near the edge boundaries are measured in the experiments, which is found to be sufficient to extract the acoustic edge dispersions.}
}
    \label{fig:figs5}
\end{figure*}

In the main text, we present the measured acoustic dispersions for the zigzag edge boundaries in Fig. 5. We focus on the edge states in gap I since it is much larger than gap II, and the edge states in gap I is more accessible in experiments in most of the six acoustic kagome metamaterials. In addition, we show in the main text that the topological edge states demonstrate the bulk-edge correspondence as dictated by the band nodes and the Dirac strings. In particular, for the wavevector regions with an even number of projected Dirac strings, topological edge states emerge. In contrast, if there are an odd number of projected Dirac strings, there is no edge state.

Here, we present the measured acoustic edge dispersions in Supplementary Fig.~\ref{fig:figs6} and discuss the underlying mechanisms and the bulk-edge correspondence. As shown in Supplementary Fig.~\ref{fig:figs6}A, for the armchair boundary, the Dirac nodes at both the $K$ and $K^\prime$ points are projected into the $k_x=0$ point. According to the bulk-edge correspondence, topological edge states emerge for all $k_x$. This is essentially because the overlapped projection of the Dirac nodes does not change the Zak phase. Besides, the Dirac strings are projected even times onto the $k_x$ axis. These observations lead to the conclusion that there should be edge states for all $k_x$. Similar conclusions can be arrived at for Supplementary Figs.~\ref{fig:figs6}B-D, where edge states are found for all $k_x$. In Supplementary Fig.~\ref{fig:figs6}E, according to the same bulk-edge correspondence, we find that the edge states emerge for large positive and negative $k_x$’s. However, for the sample in Supplementary Fig.~\ref{fig:figs6}F, because the NN coupling vanishes, the system breaks into two independent parts. Each of them can be regarded as a triangular lattice. Therefore, additional trivial boundary states can be triggered for small $|k_x|$ between the projections of the Dirac nodes.

For all six samples, we find consistency between the calculated and measured edge dispersions. The slight deviations between the experimental data and the theoretical curves are probably due to the geometry imperfection of the resonators near the edge boundaries in some samples due to uncontrollable fabrication errors.

\begin{figure*}
 \centering
 \includegraphics[width=\linewidth]{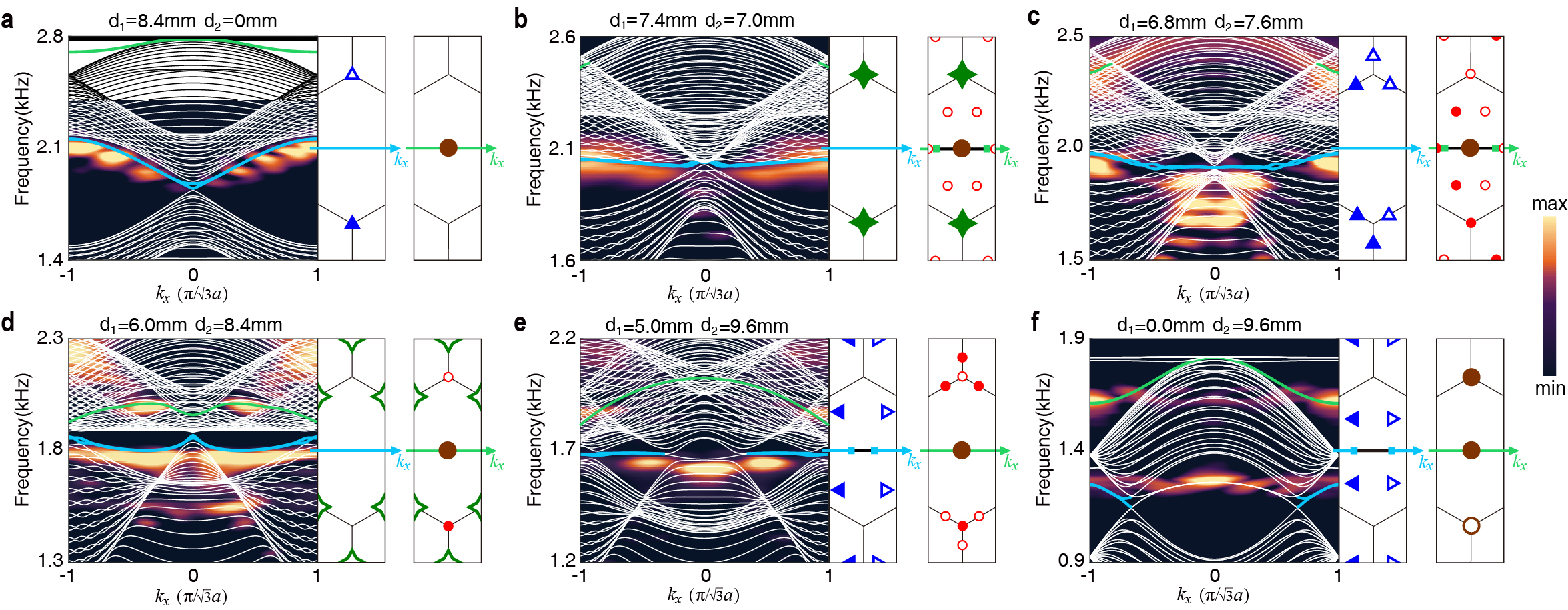}
 \caption{{\it Measured acoustic edge dispersions along the armchair boundary. A-F: Acoustic dispersions measured along the armchair boundary. White curves: calculated bulk dispersions. Green curves: calculated edge dispersions in gap I. The right-side panel in each figure depicts the band nodes and the Dirac strings in gap I. The projection of the band nodes on the  axis is shown by the red squares. The red segments on the  axis denote the regions with nontrivial Zak phase that predicts the emergence of the topological edge states.}
}
    \label{fig:figs6}
\end{figure*}


\end{document}